\documentclass[aps,prx,twocolumn,superscriptaddress,floatfix]{revtex4-2}
\usepackage{amsmath,amssymb,graphicx}
\usepackage{amsthm}
\usepackage{float}
\usepackage{enumitem}
\usepackage{braket}
\usepackage{hyperref}  
\usepackage{cleveref}

\usepackage{graphicx}
\usepackage{booktabs} 
\usepackage{array}
\usepackage{makecell} 

\usepackage{algorithm}
\usepackage[noend]{algpseudocode}

\crefname{algorithm}{algorithm}{algorithms}
\Crefname{algorithm}{Algorithm}{Algorithms}

\newtheorem*{definition}{Definition}
\newtheorem{theorem}{Theorem}
\newtheorem{lemma}{Lemma}
\newtheorem{proposition}{Proposition}

\usepackage{tcolorbox}

\usepackage{enumitem}

\newcommand{\classname}[1]{\texttt{#1}}

\newcommand{\cnot}[1][cnot]{\textsc{#1}}
\newcommand{\cz}[1][cz]{\textsc{#1}}
\newcommand{\cnots}[1][cnots]{\textsc{#1}}

\makeatletter
\newenvironment{proofof}[1]{%
  \let\oldproof\proof
  \let\oldendproof\endproof
  \renewcommand{\proofname}{Proof of \cref{#1}}%
  \oldproof
}{\oldendproof}
\makeatother

\usepackage[normalem]{ulem}

\usepackage{xcolor}

\begin{document}
\title{Cost-aware Photonic Graph State Generation: A Graphical Framework}
\author{Sobhan Ghanbari}
\affiliation{Department of Physics, University of Toronto, 60 St George St., Toronto, ON, Canada}
\affiliation{Quantum Bridge Technologies Inc., 108 College St., Toronto, ON, Canada}
\author{Hoi-Kwong Lo}
\affiliation{Quantum Bridge Technologies Inc., 108 College St., Toronto, ON, Canada}
\affiliation{Department of Physics, National University of Singapore, 2 Science Drive 3, Singapore
117551, Singapore}
\affiliation{Department of Electrical and Computer Engineering, University of Toronto, 10 King's College Road, Toronto, ON, Canada}

\begin{abstract}
Photonic graph states are essential resources for quantum computation and communication. Deterministic emitter-based generation of graph states overcomes the scalability issues of probabilistic approaches; nonetheless, their experimental realization is constrained by technological demands, often expressed by the number of two-qubit gates and the depth and/or width of the quantum circuits used to model the generation process.
Here, we introduce a cost-aware framework for the generation of photonic graph states of arbitrary size and shape, built on a complete set of necessary and sufficient conditions and a universal set of elementary graph operations that govern the evolution of the state toward the target. Within this framework, we develop \textit{Graph Builder}, a deterministic generation algorithm that achieves substantial reductions---up to an order of magnitude---in two-qubit gate usage for both random and structured graphs, compared with alternative approaches. \textcolor{black}{Furthermore, we show that this framework enables the identification of elementary building blocks in specific cases, such as encoded 6-ring states.} The algorithm uses the minimum number of emitters possible for a fixed emission sequence, while also supporting the use of extra emitters for controlled trade-offs between emitter count and other cost metrics. Moreover, by systematically identifying the degrees of freedom at each stage of the generation process, this framework fully characterizes the optimization landscape, enabling analytic, heuristic, or exhaustive strategies for further cost reductions.
Our approach provides a general and versatile tool for designing and optimizing emitter-based photonic graph state generation protocols, essential for scalable and resource-efficient photonic quantum information processing.
\end{abstract}

\maketitle
\section{Introduction}
Photonic graph states \cite{briegel_persistent_2001, hein_multiparty_2004} are gaining increasing attention as promising candidates for realizing fault-tolerant measurement based \cite{raussendorf_one-way_2001,Raussendorf-measurement-based-2003, briegel_measurement-based_2009}, or fusion based \cite{bartolucci_fusion-based_2023}, quantum computing platforms as well as their use as quantum repeaters in quantum networks, QKD, and entanglement distribution systems \cite{zwerger_measurement-based_2012,zwerger_universal_2013,zwerger_measurement-based_2016,azuma_all-photonic_2015, Azuma2023}. The error and loss tolerance of such platforms depends heavily on the architecture of the entangled states that are to be consumed (measured) for quantum information processing. Additionally, various logical qubit encoding strategies have been proposed to increase the fault-tolerant thresholds with respect to different system parameters such as the probability of photon loss \cite{Chan2025, Song2024, Bell2023, cesa_hierarchical_2025}. This has led to requiring larger, more complex graphs \cite{Schlingemann_2001, Pettersson2025}, with increasingly challenging preparation processes. Advancing beyond the current state of the art in photonic graph state generation requires---in addition to improved hardware---more refined theoretical algorithms in order to keep the generation requirements within the technological constraints of the NISQ era devices and to make resource-efficient options available.

Particularly, what separates the photonic graph state generation problem from a non-photonic case is the non-interacting nature of photons in linear optics. This leads to the absence of deterministic entangling gates \cite{knill_scheme_2001, browne_resource-efficient_2005}, which are a main requirement for establishing the necessary entanglement edges between qubits in a graph state. To overcome this issue, two general categories of methods have been proposed: probabilistic and deterministic. In probabilistic methods, small entangled states, such as Bell pairs or GHZ states, are merged together using non-deterministic fusion gates \cite{browne_resource-efficient_2005, grice_arbitrarily_2011}. Such fusion operations are repeated until the target shape is acquired. The resource states used in such probabilistic generation methods can be obtained, for instance, using nonlinear down-conversion sources \cite{kwiat_new_1995}, the biexciton decay cascade in quantum dots \cite{akopian_entangled_2006}, or single-photon sources employed within probabilistic interferometric setups to produce EPR pairs or GHZ states \cite{varnava_how_2008, zhang_demonstration_2008}. Scalability is the main issue with the probabilistic approach. In particular, the non-unity success probability of a fusion operation \cite{ewert_34-efficient_2014} means the overall chance of obtaining a graph state decreases exponentially with its size; therefore, significant multiplexing would be necessary for a practical generation rate. 

The idea of emitter-based generation of photonic graph states \cite{schon_sequential_2005} where each photon is emitted entangled to its emitter was proposed to address the scalability challenges. In such systems, the emitters (matter qubits) are used as mediators to create and maintain entanglement between emitted photons without relying on direct photon-photon interactions, thereby enabling a deterministic process. The deterministic nature of this process leads to significant reductions in resource overhead compared to the probabilistic generation methods \cite{pant_rate-distance_2017, li_resource_2015}. As a result, the deterministic approach has gained increasing attention in recent years, with quantum dots \cite{schwartz_deterministic_2016, cogan_deterministic_2023}, neutral atoms \cite{yang_sequential_2022, thomas_efficient_2022, Aqua2025}, and ions \cite{Inlek2017, Drmota2023} as potential emitter platforms. Various generation schemes have been proposed and demonstrated for simple states such as GHZ or linear cluster states \cite{lindner_proposal_2009}, and later generalized to accommodate other types of graphs \cite{economou_optically_2010, russo_generation_2019, huet_deterministic_2025}. The emitter-based generation recipe for a target graph is commonly represented as a quantum circuit composed of emitter and photonic registers. The circuit consists of a sequence of quantum operations applied to the emitter qubits between photon emission events, such that the desired target state is ultimately established on the photonic registers. For instance, an obvious generation recipe for any graph is to use as many emitters as there are photons in the graph, apply deterministic entangling gates between the emitter qubits to establish the target entanglement structure on them, and then emit one photon from each corresponding emitter to transfer the entanglement pattern to the photons.

In practice, however, realizing such deterministic circuits is not without challenge. One of the key requirements is the availability of an array of emitter qubits on which quantum logic gates, including two-qubit interactions, can be reliably applied. Yet, implementing each of these two-qubit operations, typically via electronically coupled emitters \cite{carlson_theory_2019,qiao_coherent_2020}, waveguide or cavity-assisted interactions \cite{hurst_generating_2019,chu_independent_2023,wei_cavity-assisted_2025}, or photon-mediated entanglement swapping \cite{kaur_resource-efficient_2024,dhara_entangling_2023, levonian_optical_2022}, remains a significant challenge considering the gate fidelity and duration, compared to single qubit gates even for one pair of emitters. Besides, the limited coherence time of the emitter qubits constrains both the number of gates that can be applied on each emitter and the number of photons that can be emitted into a graph state by each emitter before decoherence renders the emitter ineffective as a mediator qubit. Consequently, a generation recipe should be designed with practical limitations in mind, preferably using the least possible number of emitters and minimizing the number of gates applied on and between them. The quantum circuit picture offers a suitable platform for quantifying the generation cost, which can be expressed, for instance, in terms of parameters such as circuit depth and the number of two-qubit gates, making it a useful tool for benchmarking generation strategies.

We note that for the same target graph state, distinct generation circuits may exist that potentially differ in cost. This non-uniqueness gives rise to an extensive optimization landscape, and finding optimal generation recipes is generally non-trivial.
Nevertheless, identifying resource-efficient solutions is crucial, as experimental and technological limitations significantly constrain the size, generation rate, and types of achievable graph states \cite{schwartz_deterministic_2016, cogan_deterministic_2023,yang_sequential_2022, huet_deterministic_2025}. Previous optimization attempts include minimizing the required number of emitters \cite{li_photonic_2022}, minimizing the circuit depth \cite{kaur_resource-efficient_2024}, and decreasing the required number of two-qubit gates between emitter qubits \cite{ghanbari_optimization_2023, Takou2024, ren2025} in generation circuits of arbitrary graphs. Numerous studies have also focused on finding optimized circuits of specific graphs with certain use cases such as quantum communications \cite{buterakos_deterministic_2017, cesa_hierarchical_2025} or fusion based quantum computing \cite{wein_minimizing_2024, Chan2025}. Despite these efforts, the optimization of the generation circuit remains an open problem. The solution is heavily dependent on the algorithm used to find a circuit for the target state and the cost metrics considered, with current approaches mostly relying on heuristic or brute-force search strategies. Importantly, the degrees of freedom available for optimization have yet to be systematically characterized, and there is clear room for exploring, formalizing, and utilizing such flexibilities.

In this work, we introduce a new approach to photonic graph state generation based on a cost-aware framework that leverages structural and physical constraints in the inherently sequential generation process of photonic states \footnote{As photonic qubits do not exist prior to their emission, photonic graph state generation is a sequential process, where photons are created one by one to represent the nodes of the target graph \cite{schon_sequential_2005}. If multiple photons are emitted simultaneously or in parallel using multiple emitters, an arbitrary ordering can still be assigned to define the structure of the graph state.}. First, we identify a set of necessary and sufficient conditions that must be satisfied for the system of emitter and photonic qubits, at each intermediate stage of the generation, to ensure that the desired target state can be reached from that stage. These eligibility conditions dictate the requirements for the entanglement structure among qubits at each step and, in doing so, provide guidelines for selecting operations that prepare the system for the emission of the next photon. This not only allows finding a generation circuit but also reveals the available degrees of freedom in constructing such circuits.

To formalize this process, we develop a graphical framework built on a proposed set of elementary graph operations. The chosen elementary set is cost-efficient---particularly in minimizing two-qubit gate usage---and universal for emitter-based photonic graph generation, as any transformation achievable by the allowed quantum operations on the qubits can be realized using this set.
The graphical framework enables us to track the full evolution of the quantum state during generation, where each intermediate state is a graph obtained from its predecessor via elementary graph operations. This framework, in conjunction with the set of eligibility conditions, allows for a systematic characterization of the degrees of freedom in identifying the next operation in the generation process.  Specifically, at each intermediate step, there can be multiple eligible graphs that can evolve into the target shape, offering freedom in selecting the shape for the next intermediate graph. Once selected, this graph can be obtained in various ways from the preceding graph using the allowed graph operations, adding more degrees of freedom to the process.

Building on these tools, we present \textit{Graph Builder}, an algorithm for generating arbitrary photonic graph states using deterministic emitter-based platforms. The algorithm achieves substantial reductions in the use of two-qubit gates for preparing random graphs, achieving average (maximum) reductions of up to 61\% (80\%) and 52\% (70\%) compared to the two alternative methods presented in Refs.~\cite{GraphiQ} and \cite{Takou2024} (both based on the algorithm introduced by Ref.~\cite{li_photonic_2022}), respectively. We also show consistent improvements in the use of two-qubit gates in preparation of structured graphs, such as Raussendorf--Harrington--Goyal lattices \cite{Raussendorf-fault-2006, Raussendorf-topological-2007}, tree graphs \cite{Varnava-loss-2006}, and parity-encoded 6-ring states \cite{bartolucci_fusion-based_2023}, \textcolor{black}{for which we also identify the explicit elementary generation circuit blocks.} We remark that the reported improvements arise from the inherent cost-awareness built into the algorithm's design, enabled by the new framework, rather than from dedicated heuristic or exhaustive optimization efforts. In fact, further cost reductions are possible by leveraging the relevant degrees of freedom available in the algorithm. As a key feature, the introduced algorithm enables a modular optimization framework in which explicit cost functions can be assigned to each decision point whenever there is freedom in the choice of operations. This facilitates the implementation of analytic, exhaustive, or heuristic optimization strategies that can be applied on top of the base generation algorithm to obtain further improvements. The usefulness of such secondary optimizations has been demonstrated in previous works \cite{ghanbari_optimization_2023, Takou2024, ren2025}.

Another important advantage of the proposed algorithm is that, being embedded in a fully graphical framework, it directly outputs a generation recipe as a sequence of graph operations, which can then be transpiled into a quantum circuit. Notably, the intermediate graphical representation of the generation recipe allows us to identify simplification opportunities---such as merging or canceling of operations---before circuit translation, thereby avoiding the challenges of direct optimization on constrained quantum circuits. We demonstrate this by devising and employing such a simplification approach to reduce the redundancies in the final circuit.

Lastly, while our algorithm has the advantage of operating with the minimum number of emitters required to generate a target photonic graph state under a fixed emission order, this constraint can be relaxed. Prior work \cite{kaur_resource-efficient_2024} has shown that introducing additional emitters during generation can reduce other costs, such as gate count and circuit depth, which may be more relevant for certain experimental platforms. To support such trade-offs, we provide a straightforward generalization of our algorithm that accommodates extra emitters when the minimum-emitter constraint is lifted. \textcolor{black}{We study the effect of using extra emitters on different cost metrics and observe reductions of up to 72\% in circuit depth and 45\% in two-qubit gate count for random graphs.}

Here is the outline of the paper. Section II provides the preliminaries, including background on graph states, the photonic state generation problem, and local complementation in graphs. In Section III, we introduce the concept of \textit{generative set}, the necessary and sufficient eligibility conditions, and an inductive procedure for the generation. Section IV presents the graphical framework that forms the basis of our generation algorithm, including edge manipulation operations, emission modes, and qubit decoupling. The \textit{Graph Builder} algorithm and its output is described in Section V through an exhaustive case analysis for all possible generation scenarios. Section VI provides a performance analysis of two-qubit gate count across varying graph sizes for random graphs and specific states of interest. Section VII discusses an optimization framework, lists the associated degrees of freedom, and highlights how different strategies can be employed to reduce generation cost according to various metrics. Section VIII provides formal proofs for the necessity and sufficiency of the eligibility conditions. Finally, Section IX concludes with a summary and discussion of potential future directions, emphasizing the implications of our algorithmic approach for photonic graph state generation.

\begin{figure*}[t]
\centering
\includegraphics[width=\textwidth]{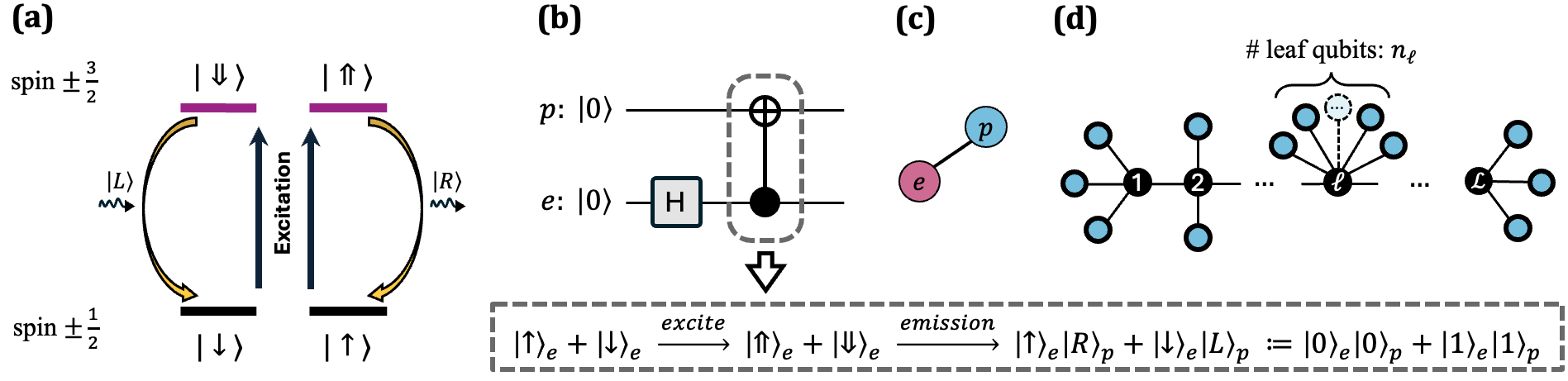}%
\caption{Generation of photonic graph states with a quantum emitter. \textbf{(a)} A proposed energy level structure \cite{lindner_proposal_2009}, made up of doubly degenerate ground and excited states of spin 1/2 and 3/2, respectively. The selection rules then allow the emitter's spin to become entangled with the polarization of the emitted photon \textbf{(b)} A quantum circuit representation of the process. Initially, the emitter and photonic qubits are considered to be in $\ket{0}$ state, a Hadamard gate on the emitter brings it to a superposition state and a \cnot[cnot] is used to model the emission process as described in the dashed box. \textbf{(c)} The outcome of the emission circuit in graph representation: the emitter and the photonic node connected with an edge. The emitter can continue emitting and grow the graph by adding photonic nodes to it. \textbf{(d)} Caterpillar graphs. A class of entangled states obtainable with a single quantum emitter and local operations. A general caterpillar state consists of a main path of $\mathcal{L}$ entangled qubits, each having an arbitrary number ($n_\ell$) of leaf qubits attached to them.}
\label{fig:0}
\end{figure*}

\section{Preliminaries}
\subsection{Graph States and Stabilizers}
A graph state \cite{briegel_persistent_2001, hein_multiparty_2004} is a multi-qubit quantum state with an entanglement structure characterized by a graph. For any graph $G=(V, E)$ consisting of a set of nodes $V=\{0, \dots, N-1\}$ and edges between them $E\subseteq\{(u,v)\in V\times V\mid u\neq v\}$, the quantum state can be obtained by initializing a set of qubits each corresponding to one of the nodes in a $\ket{+}$ state, followed by applying a controlled-Z (\cz[cz]) gate on any two nodes connected by an edge. 
\begin{align}\label{eq:graph_state}
    \ket{G} =\left( \prod_{(u, v) \in E} CZ_{uv} \right) \hspace{1mm} \ket{+}^{\otimes N}
\end{align}

Graph states are a subclass of stabilizer states \cite{GottesmanThesis}; each graph state can be uniquely defined as the common eigenstate of a stabilizer group with the eigenvalue of +1, where each stabilizer operator is an n-qubit Pauli operator. The generators of the corresponding stabilizer group can be found by looking into the adjacency matrix $(A)$ of the graph with elements $A_{ij}\in\{0,1\}$ showing the connectivity between the nodes $i$ and $j$. By considering row $i$ of this matrix and replacing the diagonal element $A_{ii}$ with a Pauli $X_i$ operator acting on qubit $i$ and any non-zero elements $A_{ij}$ with a Pauli $Z_j$, we obtain a multi-qubit Pauli operator for each row. This operator is one of the generators of the stabilizer group for the given graph:
\begin{align}
    g_i= X_i\prod^{N-1}_{j=0} Z^{A_{ij}}_j  
\end{align}
Having all $N$ generators for an $N$-qubit graph, the respective stabilizer state is uniquely defined. Therefore, all the required information about the entanglement structure between the nodes of the graph can be extracted from the adjacency matrix alone.

\subsection{The Photonic State Generation Problem}\label{subsec:generation_problem}
Deterministic entangling interactions, such as \cz[cz] gates, are not available for photonic graph state preparation in linear optical setups \cite{knill_scheme_2001, browne_resource-efficient_2005}. Therefore, the direct creation of edges between photonic qubits, as described in \cref{eq:graph_state}, is not a viable option. To address this, quantum emitters with an internal qubit, such as spin qubit in quantum dot excitations \cite{schwartz_deterministic_2016, cogan_deterministic_2023}, have been proposed to play the role of entanglement mediators. With an appropriate energy level structure \cite{lindner_proposal_2009} for the emitter (see Fig.~\ref{fig:0}), an emitted photon would remain entangled to its emitter \cite{lindner_proposal_2009}. Such a system, in conjunction with single qubit operations, enables the generation of a certain class of graphs known as caterpillar states \cite{caterpillar,huet_deterministic_2025} with a single emitter (see Fig.~\ref{fig:0}d). Graphs of arbitrary size and shape can be generated in a similar manner by utilizing a sufficiently large pool of interacting emitters \cite{schon_sequential_2005}. In such a setup, entanglement is first established deterministically between emitters and then transferred to photons through emission. However, it is important to note that quantum operations on emitter qubits---especially the two-qubit gates---are often experimentally challenging and more error-prone or time-consuming than photonic operations. Thus, while this deterministic approach avoids the need for direct photon-photon entanglement and offers broad flexibility, it is still preferable to minimize both the number of emitters and the use of such inter-emitter gates.

The emission of the target state is a sequential process, and necessary quantum operations must be applied to prepare the overall state---consisting of emitter and photonic qubits---before the emission of each new photon, adding one qubit to the state at a time. The operations on emitter qubits dynamically connect the photonic nodes together by creating the required emitter-emitter and emitter-photon entanglement edges during the generation process. A quantum circuit is commonly used to model this process, in which an emission corresponds to a \cnot[cnot] gate between an emitter (control qubit) and a non-emitted photonic register, initialized in $\ket{0}$, as the target \cite{lindner_proposal_2009} (Fig.~\ref{fig:0}b). Such a circuit representation offers a platform-agnostic benchmarking framework for quantifying generation cost and limitations, namely, the total number of emitters qubits used, the number of two-qubit gates between emitters, and circuit depth. These circuit parameters can be related to physical properties of the system, such as the coherence time of emitter qubits and quantum operation fidelities, thereby characterizing the resource requirements for the generation process.

Here, we establish a set of constraints to provide a clear definition for the problem of finding a generation recipe for graph states while accounting for physical limitations, such as non-interacting photonic qubits.
\begin{itemize}
[itemindent=9pt, leftmargin=0pt]  
    \item \textbf{Sequential generation:} This is a natural constraint for any photon emission platform. The photonic nodes are emitted over time, and one can always label the nodes of an $N$-qubit graph state from $0$ to $N-1$ corresponding to the order of emission.
    
    \textit{Note:} In cases where multiple emitters can emit in parallel and at the same time, the ordering of emission is generally not unique. However, it is always possible to assume an arbitrary sequence for the photons whose order can commute with one another.
    
    \item \textbf{Two-qubit connectivity:} To keep the circuit deterministic, we only allow multi-qubit operations on matter qubits (emitters) and keep the photonic gate set limited to single-qubit ones. Emitter-photonic two-qubit interactions are limited to the cases that represent emission events. 
    
    \item \textbf{Number of emitters:} We first restrict ourselves to using the minimum number of emitters necessary to generate the target state. Later, we show our framework is compatible with lifting this constraint and allows for more than minimal emitter cases to, for instance, exploit this extra degree of freedom to reduce circuit depth or the number of two-qubit gates.
    
    \item \textbf{Gate set:} We restrict the circuit to Clifford group operations, Hadamard (H), phase (P), and \cnot[cnot]/\cz[cz] gates, and single qubit measurements (M) in computational basis, as they are enough for generation of any arbitrary graph state \cite{schon_sequential_2005}. 

\end{itemize}
Note that while it is possible to emit a photon and measure it in a subsequent step, one does not need to consider such cases explicitly, as for any emitted photon not present in the output state, the corresponding qubit in the generation circuit is replaceable by an emitter qubit. This is possible because any gate on a photonic qubit---including its emission \cnot[cnot]---is applicable to an emitter as well. As a result, and without loss of generality, we consider every emitted photon to represent a qubit of the target state.

It is worth noting that the two-qubit connectivity constraint forbids photons from interacting with emitter qubits after emission. Although such interactions are possible in certain physical systems that employ quantum feedback delay lines in a generation scheme \cite{Pichler2017, Ferreira2024}, the range of graphs achievable in this way is limited to special cases such as 2D cluster states. Since our focus here is on generating graphs of arbitrary types and shapes, we do not include photon-emitter re-interactions in the present framework.

Our goal is to develop a method for finding algorithms that identify generation recipes for any given target graph under these constraints.

\subsection{Local Complementation}
\begin{figure}[b]
\centering
\includegraphics[width=\columnwidth]{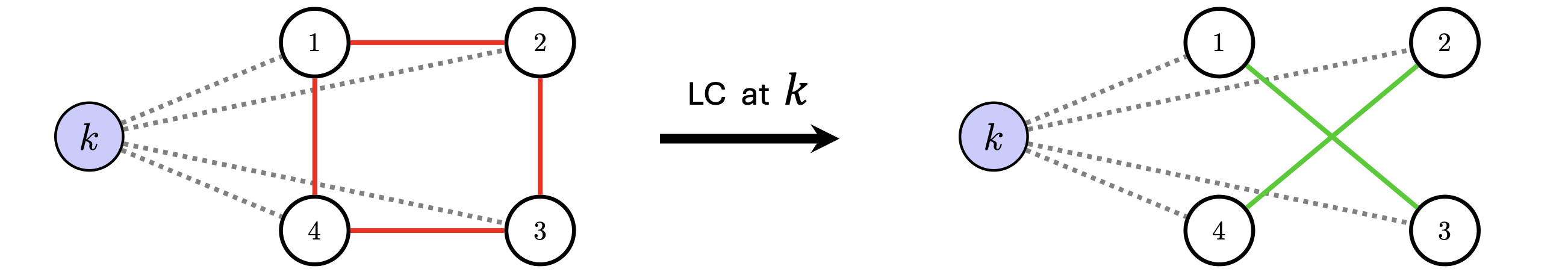}%
\caption{An example of local complementation. The transformation is applied at node $k$ with the neighborhood $N(k)=\{1,2,3,4\}$ whose subgraph is replaced by its complement. }
\label{fig:lc}
\end{figure}

Local complementation (LC) is a specific type of graph transformation. To locally complement a graph at one of its nodes $k$ is to replace the subgraph induced by its neighborhood $N(k)$ (consisting of all vertices adjacent to  $k$) with its complement while keeping other parts of the graph unchanged (See Fig.~\ref{fig:lc}). Locally equivalent graphs, defined as the graphs related to one another by a sequence of local complementations, correspond to quantum states that are equivalent under local Clifford operations \cite{van_den_nest_graphical_2004}. The converse is also true: if a graph state is obtained from another via local Clifford operations, then the two graphs are connected by a sequence of local complementation transformations \cite{van_den_nest_graphical_2004}. Applying LC to qubit $k$ of a graph state corresponds to the following local unitary operation \cite{hein_multiparty_2004}: 
\begin{align}\label{LC_unitary}
         U^{\text{\tiny{LC}}}_k = \sqrt{-i X_k}\prod_{j\in N(k)}\sqrt{i Z_j}
\end{align}
where $X_k$ and $Z_j$ are Pauli operators acting on qubits $k$ and $j$, and $N(k)$ denotes the neighborhood of node $k$. 
The adjacency matrix of a graph transforms under LC on node $k$ in the following way: 
\begin{enumerate}
    \item Adding (mod 2) the $k$-th row ($R_k$) of the initial adjacency matrix to the row $R_j$ for all $j$ in the neighborhood of $k$. 
    \begin{align}\label{eq:LC_row}
        \bar{R_j} = &R_j \oplus R_k,
        &\forall j \in N(k) 
    \end{align}
where $\bar{R_j}$ is the updated $j$-th row after LC. Equivalently, we can instead use the columns to update the adjacency matrix:
    \begin{align}\label{eq:LC_col}
        \bar{C_j} = &C_j \oplus C_k,
        &\forall j \in N(k) 
    \end{align}
Employing either of the updating methods shown in eqs.~(\ref{eq:LC_row}) and (\ref{eq:LC_col}) leads to the same final adjacency matrix as the initial matrix is always symmetric.
    \item Resetting all diagonal elements to zero, ensuring self-edges (loops between a node and itself) do not appear after an LC operation. 
 \end{enumerate}

Local complementation can thus be used as a tool to create/remove an edge between two nodes as long as they have a common neighbor. However, because LC affects the whole neighborhood of a node, this cannot be used as a standalone arbitrary edge manipulation method and possible unwanted changes should be carefully avoided by extra operations before or after an LC. In \cref{sec:graphical}, we propose a set of graph operations based on LCs that affect only a desired subset of edges, leaving the rest of the graph intact. 

\begin{figure*}[t]
\centering
\includegraphics[width=0.9\textwidth]{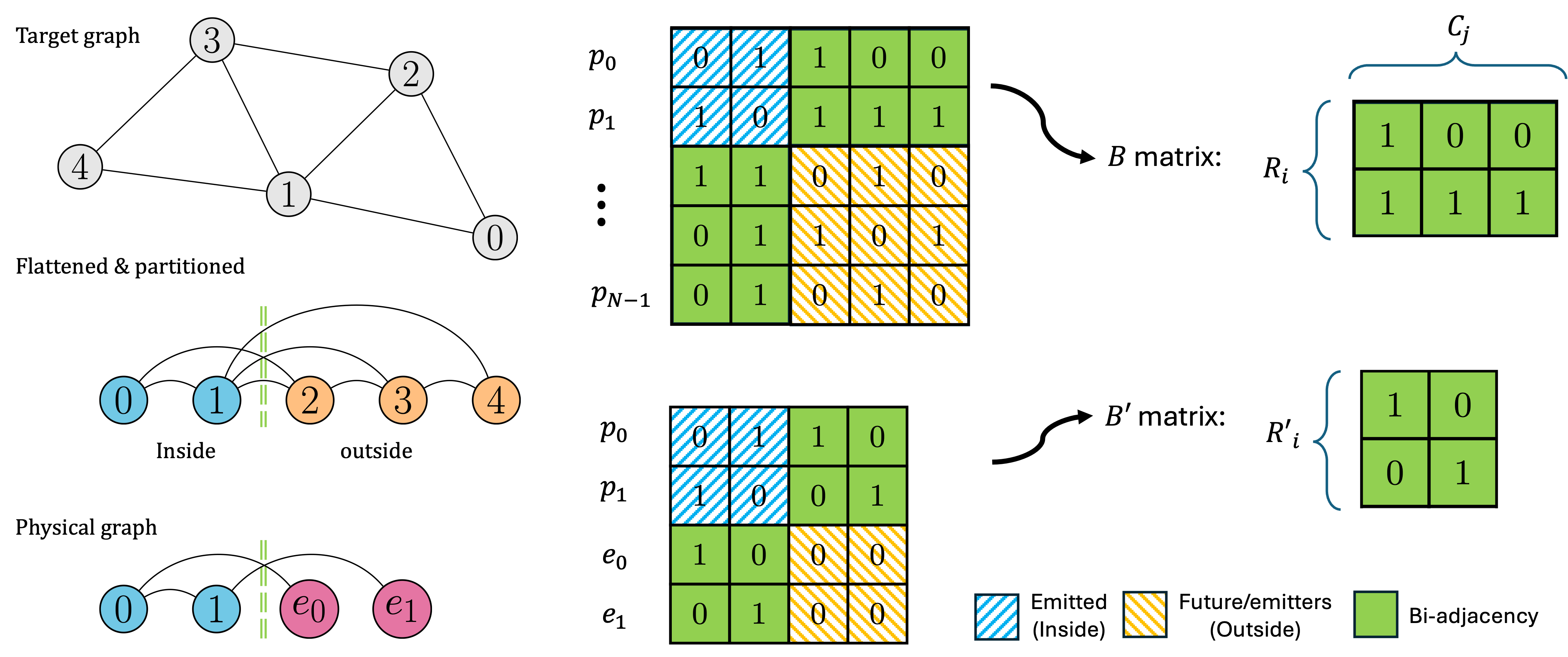}%
\caption{An example target graph is shown in its original form (top), in its bi-partitioned form with the first two photons considered as emitted (middle), and as the intermediate physical graph with two emitted photons and two emitters (bottom). The adjacency matrices corresponding to the target and physical graph states, together with the associated biadjacency matrices, $B$ and $B'$, are also presented.
The rows of the biadjacency matrix $B$ represent the connectivity vectors ($R$) of the emitted photons to the future nodes according to the target state. The physical biadjacency matrix $B'$ includes the connectivity vectors ($R'$) for each photon to the current set of emitters at this step of the generation.}
\label{fig:1}
\end{figure*} 

\section{The generative set}\label{sec:gen_set}
Consider an intermediate step in the state generation process of an $N$-qubit graph state $\ket{G}$, where the current physical graph state $\ket{G'(n)}$ consists of a number of emitter qubits and the $n$ photonic qubits emitted thus far. Our objective is to ensure that after the emission of all $N$ photons we get $\ket{G'(N)}=\ket{G}$ up to local operations. 
To that end, a prerequisite assumption is that the intermediate state $\ket{G'(n)}$ can evolve into the final target using the allowed operations as indicated in \cref{subsec:generation_problem}. This means the current intermediate state must belong to the generative set of the target graph $G$ as defined below: 
\begin{definition}\label{gen_set}
For any target graph $G$, the generative set \textbf{GenSet$_T(G, n)$} is the set of all graphs $G'(n)$ consisting of $n\leq N$ photonic nodes and any number of emitter nodes such that $G'$ can be transformed into $G$ under the transformation set $\mathcal{T}$.
    \begin{align}  
    GenSet_\mathcal{T}(G, n) = \left\{G'(n)\;\middle|\;\exists \;\tau \in \mathcal{T^*}\; \text{:}\;G'(n)\xrightarrow{\tau} G \right\}
    \end{align}
\end{definition}
\noindent Here the transformation set $\mathcal{T}$ consists of the operations described in \cref{subsec:generation_problem}, i.e, photon emissions, local Clifford gates (whose action can be realized by a sequence of LC unitaries defined in \cref{LC_unitary}), two-qubit Clifford gates on emitter qubits, and measurements:
\begin{align}\label{T}
\mathcal{T} = \{ \text{\cnot[cnot]}_{e-p}, \,{U^{\text{\tiny{LC}}}_{e/p}},\,\text{\cz[cz]}_{e-e},\,\text{M}_{e} \}
\end{align}
\noindent The subscripts indicate the qubit type on which the gates act, with $e$ denoting an emitter and $p$ denoting a photon, e.g., $\text{\cnot[cnot]}_{e-p}$ represent an emission where the control qubit is an emitter and the target is a photonic one. $\mathcal{T^*}$ is the set of finite sequences (compositions) of transformations in $\mathcal{T}$. From now on, we drop the subscript $\mathcal{T}$ and the index \textit{G} for brevity and refer to the $n$-photon generative set of the target state as \text{GenSet($n$)} when there is no ambiguity. 

Assuming $G'(n)\in$ GenSet(n) for an intermediate state, if by emitting the next photon we find a way to ensure $G'(n+1)\in$ GenSet(n+1) for arbitrary $n$, then by induction, one can start from an initial state $G'(n=0)$ consisting of only emitters and obtain the target graph after $N$ steps, $G'(n=N) = G$:
\begin{align}
G'(0)\xrightarrow{\mathcal{T}}\ldots G'(n)\ldots \xrightarrow{\mathcal{T}} G'(N) = G
\end{align}
As we will explicitly show in the following sections, it is always possible to find operations in $\mathcal{T}$ such that all intermediate quantum states are graph states.  
Note that photon emission is considered an \textit{irreversible} process in this setting, i.e., a photon cannot be absorbed back to its emitter, and although measuring a photon removes it from the system, emitting one only to discard it by measurement would be a redundant process that one must avoid in the generation recipe. Therefore, no intermediate $G'(n)$ can be allowed to fall outside of the generative set at any step in the process, from the emission of the first photon until all photons have been emitted. 

\begin{algorithm}[H]
\caption{Inductive Generation Procedure}
\label{alg:inductive_generation}

\begin{algorithmic}[1]
\Statex \textbf{Base Step} (\textit{Step 0}): Emit the first photon (labeled $0$) entangled with its emitter.  
The eligibility conditions are trivially satisfied, as any state generation process can begin with this emission. \label{base}

\Statex \textbf{Inductive Step} (\textit{Steps $n \geq 1$}): 
Assume the first $n$ photons, labeled $0$ through $n-1$, have been emitted, and the current physical graph $G'(n)$ satisfies the eligibility conditions. Now determine: \label{protocol:1}
\begin{enumerate}[label={(\roman*)}]   

     \item the required neighborhood for the next photon once it is emitted, and
     \item the required adjustments to the connectivity of the remaining nodes in $G'(n)$,
\end{enumerate}
\Statex such that the eligibility conditions still hold for the graph $G'(n+1)$ obtained after the emission of photon $n$.
\Statex \hrulefill

\For {$n=1$ to $N-1$} 

\State Find adjacency requirements (i) and (ii) for step $n$. \label{alg:first}

\State Select an eligible emitter for the next photon.\label{alg:emitter}

\State Apply required operations to realize $G'(n+1)$ as indicated by (i) and (ii).\label{alg:ops}

\State Identify and recycle redundant emitters, if any.\label{alg:decouple}

\EndFor

\end{algorithmic}
\end{algorithm}

To ensure the feasibility of the induction step, it is \text{first} crucial to determine the conditions under which the assumption of $G'(n)\in$ GenSet(n) holds true. \text{Next}, we need to identify the operations to be used such that these conditions remain satisfied with the emission of a new photon, going from $n \to n+1$. The instructions that ensure $G'$ remains a member of GenSet as photons are added to the system make up the generation recipe. Finding such a recipe is the task of a generation algorithm.  

For now, let us assume that the eligibility conditions for the generative set are known. We then propose a step-wise inductive procedure, in which step $n$ is defined as the process of finding and applying the operations that take $G'(n)$ to $G'(n+1)$, as shown in \cref{alg:inductive_generation}. Note that the entanglement between each emitted photon and the rest of the state is heavily restricted and dictated by its emitter. This is because once the photon is created, it cannot interact with any other qubit in the system, hence the necessity of lines~\ref{alg:emitter} and \ref{alg:ops} in the algorithm.

In the rest of this section we introduce explicit forms for the eligibility conditions for the intermediate physical graph $G'(n)$ to be in the generative set. Next, we determine how the connectivity between nodes should be adjusted in order for those conditions to remain satisfied through the creation of a new photonic node and completing the $n\to n+1$ inductive step.

\subsection{Eligibility Conditions}\label{subsec:eligibility}
In this part, we will derive the required eligibility conditions at each inductive step for the graph under evolution to be a member of the generative set of the target state. 

Let us consider a two-subsystem description for the intermediate physical graph state $\ket{G'(n)}$ where $n$ out of total of $N$ photons are created. The first subsystem, which we refer to as the ``inside", consists of the emitted photons labeled from $0$ to $n-1$. The second subsystem consists of the emitters, here referred to as the ``outside". At the same time, a corresponding partition can be applied to the target graph $\ket{G}$, which is a fixed graph (in contrast to the evolving physical state) composed of only $N$ photonic nodes, dividing it into the emitted (inside) nodes $\{0,\dots n-1\}$, and the not yet emitted (outside) ones $\{n,\dots N-1\}$. The connectivity between these two subsystems, representing the entanglement required between the \textit{current} photons and the \textit{future} ones, is reflected in a portion of the target graph's adjacency matrix which is obtained by intersecting the first $n$ rows (corresponding to the emitted photons) with the last $N-n$ columns (of the non-emitted photons) as seen in the Fig.~\ref{fig:1}. Let us name this submatrix the biadjacency $B(n)$ with $n$ indicating the number of photons in the emitted (inside) subsystem. A similar submatrix---denoted as physical biadjacency matrix $B'(n)$---can be defined over the physical graph's adjacency matrix as well (see Fig.~\ref{fig:1}) that stores the information on the connectivity between the inside and outside partitions, i.e., how the emitted photons are connected to the emitter nodes in the intermediate graph $G'(n)$. Isolated nodes (nodes with no edges) in the graph are not considered in the biadjacency matrix.

We first note that the bipartite entanglement across the inside-outside partition on the physical graph cannot be increased in future steps, even though additional edges across the partitions may be created in the process. In more general and concrete terms, one can state:%
\begin{proposition}\label{prop1} 

Let $\mathcal{P}$ be a fixed subset of the photonic nodes in the graph state $G'(n)$, then the bipartite von Neumann entanglement entropy ($S_\mathcal{P}$) between the subsystem $\mathcal{P}$ and the rest of the qubits in the complement set $\mathcal{Q}$ in $G'$ does not increase under transformation set \hyperref[T]{$\mathcal{T}$}.
\end{proposition}%
\noindent Here $S_P$ is defined as 
\begin{align}
    S_\mathcal{P} = -{\text{Tr} \hspace{1mm}} \left(\rho_\mathcal{P} \log \rho_\mathcal{P}\right)
\end{align}
\noindent and $\rho_\mathcal{P} = \text{Tr}_{\mathcal{Q}}\ket{G'}\bra{G'}$. Such conservation of entropy originates from prohibiting non-local operations (two-qubit gates) between the emitter qubits and emitted photons. For a detailed proof of \cref{prop1} refer to \cref{sec:proofs}. As a consequence, the bipartite entanglement in the intermediate physical graph state $\ket{G'(n)}$ does not increase through the evolution of the state during the generation process and so it must be compatible with the entanglement structure of the final state as encoded in the biadjacency matrix $B(n)$, i.e., we must have
\begin{align}\label{entropy_req}
    S_\mathcal{P}(G') \geq S_\mathcal{P}(G)
\end{align}
otherwise, the final output cannot be equal to the target state as their bipartite entanglement would not match.
In other words, since the emitters in the intermediate state should be able to act as representatives for all the future photons, the inside-outside entanglement at each step must match that of the final state, if $G'(n)$ is to be a member of {GenSet($n$)}. The following theorem helps us formulate a hard limit on the minimum number of emitters and the way they need to be connected to the set of emitted photons at the beginning and end of each step of the generation.
\begin{theorem}\label{theo1}
The bipartite entanglement entropy of a graph state for any bipartition is equal to the rank of the corresponding biadjacency matrix (over the field $\mathbb{Z}_2$). 
\end{theorem} 
\noindent See \cref{sec:proofs} for a proof of theorem 1.

Knowing from linear algebra, the rank of an \( x \times y \) matrix is upper-bounded by \( \min\{x, y\} \). Let us denote the rank of the \( n \times (N - n) \) biadjacency matrix \( B(n) \) of the target graph as $m'$:
\begin{align}
    \mathrm{rank}[B(n)] =  m' \leq \min\{n, N - n\}
\end{align}
In addition, by using \cref{theo1} and following \cref{entropy_req}, we see that the physical biadjacency matrix \( B'(n) \), which is by definition an \( n \times m \) matrix associated with $G'(n)$ that has \( n \) emitted photons and \( m \) emitters, must also have at least the same rank, i.e., 
\begin{align}
 m' \leq \mathrm{rank}[B'(n)] \leq \min\{n, m\}
\end{align}
The number of emitters, $m$, in the physical state is thus lower-bounded by the rank of $B(n)$, i.e., \( m \geq m' \), if $G'(n)\in$ GenSet($n$). Here, we consider the minimal use of emitters and therefore use the case \( m = \mathrm{rank}[B(n)] \), which corresponds to a full column rank physical biadjacency matrix \( B'(n) \) at each step. Therefore, when considering all generation steps from $0$ to $N-1$, the minimum number of emitters $M$ needed to complete the process would be
\begin{align}\label{eq:min_emitter}
M=\max \{ \mathrm{rank}[B(n)] \mid 0 \leq n < N \}
\end{align}
As a result, for $G'(n)$ to be a member of the GenSet($n$) at each intermediate step, we need $M$ total emitters, where at least $m=\mathrm{rank}[B(n)]$ of which are active (non-isolated) in the graph $G'(n)$, i.e., they have non-zero contribution in the biadjacency matrix $B'(n)$. The rest of the emitters are isolated qubits/nodes in the physical graph state and are not represented in the biadjacency matrix at step $n$. The requirement on the overall number of emitters stated in \cref{eq:min_emitter} is also consistent with the findings of Ref.~\cite{li_photonic_2022}. From now on, we assume the system always has the minimum sufficient total number of emitters ($M$), and focus only on the number of \textbf{active} emitters ($m$) required at each step. 

In addition, by the definition of the rank, the emitters must be connected to the photonic nodes in a way that ensures the biadjacency matrix contains at least \( m \) linearly independent rows in $\mathbb{Z}_2^m$ that span the row space of $B'(n)$. In other words, one must be able to select \(m\) linearly independent rows such that the corresponding \(m \times m\) sub-matrix is invertible. From $B'(n)$ being a full column rank matrix, it also immediately follows that for every emitter node \(e_j\), associated with the column $j$ of \(B'(n)\), there is at least one row \(R'_i\) among our chosen \(m\) mutually independent rows such that \(B'_{ij}(n) = 1\). Therefore, we can always find a bijective map from the set of emitters $\{e_j\mid 0\leq j< m \}$, to the set of independent rows $\{R'_i\}$, such that if $e_j\to R'_i$, then \(e_j\) is connected to the photon \(i\) in the graph $G'(n)$. Given such a bijection, we define a new parameter, referred to as the “emitter row,” to help formalize the eligibility conditions.
\begin{definition}
    For each emitter \(e_j\), let \(i\) be the index of its assigned row in the physical graph's biadjacency matrix \(B'(n)\), then the \textbf{emitter row} of $e_j$ at the beginning of the step n, denoted by $R_{e_j}(n)$, is defined as
\begin{align}
  R_{e_j}(n) :=  R_{i}(n)
\end{align}
where \(R_{i}(n)\) is row $i$ of target graph's biadjacency matrix \(B(n)\).
\end{definition}
\noindent Note that the index $i$ for each emitter row is selected based on the adjacency relations in $B'(n)$ but the row itself is picked from the matrix $B(n)$. Since each element of the row $i$ in $B(n)$ indicates a connection between the photon $i$ and a later photon, one interpretation for the emitter row \(R_{e_j}\) is that the emitter \(e_j\) is responsible for generating all future edges encoded in that row of the target's biadjacency matrix, .

We now claim the necessary eligibility conditions for an intermediate graph state $G'$ to be a member of the generative set of $G$ at the start of step $n$ are as follows (the step index $n$ is suppressed for clarity):
{
\renewcommand{\theenumi}{(\Roman{enumi})} 
\renewcommand{\labelenumi}{}                        

\begin{enumerate}[itemindent=0pt, leftmargin=20pt]  
    \item \label{I}
     \textcolor{black}{Condition \textbf{(I)}:} If rank$\left( B\right)=m$, then $B'$ must be a full column rank matrix with at least $m$ columns (active emitters) and we can always form the linearly independent set of emitter rows:
\begin{align}
\left\{R_{e_j}\;|\; 0\leq j< m\right\}
\end{align}
\noindent that forms a basis for the row space of $B$.

    \item \label{II} 

    \textcolor{black}{Condition \textbf{(II)}:} Let $\mathcal N(i)=\{j\;|\; B'_{ij}=1\}$ denote the set of indices corresponding to the emitters that are neighbors (connected) to photon $i$ in the physical graph. Then, for every \(0\le i< n\), the following must hold for the corresponding row $R_i$ in \(B\): 
\begin{align}\label{II_eq}
R_i=\sum_{j\in\mathcal{N}(i)} R_{e_j}
\end{align}

\end{enumerate}
}

\noindent The proof of necessity for these conditions can be found in \cref{sec:proofs}.  As a simple example of employing conditions (I) and (II), let us consider the target graph and the physical intermediate state (with $n=2$ emitted photons and $m=2$ emitters) depicted in Fig.~\ref{fig:1}. The rank of the biadjacency matrix $B(n=2)$ is clearly equal to two as it has two linearly independent rows:
\begin{equation}
    R_0 = \{1, 0, 0\}, \quad
    R_1 = \{1, 1, 1\}
\end{equation}
As required by the first condition, this rank determines the number of emitters needed in the physical graph. We can now determine the set of emitter rows. Since photon $0$ is only connected to $e_0$ and photon $1$ to $e_1$, then the index of the assigned rows in physical biadjacency matrix $B'(n=2)$ for them is also $0$ and $1$, respectively. As a result for the emitter rows we get $R_{e_0}=R_0$ and $R_{e_1}=R_1$. Let now now check the validity of the second condition. For $i=0$, we have $\mathcal N(0) = \{0\}$ and thus according to \cref{II_eq} we must have $R_{i=0} = R_{e_0}$, which is the case. Similarly for $i=1$, we have $\mathcal N(1) = \{1\}$ and according to \cref{II_eq} we must have $R_{i=1} = R_{e_1}$, which is true as well. As a result, the intermediate state shown in Fig.~\ref{fig:1} is a valid member of the generative set of the target.

The first condition sets the required number of emitters, and the second one dictates how those emitters must be connected to the emitted photons. For both conditions, the validity can be determined solely from the target graph's biadjacency matrix \(B(n)\) which is readily available once a target is given. Note that these two conditions are invariant under local operations as the effect of LC transformations on the biadjacency matrix is a subset of row addition operations (\cref{eq:LC_row}) which cannot affect the linear dependencies of the rows. 

Conditions \ref{I} and \ref{II} only concern the bipartite entanglement across the inside-outside partition. We need to also ensure that the ``inside" subsystem's internal edge pattern at any step allows $G'(n)$ to be a member of the GenSet($n$). As non-local interactions are forbidden between emitted photons, the changes that can be made in future steps in this internal entanglement structure are limited. Consequently, we need to add the following to the set of necessary conditions for $G'(n)\in$ GenSet($n$) at step $n$. 
{
\renewcommand{\theenumi}{(\Roman{enumi})} 
\renewcommand{\labelenumi}{}                        

\begin{enumerate}[itemindent=0pt, leftmargin=20pt]  
    \setcounter{enumi}{2}
    \item\label{III} 

    \textcolor{black}{Condition \textbf{(III)}:} For any graph $H$ and a subset of its nodes $\mathcal{V}$, let \(H[\mathcal{V}]\) be the induced subgraph of $H$ on $\mathcal{V}$. Then for \(\mathcal{P}=\{i \mid 0 \le i < n\}\), the subgraph $G'[\mathcal{P}]$  must be equal to $\tilde{G}[\mathcal{P}]$ where $\tilde{G}$ is some member of the Partial Local Clifford equivalence class PLC($G$, $\mathcal{P}$) as defined below.
\end{enumerate}
}
\begin{definition}\label{PLC}
    Let \( V \) be the set of all nodes in the graph \( G \). The \textbf{Partial Local Clifford} equivalence class, denoted by \textbf{PLC}(G, $\mathcal{P}$), over a subset of nodes \( \mathcal{P} \subseteq V \), is the set of all graphs that can be obtained from \( G \) by applying sequences of local Clifford (LC) transformations on the full set \( V \), and two-qubit Clifford gates restricted to the complement set \( V \setminus \mathcal{P} \), i.e.,
\end{definition}
\[
PLC(G, \mathcal{P}) = \left\{ H \,\middle|\,\exists \; \tau \in \left( {LC}_V \cup \mathcal{C}_{V \setminus \mathcal{P}}^{(2)} \right)^* :\; G \xrightarrow[]{\tau} H\  \right\}
\]
\noindent
\text{where } 
\begin{itemize}
  \item \( {LC}_V \) is the set of local Clifford operations on all nodes \( V \),
  \item \( \mathcal{C}_{V \setminus \mathcal{P}}^{(2)} \) is the set of two-qubit Clifford gates acting only within \( V \setminus \mathcal{P} \),
  \item \( (\cdot)^* \) denotes sequences (finite compositions) of such operations.
\end{itemize}

\noindent In the above statements, an induced subgraph on a subset of nodes is obtained by deleting all other nodes in the graph together with every edge incident on them. The equivalent quantum operation for obtaining the state associated with such a subgraph is to measure the rest of the qubits in the $Z$ basis and to apply the local unitary $Z^M$ on the neighbors of each node based on its measurement result $M$ \cite{hein_entanglement_2006-1}. Note that this process is distinct from tracing out the qubits not included in the subgraph, as here the measurement outcomes are considered known and can be accounted for. 

Since any subgraph of $G$ is trivially a member of its partial local Clifford equivalence class, i.e.,
\begin{align}
G[\mathcal{V'}] \in PLC(G, \mathcal{V'}), \quad  \forall \;\mathcal{V}' \subseteq V
\end{align}
\noindent for concreteness, we opt to use a less general (more restricted), but sufficient version of condition \ref{III}:

{
\renewcommand{\theenumi}{(\Roman{enumi}$'$)} 
\renewcommand{\labelenumi}{}                        

\begin{enumerate}[itemindent=0pt, leftmargin=20pt]  
    \setcounter{enumi}{2}
    \item\label{III-2} 
    \textcolor{black}{Condition \textbf{(III$'$)}:} For any intermediate physical graphs $G'(n)$, we must have $G'[\mathcal{P}]=G[\mathcal{P}]$ where $G$ is the target graph and \(\mathcal{P}=\{i \mid 0 \le i < n\}\). 
\end{enumerate}
}

\noindent In \cref{subsec:lc_freedom} we discuss how this gauge freedom in fixing an exact form of condition \ref{III} can be leveraged as one of the important optimization degrees of freedom.

\begin{figure}[hb]
\centering
\includegraphics[width=0.9\columnwidth]{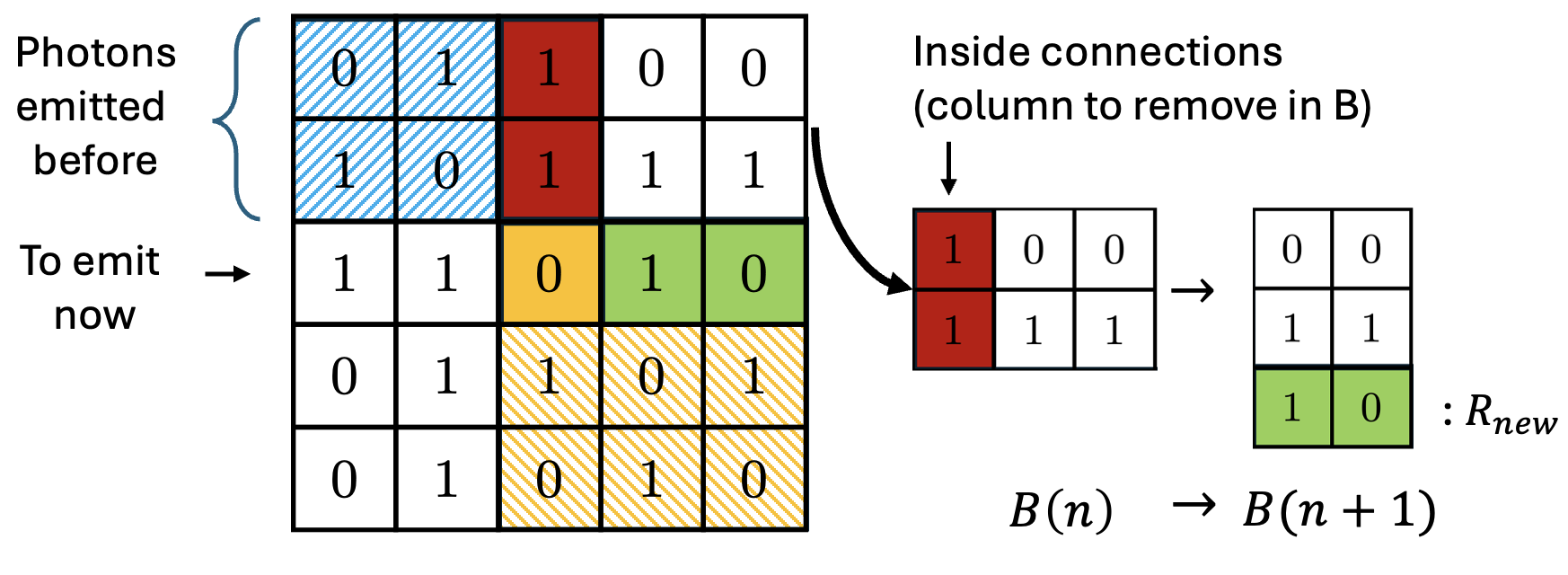}%
\caption{The evolution of the biadjacency matrix with emission of each photon at each step. The figure shows the case of $B(n)\to B(n+1)$ for $n=2$. The column to remove shows the connections of the new photons to the inside set. The new row  ($R_{new}$) represent the edges that need to be established between the new photon and the future ones to come.}
\label{fig:evolution}
\end{figure}
\subsection{Induction Step}

Assuming conditions \ref{I}–\ref{III} are satisfied for $G'(n)$, we now specify how the next emission must happen so that these conditions still hold after adding a new node (the photon labeled $n$) to the system. 

The validity of conditions \ref{I} and \ref{II} can be kept in check by tracking the evolution of the biadjacency matrix when a new photon is emitted. Specifically, the emission triggers a change in the bipartition of the target graph: photon \(n\) is moved from the outside subsystem to the inside one. As a result, the biadjacency matrix must be updated accordingly, i.e., for $B(n)\to B(n+1)$: 
\begin{itemize}
    \item The edges incident on photon $n$ from previous photons, encoded in the first column of $B(n)$, no longer contribute to the inside–outside entanglement and are not a part of $B(n+1)$. 
    \item The new photon's future connections, encoded in row $n$ in target graph's adjacency matrix, now contribute to the inside–outside entanglement and should be reflected in $B(n+1)$.
\end{itemize}
Therefore, $B(n+1)$ can be obtained from $B(n)$ by removing its first column and then appending a new row to it (see Fig.~\ref{fig:evolution}). We can now use this new biadjacency matrix for eligibility checks.

Condition \ref{I} concerns the number of emitters and since it is assumed to already hold for step $n$, i.e, rank$\left[B(n)\right]=$ rank$\left[B'(n)\right]$, we can ensure its validity by finding the changes in rank of $B$ when going from $n$ to $n+1$ and ensuring the structure of $B'$ follows accordingly. Let us define the terms ``column effect" and ``row effect" as follows:  
\begin{itemize}
    \item[--] \textbf{Column Effect:}\label{col_effect} The change in the rank of $B(n)$ after deleting its first column. This value is $-1$ if the removed column is linearly independent of the remaining columns, and 0 otherwise.
    \item[--] \textbf{Row Effect:}\label{row_effect}  The change in the rank obtained by appending the new row $R_{new}$, which makes up the last row of \(B(n+1)\), to \(B(n)\) once its first column has been removed already. This change lies in \(\{0,+1\}\), depending on whether the appended row is linearly independent of the other rows or not.
\end{itemize}
\noindent The total change in rank when a new photon is emitted equals the sum of the column and row effects and can therefore take any value in \(\{-1, 0, +1\}\). In case of no change, the current emitters are enough to handle the next emission. A total rank change of \(+1\) indicates that a new emitter must be activated and introduced into the physical graph while a change of $-1$ means one of the active emitter can be decoupled (isolated) and considered removed from the system \textbf{after} the emission. 

Besides, condition \ref{I} also requires updating the emitter rows after each emission, as in general \(R_{e_j}(n) \neq R_{e_j}(n+1)\). In particular, when the column effect is $-1$, it means one of the emitter rows \(R_{e_j}(n)\) becomes linearly dependent on the rest of them as a result of deleting the first entry of each row (removing the first column of $B$). Since emitter rows must always form a linear independent set, \(R_{e_j}(n)\) is no longer a member of this set and the corresponding emitter \(e_j\) becomes free to adopt a different row. Now if the row effect is $+1$, the same freed up $e_j$ is assigned the new row, i.e.,
\begin{align}
R_{e_j}(n+1)=R_{new}:= R_{n}(n+1)
\end{align}
where \(R_{i}(n+1)\) is row $i$ of \(B(n+1)\). In another case, if the row effect is $0$, then $e_j$ becomes a redundant emitter that can be removed from the system after the emission in this step. Alternatively, if column effect is $0$ and row effect is $+1$, the new added emitter is assigned the new row and there is no need to update any of the previous emitter rows. Lastly, the case of both row and column effects being zero also required no update as the previous set of emitter rows would still satisfy condition \ref{I}.

Having established the updated set of emitter rows, we can now employ condition \ref{II} to determine the required connectivity between the emitters and photons in $G'(n+1)$, after the emission. Based on \cref{II_eq}, for each index $0\leq i\leq n$, if we expand the corresponding row in the new biadjacency matrix in terms of updated emitter rows:
\begin{align}
\label{J}
R_i(n+1) = \sum_{j\in\mathcal{N}(i)} R_{e_j}(n+1)
\end{align} 
\noindent then photon $i$ must be connected to the emitters $\{e_j\;|\;j\in \mathcal{N}(i)\}$. Note that such expansion is always possible as the emitter rows make up a basis for the row space of the biadjacency matrix. As a result, the adjacency relations between emitters and photons across the inside-outside partition in $G'(n+1)$ can be fully determined.

\begin{figure}[b]
\centering
\includegraphics[width=0.7\columnwidth]{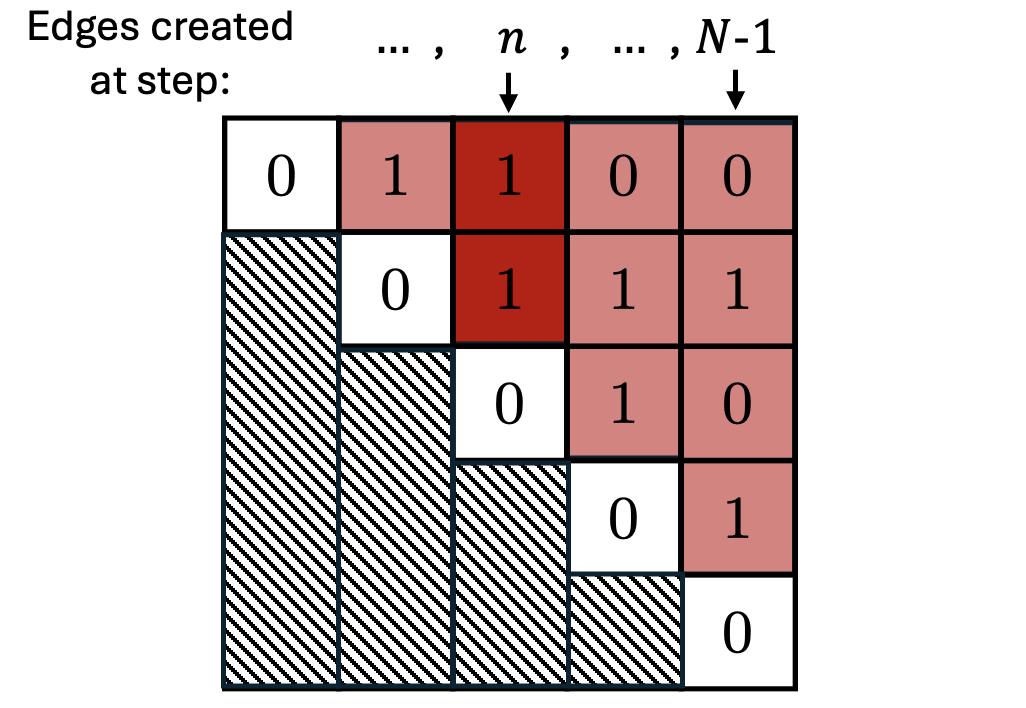}%
\caption{A representation of condition \hyperref[III-2]{$(\text{III}')$} satisfied through all steps on the adjacency matrix of the target graph. At step \(n\), the inside neighbors of the newly emitted photon are determined by the non-zero entries in the column above the \(n\)-th diagonal element. In the final step the complete adjacency matrix of the target graph is recovered. Due to symmetry of the adjacency matrix, fixing only one side of the diagonal is enough.}
\label{fig:cond3}
\end{figure}

Next, condition \hyperref[III-2]{$(\text{III}')$} helps specify the required connectivity within the inside subsystem of $G'$, i.e., to which previously emitted photons should the new photon connect. Since the condition is satisfied before the emission, the inside subgraph comprising vertices \(0,\dots,n-1\) already matches its counterpart in the target graph. To maintain this equality, it suffices to connect the newly emitted photon to the inside neighbors of its respective node (labeled $n$) in the target graph $G$. This neighborhood is determined by the non-zero entries in the first column of the biadjacency matrix (see Fig.~\ref{fig:cond3}), that is
\begin{align}\label{eq:inside_neighbors}
    \{i \mid B_{i0}(n)=1 \text{ for } 0\leq i<n\}
\end{align}

\begin{figure*}[t]
\centering
\includegraphics[width=0.9\textwidth]{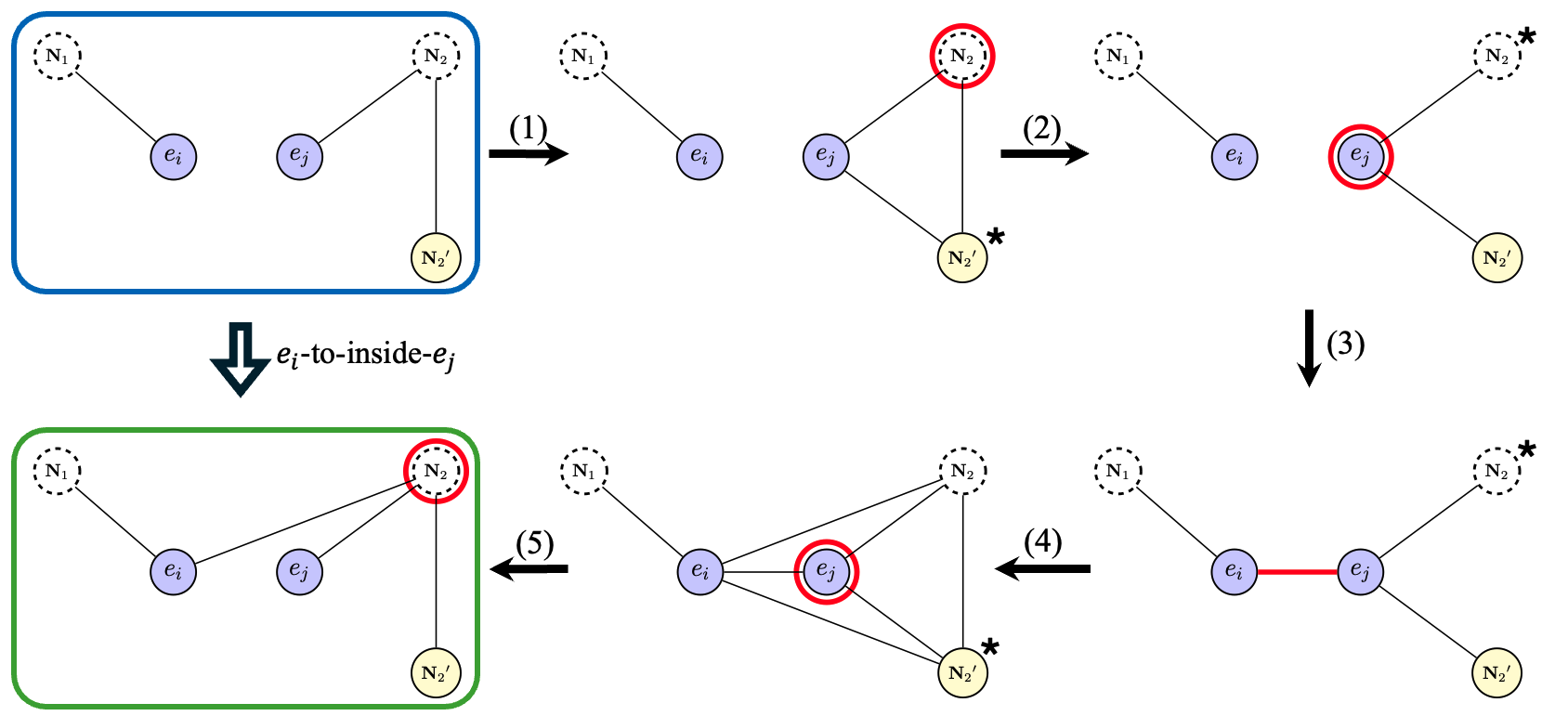}%
\caption{Stepwise representation of the e-to-inside-e operation. The nodes $N_1$ and $N_2$ are arbitrary neighborhoods of the two emitters. The LC operation was applied on the node indicated with the red circle in each step. When applied on $N_2$, it means applying LC on one of its member. $N_2'$ represents the neighborhood of this selected node in $N_2$. The starred nodes in each step indicate possible temporary alternation of the internal adjacency relations between the nodes belonging to that set as a result of the LC operations.}
\label{fig:e2in}
\end{figure*}

All the required connectivity adjustments, corresponding to \hyperref[protocol:1]{inductive step} of the algorithm, for taking the step $G'(n)\to G'(n+1)$ are now identified. Finally, to complete the induction step, we need to show how the emission and these adjustments can be implemented, which includes the choice of a suitable emitter for each step and using the correct graphical transformations that respect the physical restrictions outlined in \cref{subsec:generation_problem} to make sure $G'(n+1)\in$ GenSet is obtained as expected. To this end, in \cref{sec:protocol}, we present an algorithm that finds explicit preparation recipes that is based on an exhaustive case analysis of all possible initial states for $G'(n)$ to complete the induction step. 

Before presenting the generation algorithm, we introduce our graphical framework, which provides the foundation for applying and tracking edge-structure manipulation and evolution in the generation process.

\section{The graphical framework}\label{sec:graphical}

In this section, we introduce a graphical framework, compatible with the physical limitations of the generation problem mentioned in \cref{subsec:generation_problem}, for adding or removing edges between designated parts of a graph. Every graphical operation introduced here comes with an equivalent quantum operation in the transformation set \hyperref[T]{$\mathcal{T}^*$}, so any sequence of graphical transformations can be directly transpiled into a quantum circuit, applicable to an emitter-photonic system of qubits, that produces the same effect on the input state and yields the expected graph state as output.

\begin{figure}[h]
\centering
\includegraphics[width=\columnwidth]{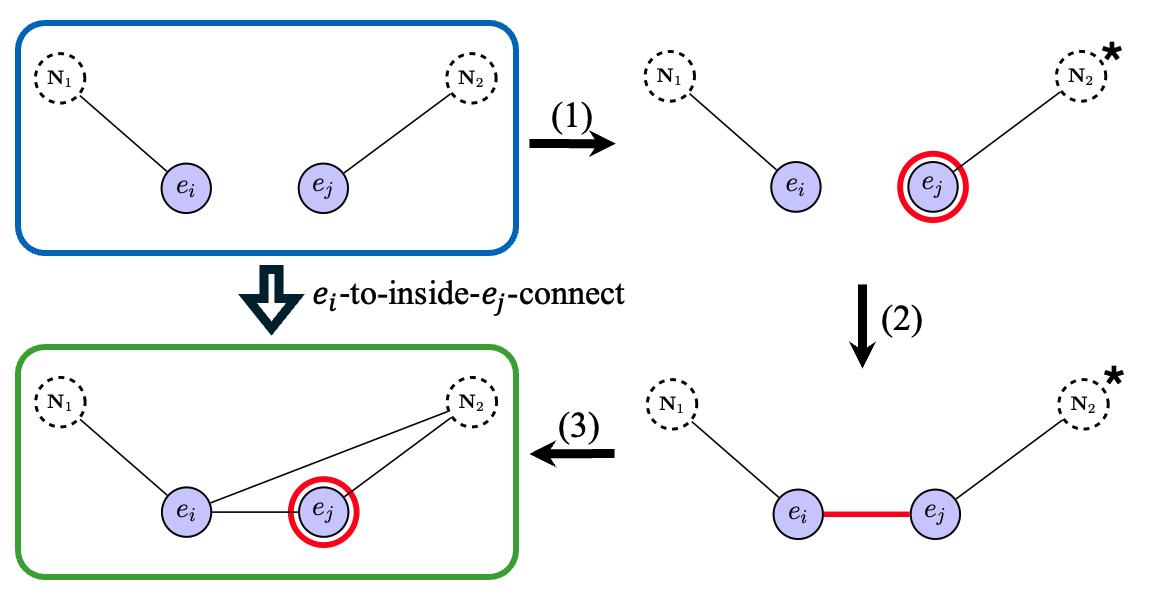}%
\caption{Stepwise representation of the e-to-inside-e-connect operation. The nodes $N_1$ and $N_2$ are arbitrary neighborhoods of the two emitters. The LC operation was applied on the node indicated with the red circle in each step. The starred nodes in each step indicate possible temporary alternation of the internal adjacency relations between the nodes belonging to that set as a result of the LC operations}
\label{fig:e2inconnect}
\end{figure}

Previously, we demonstrated that local~complementation~(LC) operations constitute a subset of row operations on the adjacency matrix; however, an LC operation always alters the entire neighborhood of a selected node, so arbitrary and targeted edge manipulations through row‑vector additions (over the field $\mathbb{Z}_2$) are not directly attainable. To overcome this limitation, in what follows we define compound operations, made up of a series of LCs and emitter–emitter \cz[cz] gates, that enable targeted modifications to a specific subset of edges while leaving the rest of the graph unchanged. We also define different emission modes that allow obtaining a desired connectivity for each newly emitted photon. The introduced set is \textit{universal} in the sense that it is enough for generation of arbitrary graphs. 

\subsection{Direct Edge-toggle}
The simplest graphical operation is toggling the connectivity between two emitter nodes, which corresponds to applying a \cz[cz] gate between them:
\begin{itemize}[itemindent=10pt, leftmargin=0pt] 
\setcounter{enumi}{0}
\item \textbf{e$_i$-to-e$_j$}: Creating an edge between $e_i$ and $e_j$ if none exists or deleting the edge if one is already present. 
\end{itemize}

\subsection{Batch Edge-toggle Operations}\label{sec:batch}
\begin{itemize}[itemindent=10pt, leftmargin=0pt]
\setcounter{enumi}{1}
\item\textbf{e$_i$-to-inside-e$_j$}: Inverting the connectivity of an emitter node $e_i$ to the neighborhood of another emitter $e_j$. The updated neighborhood of $e_i$ will be:
\begin{align}
    N_{new}(e_i) = N(e_i) \hspace{1mm}\Delta\hspace{1mm} N(e_j)
\end{align}
where $\Delta$ shows symmetric difference and $N(e)$ is the set of neighbors of emitter $e$.
In the case of having no common neighbors, $e_i$ get connected to all neighbors of a $e_j$. We denote this by $e_i$-to-inside-$e_j$, since we predominantly use it to connect the inside neighborhood of one emitter to another. Our proposed implementation requires only a single use of a two-qubit gate (see Fig.~\ref{fig:e2in}). In terms of quantum operations, this is equal to the following sequence:
\begin{enumerate}
    \item LC on a neighbor of $e_j$ (if any)
    \item LC on $e_j$
    \item CZ between $e_i$ and $e_j$
    \item LC on $e_j$
    \item LC on the same neighbor of $e_j$
\end{enumerate}
\noindent where each LC operation is a collection of single qubit gates as seen in \cref{LC_unitary}. 

\item\textbf{e$_i$-to-inside-e$_j$-connect}: Performing an $e_i$-to-inside-$e_j$ operation while inverting the connectivity between the two emitters as well (see Fig.~\ref{fig:e2inconnect}). The effect is the same as applying $e_i$-to-inside-$e_j$ and $e_i$-to-$e_j$ operations consecutively, but instead of using two \cz[cz] operations, we can implement this using only one two-qubit gate:
\begin{enumerate}
    \item LC on $e_j$
    \item CZ between $e_i$ and $e_j$
    \item LC on $e_j$
\end{enumerate}
\end{itemize}

\subsection{Emission Modes}

A simple emission can be represented as creating a leaf qubit--a dangling photon--for its emitter in the graph (Fig.~\ref{fig:0}c). Here we define a set of emission modes consisting of a sequence of local operations before and after this simple photon emission that allows for more flexibility in the connectivity of the emitted photon. Below, we explain the placement of the newly emitted node in the graph with respect to its emitter and the rest of the nodes.

\begin{itemize}
[itemindent=10pt, leftmargin=0pt]
\item[--] \textbf{Linear (L) Mode:} The created photon takes the place of the emitter and inherits all its neighbors, while the emitter becomes a leaf attached to the new photon. The required quantum operations are an emission \cnot[cnot] followed by a Hadamard on the emitter qubit. \\\textit{Remark}: repeated use of this mode results in the creation of a linear cluster state (see Fig.~\ref{fig:em_L}). 
\begin{figure}[h]
\centering
\includegraphics[width=0.95\columnwidth]{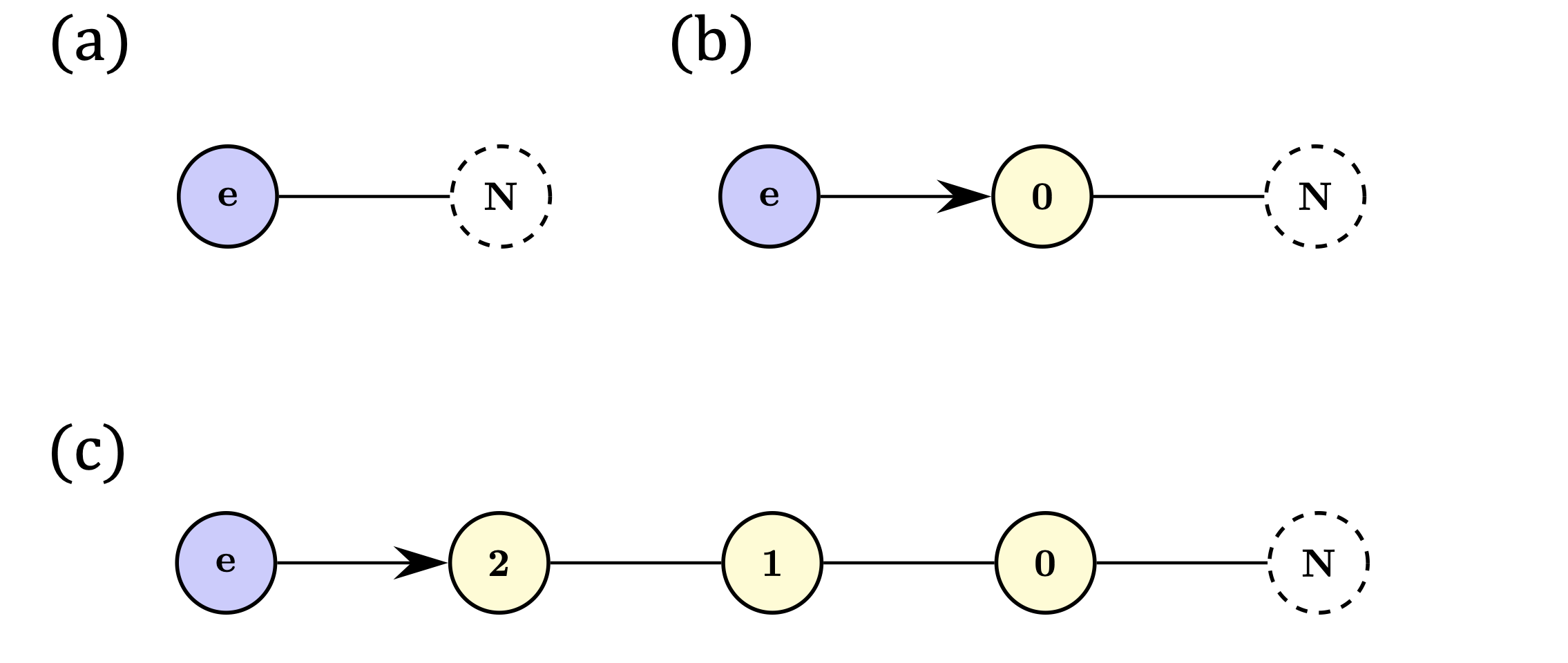}%
\caption{Linear emission. \textbf{(a)} Showing the state of the emitter and its neighborhood before the emission. \textbf{(b)} The state after emission of one photon in (L) mode. \textbf{(c)} The state after sequential emission of multiple photons using the same mode.}
\label{fig:em_L}
\end{figure}
\item[--]\label{SS} \textbf{Simple Star (SS) Mode (leaf emission):} This emission consists of a \cnot[cnot] followed by a Hadamard on the photonic qubit, creating a leaf node attached to the emitter node used for the emission. \\\textit{Remark}: repeated use of this mode creates a star graph centered at the emitter node (see Fig.~\ref{fig:em_SS}).
\begin{figure}[H]
\centering
\includegraphics[width=0.95\columnwidth]{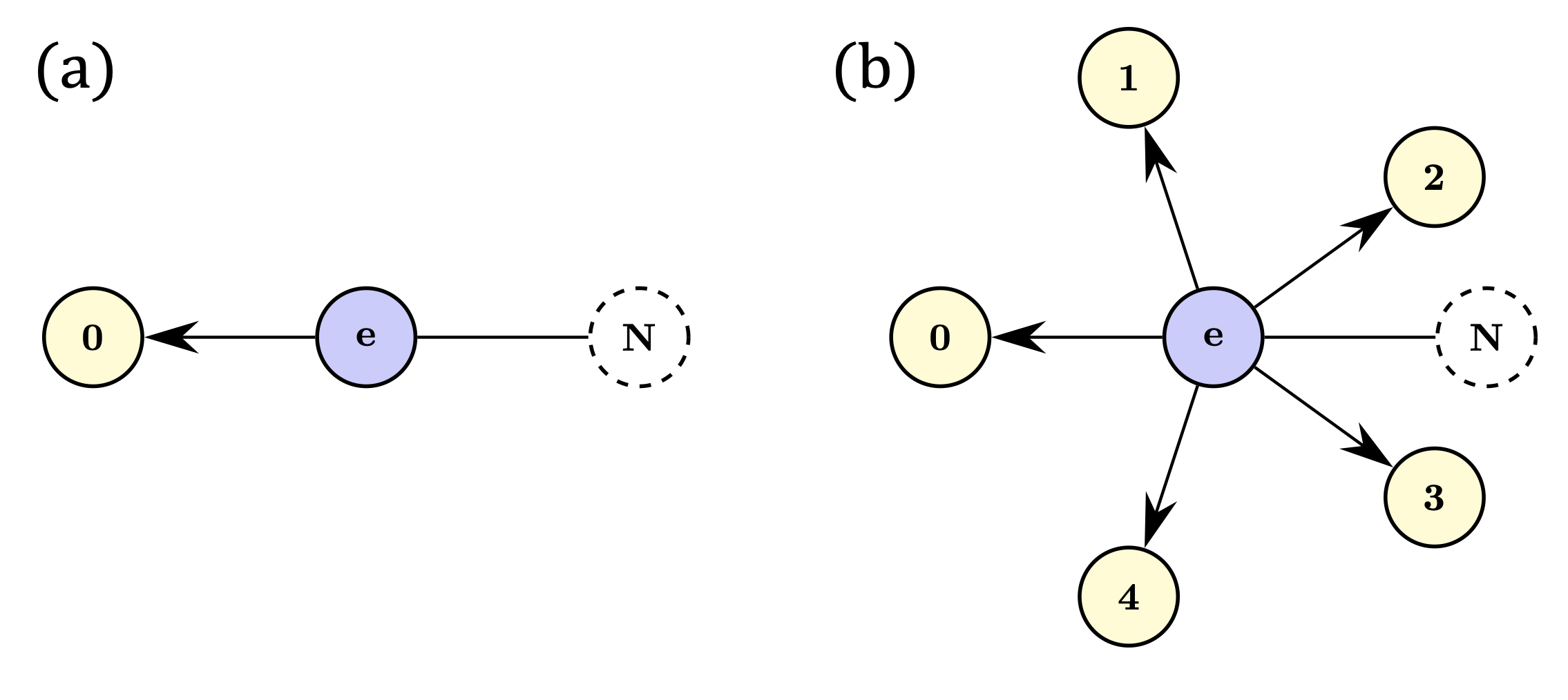}%
\caption{Simple star emission. \textbf{(a)} The state after emission of one photon in (SS) mode. \textbf{(b)} The state after sequential emission of multiple photons using the same mode.}
\label{fig:em_SS}
\end{figure}
\item[--] \textbf{Star (S) Mode:} This emission adds a photonic node to the system, which is---upon emission---only connected to all neighbors of its emitter and not the emitter itself. This happens if we apply LC on a neighbor of the emitter node, then apply LC on the emitter before a leaf emission, followed by LC on the emitter and LC on the same neighbor of the emitter. \\\textit{Remark}: repeated use of this mode results in a star graph with the emitter being one of the leaf qubits (see Fig.~\ref{fig:em_S}).
\begin{figure}[H]
\centering
\includegraphics[width=0.95\columnwidth]{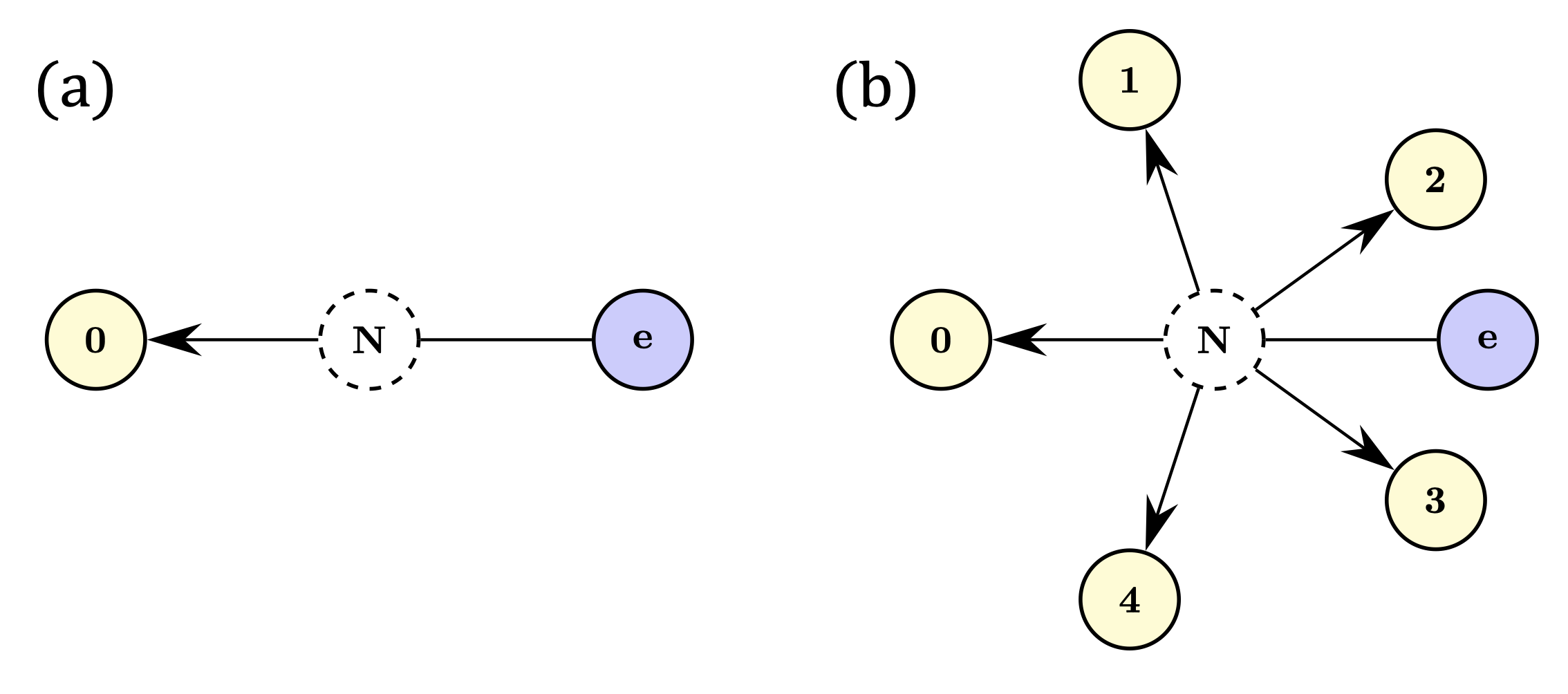}%
\caption{Star emission. \textbf{(a)} The state after emission of one photon in (S) mode. \textbf{(b)} The state after sequential emission of multiple photons using the same mode.}
\label{fig:em_S}
\end{figure}
\item[--] \textbf{Connected Star (CS) Mode:} Similar to the star mode, but the photon is also connected to its emitter (see Fig.~\ref{fig:em_CS}). A leaf emission (SS mode) followed by LC on emitter gives the desired result. \\\textit{Remark}: repeated use of this mode results in the creation of a complete graph.
\begin{figure}[H]
\centering
\includegraphics[width=0.95\columnwidth]{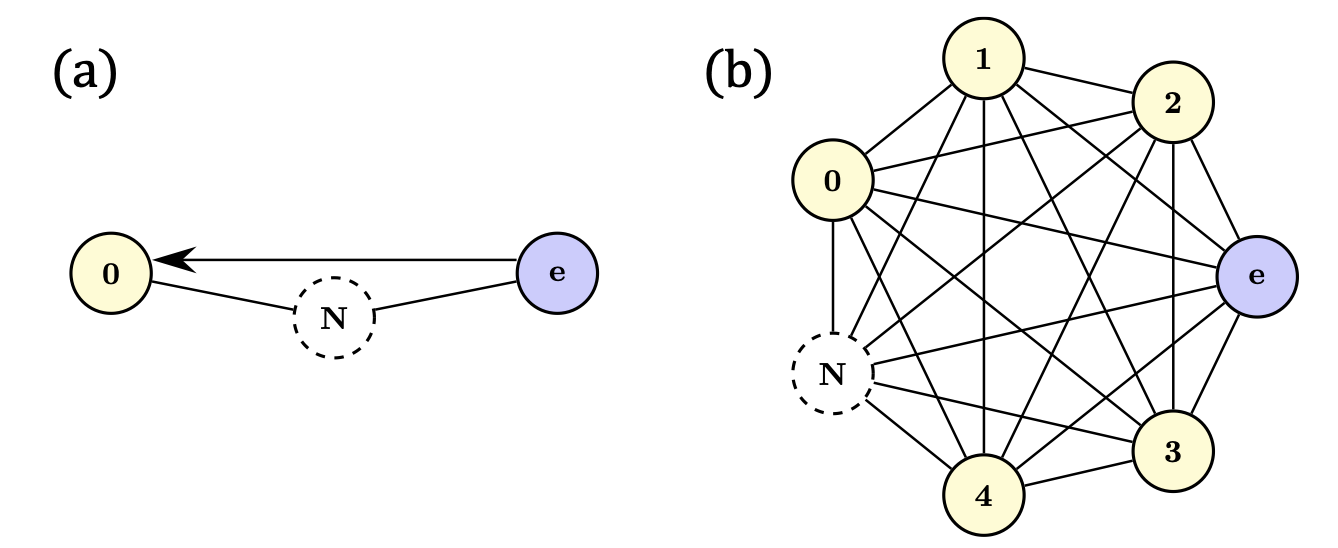}%
\caption{Connected star emission. \textbf{(a)} The state after emission of one photon in (CS) mode. \textbf{(b)} The state after sequential emission of multiple photons using the same mode.}
\label{fig:em_CS}
\end{figure}

\end{itemize}

\subsection{Decoupling}

A decoupling operation corresponds to performing a $Z$-basis measurement on a qubit and applying $Z^M$ gates on the neighbors of its corresponding node in the graph, depending on the measurement result $M\in\{0, 1\}$. The effect on the graph is simply the removal of the node and any edge incident on it.

\section{Graph Builder Algorithm}\label{sec:protocol}

We provide an exhaustive case analysis considering all possible initial configuration of the intermediate physical graph state $G'(n)$ at the beginning of step $n$, and then provide directions to complete the rest of the steps in \cref{alg:inductive_generation}. In particular, we provide explicit sequences of operations in terms of the introduced graphical transformations to make the transformation $G'(n)\to G'(n+1)\in \text{GenSet}(n+1)$ for all possible scenarios. This completes the proof (by exhaustion) of possibility of the induction step, and at the same time outputs the generation recipe when all transformation steps from $0$ to $N-1$ are put together. The result can be readily translated into a corresponding quantum circuit that when applied on a set of emitter qubits, creates the target photonic graph state.

Let us begin by proving the correctness of the \hyperref[base]{base case} in the induction. We assume the emission of the first photon (labeled $0$) by an emitter ($e_0$) using leaf emission (\hyperref[SS]{SS}) mode. After this, the inside subsystem includes a single node which is entangled to the emitter $e_0$ on the outside. The bipartite entanglement entropy is equal to 1, and the biadjacency matrix after emission $B(n=1)$ is a single row ($1\times N-1$) matrix having the future connections of the emitted photon encoded in it. Also, the physical biadjacency matrix $B'(n=1)$ is a single entry ($1\times 1$) matrix equal to 1, as there is only one active emitter in the system connected to the only photon. The first and only emitter row then will be $R_{e_0} = R_0$, which is the first row of the $B(n=1)$ matrix. The inside subgraph has only a single node and thus has no edge structure on both physical and target cases. As a result, conditions \ref{I}-\ref{III} are all satisfied for the base case (step 0).

Before finding the induction step operations, let us define two useful sets of emitters associated with the next photon (labeled $n$) in queue for emission:
\begin{itemize}[itemindent=-12pt]\label{j_set}
    \item Outside Neighbors Corresponding Emitter Set \textbf{(\textit{$\mathcal{J}$})}: 
\end{itemize}
denotes the set of emitters $\{e_j\}$ such that $R_{\text{new}}$ (as defined in Fig.~\ref{fig:evolution}) is the sum of their corresponding emitter rows, i.e., if we can write
    \begin{align}\label{j_eq_1}
     R_{new} = \sum_{j\in \mathcal{N}} R_{e_j}(n+1)
    \end{align}
    then
    \begin{align}\label{j_eq_2}
    \mathcal{J} =\left\{e_j\;\middle|\; j\in \mathcal{N} \right\}
    \end{align}
where $R_{new}$ is equal to the last row of $B(n+1)$. Note that the updated emitter rows $R_{e_j}(n+1)$ always form a basis for the row space of $B(n+1)$ and such an expansion of $R_{new}$ is always possible. These emitters that can act as representatives of the future (outside) neighbors of photon $n$. 
\begin{itemize}[itemindent=-12pt]\label{k_set}
    \item Inside Neighbors Corresponding Emitter Set \textbf{(\textit{$\mathcal{K}$})}: 
\end{itemize}
    is the set of all emitters $\{e_k\}$ whose rows indicate connectivity to the photon that is to be emitted in the current step, i.e., the first element of the emitter row is equal to one.
    \begin{align}\label{k_eq}
    \mathcal{K} = \left\{ e_k \;\middle|\; \big[ R_{e_k}(n) \big]_0 = 1 \right\}
    \end{align}
\noindent These are the emitters that can act as representatives of the new photon's required neighborhood---according to the eligibility conditions---among the prior emitted (inside) photons.

We can now look back at our \hyperref[base]{inductive algorithm}; building upon the base case and utilizing the discussed eligibility conditions, and in particular eqs.~(\ref{J}) and (\ref{eq:inside_neighbors}), we provide the following requirements in accordance with the \hyperref[protocol:1]{inductive step}:

\begin{enumerate}[label={(\roman*)}]
    \item The required neighborhood of the next photon consists of all emitters in $\mathcal{J}$, plus all photons present in the symmetric difference of the neighborhoods of all emitters in $\mathcal{K}$.
    \item For the rest of the nodes, from each photon $i$, there must be an edge to the emitter $e_j$ if $j\in \mathcal{N}(i)$, where $\mathcal{N}(i)$ is found according to \cref{J}.
\end{enumerate}

\noindent The implementation of the above adjacency requirements is expanded in detail for all possible scenarios in the following case analysis, which corresponds to the operations required in \cref{alg:inductive_generation} for each $n\in\{1,\dots,N-1\}$.

\subsection{Case Analysis}
The breakdown of possible scenarios for an intermediate physical graph $G'(n)\in\;$GenSet(n) at some step of the generation can be done based on the biadjacency matrices $B(n)$ and $B(n+1)$, both known to us for any given target graph $G$ and step $n$. In particular, we consider four overall scenarios corresponding to possible values for \hyperref[col_effect]{column effect}, \hyperref[row_effect]{row effect} being equal to (0, +1), (0, 0), (-1, +1), or (-1, 0) respectively. The operations required  to complete transition from step $n$ to $n+1$ are then prescribed.

\begin{enumerate}[itemindent=44pt, leftmargin=0pt]  
    \renewcommand{\labelenumi}{\textbf{Case \Alph{enumi}:}}
    \item column effect $= 0$, row effect $= +1$
    \label{sec:case_analysis}
\end{enumerate}

    The new row $(R_{new})$ is linearly independent and the rank is increased by 1.
    \begin{itemize}
        \item A new (isolated) emitter $e_i$ is be used for the next emission.
        \item Connect $e_i$ to inside of all $e_{k \neq i}$ $\in$ \textit{$\mathcal{K}$} \\(using e-to-inside operations).
        \item Emit the photon in (L) mode.
    \end{itemize}

\begin{figure}[h]
\centering
\includegraphics[width=.95\columnwidth]{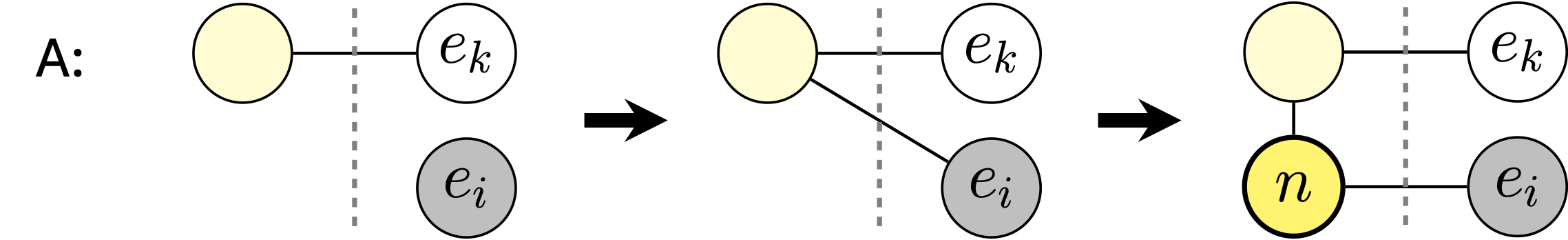}%
\caption{Case A. From left to right, the state of the physical graph at the beginning of the step (only showing the affected region), just before emission, and after emitting the new photon. The node $e_i$ is the chosen emitter. $e_k$ is a representative of the emitters in the set $\mathcal K$. Node $n$ is the new emitted photon. }
\label{fig:A}
\end{figure}

\begin{tcolorbox}[colback=gray!5, colframe=black, arc=2mm, boxrule=0.1mm]\label{rationale: A}

                \textbf{Rationale:} Since rank is increased by 1, we need a new emitter row to be added to the previous set which forms a basis for the row space of $B$ matrix. This is done by introducing a new active emitter $(e_i)$ to the system and setting the new photon's row $R_{new}$, which we know is linearly independent from the rest of the rows, to be the row assigned to the new emitter $(R_{e_i})$. Conditions \ref{I} and \ref{II} are thus satisfied. The required inside connection of the new photon are established by connecting $e_i$ to the inside of all $e_k\in \mathcal{K}$ before an emission in L mode (Condition \hyperref[III-2]{$(\text{III}')$} satisfied). See Fig.~\ref{fig:A} for the process.
\end{tcolorbox}
\noindent For the remaining cases, the rationale can be found in appendix \ref{app: rationale}, explaining how the prescribed operations result in a state that satisfies conditions \ref{I}-\ref{III}.
    \begin{enumerate}[itemindent=44pt, leftmargin=0pt]  
    \renewcommand{\labelenumi}{\textbf{Case \Alph{enumi}:}}
    \setcounter{enumi}{1}
    \item column effect $= 0$, row effect $= 0$
    \end{enumerate}
    
    No new emitter is needed in this case since rank is unchanged. Since the new row $R_{new}$ is not linearly independent, there are two possible scenarios: it is either a zero vector, $R_{\text{new}}=0$, or else it can be written as a linear combination of other rows 
    $R_{\text{new}} = \Sigma R_j$.
\begin{enumerate}[itemindent=55pt, leftmargin=0pt]    
    \renewcommand{\labelenumi}{-- Case B\arabic{enumi}:}
    \item{$R_{\text{new}}=0$}
        \label{case:B1}
        \begin{itemize}
            \item The emitter $e_i$ is chosen from the set \textit{$\mathcal{K}$}.  
            \item Connect $e_i$ to inside of all emitters $e_{k\neq i}$ $\in$ \textit{$\mathcal{K}$}. \\(e-to-inside operations)
            \item Emit the photon in (S) mode.
            \item Disconnect $e_i$ from inside of $e_{k\neq i}$ $\in$ \textit{$\mathcal{K}$}. \\(e-to-inside operations)
        \end{itemize}

    \item{$R_{\text{new}} \neq 0$}
        \begin{enumerate}[label=(\roman*), itemindent=16pt, leftmargin=8pt]
            \item \textbf{{if \textit{$\mathcal{K}$} $=\O$}}

            \begin{itemize}
                \item The emitter $e_i$ is chosen from the set \textit{$\mathcal{J}$}.
                \item Connect all $e_{j \neq i} \in \,$\text{\textit{$\mathcal{J}$}} to the inside of $e_i$.
                \item Emit with (SS) mode.
            \end{itemize}

            \item \textbf{else if \textit{$\mathcal{K}$} $\cap$ \textit{$\mathcal{J}$} $=\O$}

            \begin{itemize}
                \item The emitter $e_i$ is chosen from the set \textit{$\mathcal{K}$}.
                \item Connect $e_i$ to inside of all $e_{k \neq i}$ $\in$ \textit{$\mathcal{K}$}.
                \item Connect $e_i$ to all $e_{j}\,\in$  \textit{$\mathcal{J}$}.
                \item Emit with (S) mode.
                \item Disconnect $e_i$ from all $e_{j}$ $\in$ \textit{$\mathcal{J}$}
                \item Disconnect $e_i$ from inside of all $e_{k \neq i}$ in \textit{$\mathcal{K}$}
            \end{itemize}

            \item \textbf{else if \textit{$\mathcal{K}$} $\cap$ \textit{$\mathcal{J}$} $\neq \O$}

            \begin{itemize}
                \item The emitter $e_i$ is chosen from the \textit{$\mathcal{K}$} $\cap$ \textit{$\mathcal{J}$}.
                \item Connect $e_i$ to inside of all $e_{k \neq i}$ in \textit{$\mathcal{K}$}.
                \item Connect $e_i$ to all $e_{j \neq i}$ $\in$ \textit{$\mathcal{J}$}.
                \item Emit with (CS) mode if $|\textit{$\mathcal{K}$}$ $ \cap$ $\textit{$\mathcal{J}$}|$ is odd and mode (S) if it is even.
                \item Disconnect $e_i$ from all $e_{j \neq i}$ $\in$ \textit{$\mathcal{J}$}.
                \item Disconnect $e_i$ from inside of all $e_{k \neq i}$ $\in$ \textit{$\mathcal{K}$} (*)
            \end{itemize}
      \end{enumerate}
    \end{enumerate}
    
\begin{enumerate}[itemindent=44pt, leftmargin=0pt]  
    \renewcommand{\labelenumi}{\textbf{Case \Alph{enumi}:}}
    \setcounter{enumi}{2}
    \item column effect $= -1$, row effect $= +1$
    \end{enumerate}
 
Initially, the emitter rows form a basis for rows of biadjacency matrix $B$. Since column effect $= -1$, the rank of $B$ matrix is reduced after the column removal; in this updated matrix we can always find one linearly dependent set of emitter rows $\left\{R_{e_m} \mid e_m \in \mathcal{M} \right\}$ whose sum of members is equal to zero. Here, $\mathcal{M}$ is the set of emitters corresponding to these linearly dependent rows.

    \begin{itemize}
        \item The emitter $e_i$ is chosen from the set $\mathcal{K} \cap \mathcal{M}$.
        
        \item Connect all $e_{m \neq i} \in\mathcal{M}$ to inside of $e_i$.
        \item Connect $e_i$ to inside of all $e_{k \neq i}$ $\in$ \textit{$\mathcal{K}$}.
        \item Emit with (L) mode.
        \end{itemize}

\begin{enumerate}[itemindent=44pt, leftmargin=0pt]  
    \renewcommand{\labelenumi}{\textbf{Case \Alph{enumi}:}}
    \setcounter{enumi}{3}
    \item column effect $= -1$, row effect $= 0$
    \end{enumerate}    
The same arguments of previous case (\textbf{C}) regarding the column effect and the linearly dependent set $\left\{R_{e_m} \mid e_m \in \mathcal{M} \right\}$ holds true. In addition, since row effect $= 0$, the new row $R_{new}$ can be written as a combination of the emitter rows. We study two sub-cases based on the size of the dependent set:
\begin{enumerate}[itemindent=55pt, leftmargin=0pt] 
        \renewcommand{\labelenumi}{-- Case D\arabic{enumi}:}
        \item{$|\mathcal{M}| = 1$}
        
            One of the emitter rows $R_{e_m}$ must have become a zero vector after the column removal. 
            \begin{itemize}
            \item The emitter $e_{i}$ is chosen to be the same emitter $e_m$ affected by the column removal.
            \item Connect $e_i$ to inside of all $e_{k \neq i} \in$ \textit{$\mathcal{K}$}.
            \item Connect $e_i$ to all $e_{j} \in$ \textit{$\mathcal{J}$}.
            \item Emit with (L) mode.
            \item Decouple the emitter $e_i$.
            \end{itemize}

    \item{case $|\mathcal{M}| > 1$}
        \begin{enumerate}[label=(\roman*), itemindent=16pt, leftmargin=8pt]
            \item \textbf{if $R_{new}$ = 0}\\
            
                Since $\Sigma R_{e_m} = 0$ for $e_m$ $\in$ $\mathcal{M}$ over the field $\mathbb{Z}_2$, one can always choose an arbitrary $e_i$ $\in$ $\mathcal{M}$ such that $\Sigma R_{m\neq i} = R_i$. 
                Therefore, one of the previous emitter rows can now be written as a sum of others.
            \begin{itemize}
                \item The emitter $e_{i}$ is chosen from the set $\mathcal{M}$.
                \item Connect all $e_{m \neq i}$ $\in \mathcal{M}$ to inside of $e_i$.
                \item Connect $e_i$ to inside of all $e_{k \neq i} \in$ \textit{$\mathcal{K}$}.
                \item Emit with (L) mode.
                \item Decouple the emitter $e_i$.
                \end{itemize}

           \item \textbf{else if $R_{new} \neq 0$} 
           
            \begin{enumerate}[label=(\alph*)]
                \item if $\mathcal{M} \cap \mathcal{K} \not\subset \mathcal{J} $: 
                     \begin{itemize}
                     \item The emitter $e_{i}$ is chosen from the set $(\mathcal{M} \cap \text{\textit{$\mathcal{K}$}}) - \text{\textit{$\mathcal{J}$}}$ 
                    \item Connect all $e_{m \neq i}$ $\in \mathcal{M}$ to inside of $e_i$.
                    \item Connect $e_i$ to inside of all $e_{k \neq i} \in$ \textit{$\mathcal{K}$}.
                    \item Connect $e_i$ to all $e_{j} \in$ \textit{$\mathcal{J}$}.
                    \item Emit with (L) mode.
                    \item Decouple the emitter $e_i$.
                    \end{itemize}

                \item if $\mathcal{M} \cap \text{\textit{$\mathcal{K}$}} \subset \text{\textit{$\mathcal{J}$}}$:
                \begin{itemize}
                    \item The emitter $e_{i}$ is chosen from the set $\mathcal{M} \cap \text{\textit{$\mathcal{K}$}}$.
                    \item Connect all $e_{m \neq i}$ $\in \mathcal{M}$ to inside of $e_i$.
                    \item Connect $e_i$ to inside of all $e_{k \neq i} \in$ \textit{$\mathcal{K}$}
                    \item Connect $e_i$ to all $e_{s \neq i} \in \text{\textit{$\mathcal{J}$}}\hspace{1mm}\Delta \hspace{1mm}\mathcal{M}$ (symmetric difference of the two sets).
                    \item Emit with (L) mode.
                    \item Decouple the emitter $e_i$.
                    \end{itemize}
\end{enumerate}
\end{enumerate}
\end{enumerate}

\subsection{Output}
\begin{figure*}[t]
\centering
\includegraphics[width=1\textwidth]{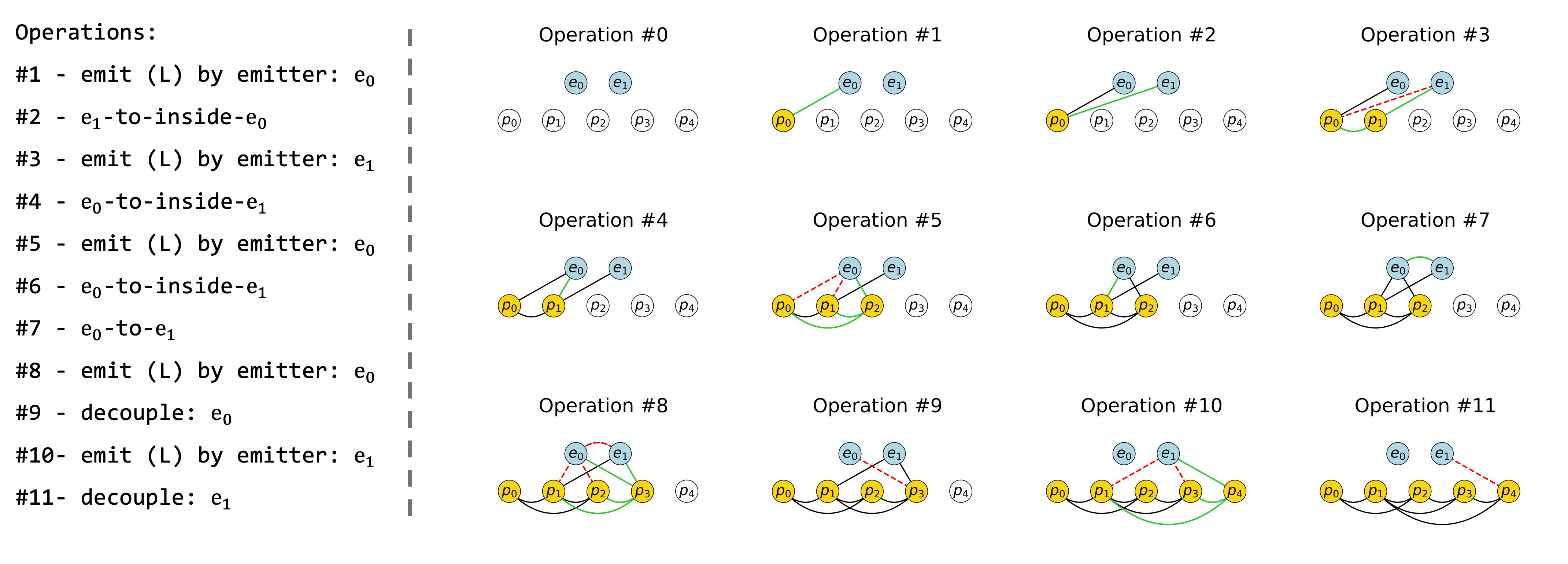}%
\caption{The generation recipe in form of a sequence of the elementary operations which is the final output of the algorithm. The emission mode for each photon is specified in the parentheses; in this case all emission are in the ``Linear" mode. The corresponding sequential evolution of the physical graph state under each operation is demonstrated. The nodes are placed as a linear array with emitter node (blue) on top and photonic nodes on the bottom. Each photonic node is shaded (yellow) once emitted. For each graph operation, the highlighted (green) edges indicate the new connections created and the dashed (red) edges indicate the removed edges with respect to its previous state. }
\label{fig:evol}
\end{figure*}

The case analysis outputs the list of operations needed to accomplish the emission of a new photon while complying with the outlined eligibility conditions for each step. Figure~\ref{fig:evol} shows the implementation of the generation algorithm for the same target graph (shown in a flattened form) used in Fig.~\ref{fig:1}. The generation recipe, written in terms of the elementary operations, and the corresponding intermediate graphs are presented. The recipe can also be represented as a quantum circuit if each graph operation is replaced by its equivalent gates provided in \cref{sec:graphical}.

We remark that one can simplify such recipes if two or more of the operations can be merged or canceled together. For instance, in Fig.~\ref{fig:evol}, operations \#6 and \#7 can be merged together and be replaced by a single \textit{$e_0$-to-inside-$e_1$-connect} operation, reducing the number of emitter-emitter interactions by one unit in this process. As the recipe is in term of a sequence of graph operations, it is straightforward to find commutation rules for these elementary operations (due to the limited size of this set and the simplicity of tracking the graphical transformations) and search for potential simplifications by checking the possibility of bringing together the operations that can be merged/canceled. We have implemented this simplification algorithm to be applied on the output recipe before compiling it to a quantum circuit. The details can be found in appendix \ref{app: simplification}.

\section{performance results}
In this section, we evaluate the performance of the proposed generation algorithm, focusing on the number of emitter-emitter two-qubit gates required during the generation process. Benchmarking is carried out against the time-reversed generation method introduced by Li et al.~\cite{li_photonic_2022}, which is another algorithm that also uses the minimum number of emitters for the generation of arbitrary graph states. Two implementations of this algorithm are used as the baselines for comparison: the version available in the GraphiQ package \cite{GraphiQ} called ``\classname{TimeReversedSolver}'', here referred to as \textbf{\textit{time-reversed-1}}, and a more recent implementation by Takou et al.~\cite{Takou2024} called the ``Naive approach'' in Ref.~\cite{Takou2024}, here referred to as \textbf{\textit{time-reversed-2}}. The latter includes an additional algorithmic step that checks, for each photon, whether an emission can proceed without applying any two-qubit gates to the emitters and opts for such cases when possible.

Notably, optimization efforts have resulted in a number of generation schemes derived from Li et al.~\cite{li_photonic_2022}'s algorithm, either in its original form or with minor modifications. These approaches, employing heuristic or brute-force, target reductions in preparation cost, such as the number of \cnot[cnot] gates \cite{ghanbari_optimization_2023, Takou2024, ren2025}, or in circuit depth \cite{kaur_resource-efficient_2024}. In this section, the reported values reflect the performance \textit{prior} to allowing for any optimization over internal algorithmic degrees of freedom. Such aspects will be addressed in \cref{sec:opt}, where we discuss how our proposed algorithm can serve both as an alternative baseline for the already available optimizing methods and also as a foundation for developing new optimization strategies---by providing better starting points and a wider range of degrees of freedom.

The details on the explicit forms of the graphs and emission orders used in the analyses presented here can be found in the supplemental materials \cite{supp}. 

\subsection{Random Graphs}
\begin{figure}[ht]
    \centering      \includegraphics[width=0.98\columnwidth]{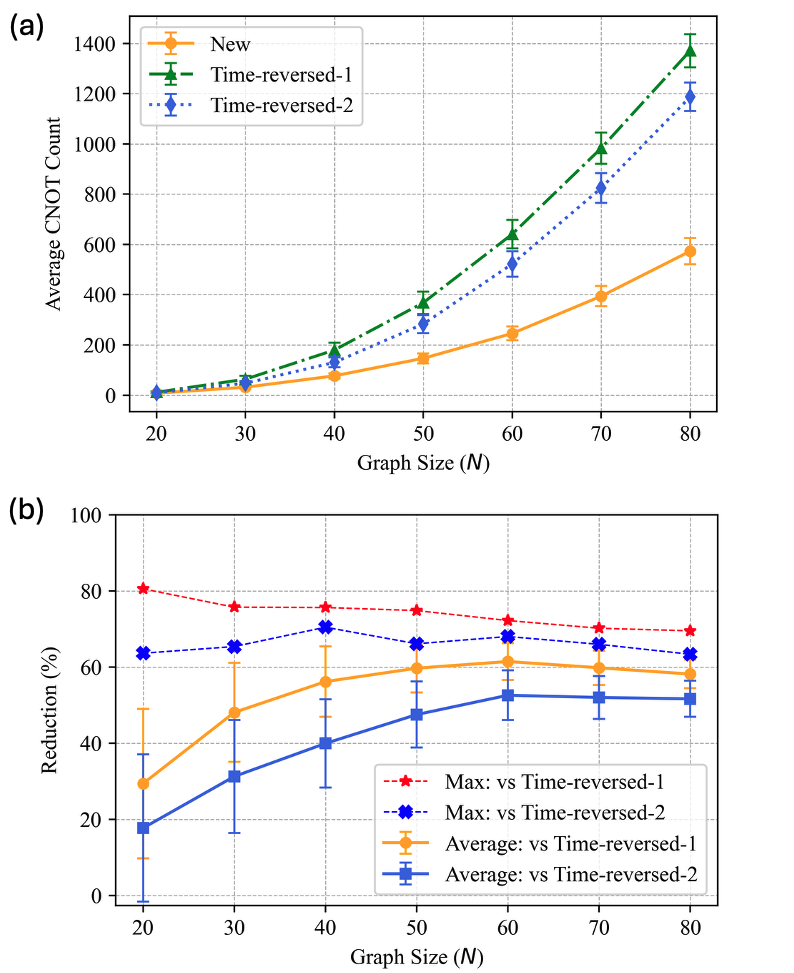}
    \caption{Performance comparison in terms of the number of \cnot[cnot] gates required to prepare random graphs. \textbf{(a)} Average \cnot[cnot] gate count for different graph sizes, from $N=20$ to $N=80$, using the proposed algorithm (solid yellow) and two alternatives, \textit{time-reversed-1} and \textit{time-reversed-2}, as defined in the text. The average is taken over 1000 random graphs for each $N$, with error bars indicating one standard deviation. \textbf{(b)} Relative reductions in \cnot[cnot] gate usage when employing the proposed algorithm compared to the alternatives. Average (solid) and maximum (dashed) reductions are shown when the proposed algorithm is compared against the two alternatives for each $N$, over a sample of 1000 random graphs.}
    \label{fig:random_vs_n}
\end{figure}

The number of two-qubit gates utilized to generate random connected graphs for a range of graph sizes is provided in Fig.~\ref{fig:random_vs_n}(a). Each data point represent the average value over a sample of 1000 random graphs of the same size with an edge density---the ratio of the number of present edges to the maximum possible number of edges in a graph---of 10\%. 
Figure~\ref{fig:random_vs_n}(b) shows the average and maximum of achieved reductions of our method relative to the alternative algorithms for each size. Average reductions of up to 61\% and 52\%, and maximum reductions of up to 80\% and 70\%, are observed relative to the \textit{time-reversed-1} and \textit{time-reversed-2} methods, respectively. 
It is worth noting that, on average, the number of two-qubit gates also depends on the edge density in random graphs. This behavior and performance as a function of edge density are analyzed in appendix \ref{app:edge_dep}. The scaling of runtime with both size and edge density is also discussed in appendix \ref{app:runtime}.

\begin{figure*}[ht]
    \centering      \includegraphics[width=0.98\textwidth]{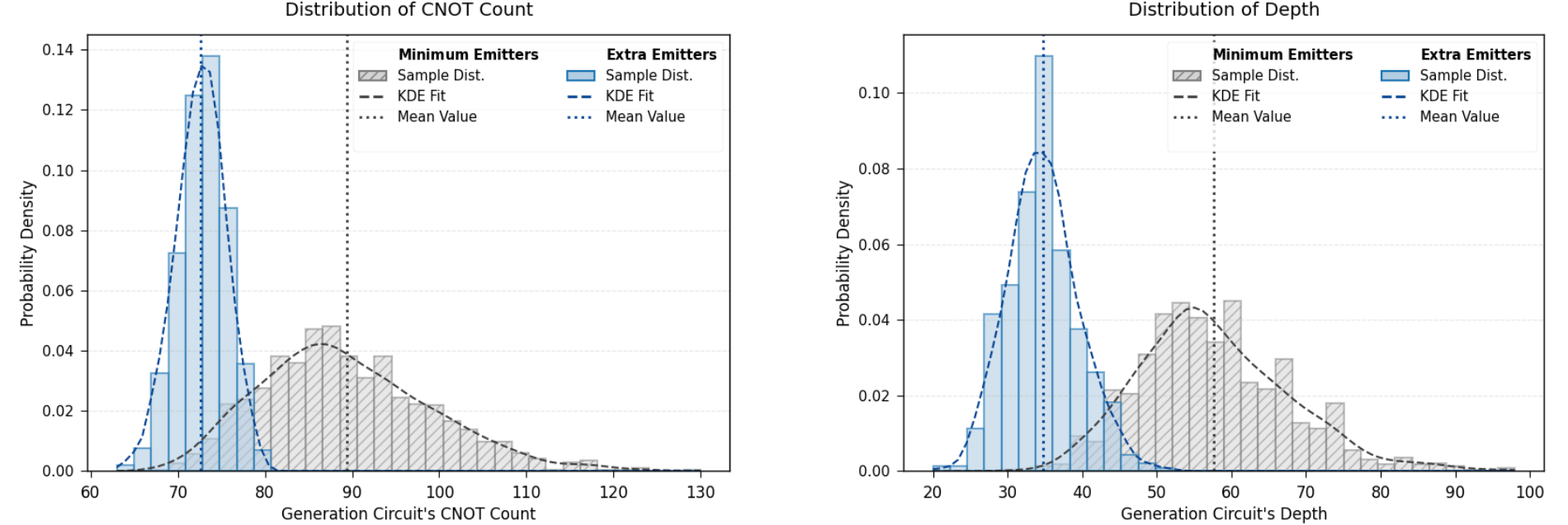}
    \caption{\textcolor{black}{Distribution of \cnot[cnot] gate count (left) and circuit depth (right) compared between using minimum emitters versus allowing for extra emitters over a sample of random graphs. A clear shift to the lower cost regime is visible for both metrics with a lower mean value and a tighter distribution when extra emitters are employed. The average number of emitters is increased from 14 to 17 between the two methods. The demonstrated results are for a sample of 1000 random 30-node graphs with a 20\% edge density.}}
    \label{fig:extra_emitter_dist}
\end{figure*}

\textcolor{black}{
Note that the presented results are obtained using the minimum number of emitters possible to generate each graph. This constraint on the number of emitters used in the generation process can be relaxed to allow for potential reductions in other metrics of interest \footnote{For more details on this modification of the Graph Builder algorithm refer to \cref{sec:morethanminimal}.}. To showcase this, we have employed a greedy strategy to use a new emitter in each step of the generation whenever the introduction of the extra emitter results in a reduction in the number of \cnot[cnot] gates used in the current step that leads to the emission of the next photon. The outcome of such an analysis can be found in Fig.~\ref{fig:extra_emitter_dist}, where the generation circuit for a sample of random graphs were found once with the Graph Builder algorithm under the minimum emitter constraint and another time under the extra emitter strategy. A simultaneous reduction in the usage of the \cnot[cnot]s and two-qubit depth of the generation circuit (e.g. the maximum number of two-qubit gates the emitters have to go through before being measured) is evident with an average (maximum) reduction of $38$\% ($72$\%) in depth and $18$\% ($45$\%) in CNOT gates for each case. The average of the number of emitters used for each case over the random sample was moved from 14 to 17, indicating that the mentioned advantages can be acquired at the cost of using approximately $3$ extra emitters in this case. For an explicit example of reducing the \cnot[cnot] cost by using additional emitters, see also section~D of the supplementary material \cite{supp}.}

\subsection{Special Cases}
\begin{figure}[h]
    \centering      
    \includegraphics[width=0.95\columnwidth]{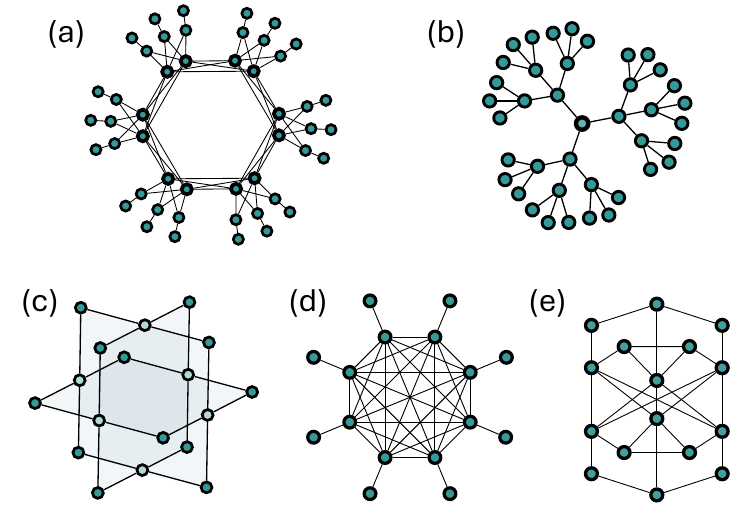}
    \caption{Examples for different classes of graphs. \textbf{(a)} 6-ring:~(4, 2) parity encoded. \textbf{(b)} Branching tree:~\{3, 3, 3\}. \textbf{(c)} RHG lattice unit cell. \textbf{(d)} 16-node~RGS. \textbf{(e)} 16-node generalized RGS.}
    \label{fig:special_graphs}
\end{figure}
In what follows, we present the \cnot[cnot] cost for preparing several classes of graphs central to many error- and loss-tolerant architectures in photonic quantum computing and communication protocols. In particular, we also study how the preparation cost scales with graph size or encoding parameters, as relevant for realistic implementations. The considered classes include: 6-ring parity encoded graphs \cite{bartolucci_fusion-based_2023} with use cases in loss-tolerant fusion-based architectures; branching tree cluster states \cite{Varnava-loss-2006}, proposed as error-correcting codes providing loss tolerance in measurement-based systems; three-dimensional Raussendorf--Harrington--Goyal~(RHG) lattice \cite{Raussendorf-fault-2006, Raussendorf-topological-2007}, designed as a fault-tolerant resource for measurement-based quantum computing; and all-photonic repeater graph state \cite{azuma_all-photonic_2015}, along with a generalized version introduced in Ref.~\cite{Bell2023}.

\subsubsection{Encoded 6-Rings}
\textcolor{black}{
Six-ring graphs have been proposed as resource states for fusion-based quantum computing \cite{bartolucci_fusion-based_2023}. Here, we consider the parity-encoded variant of these states, which offers higher loss thresholds for both computation and encoded-fusion processes \cite{Ralph-loss-2005, Lee-parity-2023, Song2024}. We analyzed the \cnot[cnot] preparation cost for an $(n,m)$-encoded six-ring graph, as defined in Ref~\cite{Lee2023}, made up of $n\times m\times 6$ physical qubits (see Fig.~\ref{fig:special_graphs}).
}
\textcolor{black}{
As with many other parameterized classes of graphs, repeated patterns can emerge as the defining parameters are varied to scale the graph size. This allows one to identify circuit blocks in the generation circuit that repeat a number of times, depending on the parameter that controls the size of the graph.
}
\textcolor{black}{
By inspecting this behavior as we change the values of $n$ and $m$, we find that the \cnot[cnot] cost depends only on the first encoding parameter $n$ and is given by $6n+4$ when using the our proposed algorithm, with only three emitters used in the generation process. In contrast, the alternative methods, \textit{time-reversed-1} and \textit{time-reversed-2}, require $8n-4$ and $9n-7$ \cnots[cnots] to prepare the same state. This is confirmed by mathematical induction, as increasing $n$ to $n+1$ results in the addition of a circuit block with a certain number of \cnot[cnot] gates to the generation circuit of the new graph in each case, leading to the linear scaling with $n$.}

\textcolor{black}{
The explicit form of such circuit blocks for each of three generation protocols, along with the numerical support for exact \cnots[cnots] relations can be found in section C of supplementary material \cite{supp}.}

\textcolor{black}{
Considering the asymptotic behavior for large $n$, which is of interest since the error-correction properties of the encoded graphs improve with increasing $n$ and $m$, we see that the $8n$ and $9n$ scaling of the alternative methods correspond to overheads of 33\% and 50\%, respectively when compared to the $6n$ case of the new algorithm.
}

\subsubsection{Branching Tree Cluster States}
The branching tree cluster states are characterized by a branching vector $\vec{b} = \left(b_1, \dots, b_d\right)$, where $b_i$ denotes the number of branches originating from each node at the $i$-th depth level. See Fig.~\ref{fig:special_graphs} for an example. These states can be used as error correction codes when attached to qubits in other graph states, providing up to 50\% loss tolerance \cite{Varnava-loss-2006}, e.g., in a 2D cluster state for universal measurement-based quantum computing (MBQC). The code can also be used to reduce the failure probability of encoded-fusion operations, for instance in the generation of 3D fault-tolerant MBQC lattices \cite{li_resource_2015}. For a homogeneous, depth~$d$, branching vector where all $b_i$ are equal to a fixed branching parameter $b$, our algorithm generates the tree graph using $b^{d-1}-1$ \cnot[cnot] gates and $d$ emitters. As seen in \Cref{tab:1}, a cost comparison with alternative methods shows possible improvements of up to 20\% compared to \textit{time-reversed-2}, for different depth and branching parameters.

\begin{table*}[ht]
\caption{Comparison of \cnot[cnot] gate counts for various graph states using the New, \textbf{TR1}, and \textbf{TR2} methods, referring to the time reversed generation algorithms defined in text. The number of emitters required ($n_e$), number of nodes ($N$), and number of edges ($|E|$) in each graph are also reported. The percentage \cnot[cnot] count reductions are calculated for our algorithm relative to the respective reference methods.}
\label{tab:1}
\centering
\small
\begin{tabular*}{\textwidth}{l @{\extracolsep{\fill}} c c c c c c c c}
\toprule
\textbf{Graph} & {\textbf{$n_e$}} & {\textbf{$N$}} & {\textbf{$\mid E \mid$}} & {\textbf{\makecell{ \# \cnots[CNOTs] \\ New}}} & {\textbf{\makecell{ \# \cnots[CNOTs] \\ TR2}}} & {\textbf{\makecell{Reduction\% \\ vs TR2}}} & {\textbf{\makecell{ \# \cnots[CNOTs] \\ TR1}}} & {\textbf{\makecell{Reduction\% \\ vs TR1}}} \\
\midrule
Tree (3,3,3) & 3 & 40 & 39 & 8 & 10 & 20.0 & {8} & {0.0} \\
Tree (4,4,4) & 3 & 85 & 84 & 15 & 17 & 11.8 & {15} & {0.0} \\
Tree (3,3,3,3) & 4 & 121 & 120 & 26 & 33 & 21.2 & {28} & {7.1} \\
Tree (4,4,4,4) & 4 & 341 & 340 & 63 & 69 & 8.7 & {65} & {3.1} \\
Tree (5,5,5,5) & 4 & {781} & {780} & 124 & 144 & 13.9 & 126 & {1.6} \\
Tree (3,3,3,3,3) & 5 & 364 & 363 & 80 & 96 & 16.7 & {83} & {3.8} \\
Tree (3,3,3,3,3,3) & 6 & 1093 & 1092 & 242 & 291 & 16.8 & {246} & {1.6} \\
\cmidrule{1-9}
RHG (1,1,1) & 4 & 18 & 24 & 14 & 14 & 0.0 & 15 & 6.7 \\
RHG (2,1,1) & 4 & 31 & 44 & 28 & 29 & 3.4 & 34 & 17.6 \\
RHG (3,1,1) & 4 & 44 & 64 & 42 & 44 & 4.5 & 50 & 16.0 \\
RHG (2,2,1) & 7 & 53 & 80 & 56 & 61 & 8.2 & 65 & 13.8 \\
RHG (3,2,1) & 7 & 75 & 116 & 84 & 96 & 12.5 & 109 & 22.9 \\
RHG (3,3,1) & 10 & 106 & 168 & 126 & 150 & 16.0 & 151 & 16.6 \\
RHG (2,2,2) & 12 & 90 & 144 & 108 & 144 & 25.0 & 151 & 28.5 \\
RHG (3,3,2) & 17 & 179 & 300 & 240 & 369 & 35.0 & 402 & 40.3 \\
RHG (3,3,3) & 24 & 252 & 432 & 354 & 610 & 42.0 & 650 & 45.5 \\
\bottomrule
\end{tabular*}
\end{table*}

\subsubsection{RHG lattice}

The RHG lattice \cite{Raussendorf-fault-2006, Raussendorf-topological-2007}, designed for fault-tolerant MBQC \cite{raussendorf_one-way_2001, Raussendorf-measurement-based-2003}, consists of cubic unit cells with nodes located at the centers of the faces and edges of each cube (see Fig.~\ref{fig:special_graphs}). Here, we consider the \((L_x, L_y, L_z)\)-RHG lattice as defined in Ref.~\cite{Lee2023}, where each parameter \(L_{x,y,z}\) specifies the number of unit cells stacked along the corresponding axis. The \cnot[cnot] cost as the lattice size scales, comparing different algorithms can be found in \Cref{tab:1}. Improvements of up to 45\% and 42\% are achieved for the \((3,3,3)\)-RHG state relative to the \textit{time-reversed-1} and \textit{time-reversed-2} algorithms, respectively.

\subsubsection{Repeater Graph States}

Photonic repeater graph states (RGSs) serve as quantum repeaters without the need for matter-based memory qubits and are considered promising candidates for enabling long-distance quantum communication \cite{azuma_all-photonic_2015}. An $N$-qubit RGS consists of an $N/2$-node complete graph, with each node having an attached leaf qubit. Using our new algorithm, the \cnot[cnot] cost for preparing the original RGS structure is $N/2 - 2$, matching the optimal cost previously found by specialized methods targeting this class of graphs in Refs.~\cite{ghanbari_optimization_2023,Takou2024}. A generalized RGS structure (see Fig.~\ref{fig:special_graphs}) was proposed in Ref.~\cite{Bell2023}, offering improved performance under lossy conditions for photonic qubits. While the optimal preparation cost for arbitrary sizes of this generalized structure is still unknown, our method prepares a 16-qubit instance using 20 \cnots[cnots] and 7 emitters. In comparison, \textit{time-reversed-2} algorithm uses 29 \cnots[cnots] to achieve the same state, representing a 31\% improvement for the new method.

\section{Optimization Framework}\label{sec:opt}
In this section, we elaborate on how the introduced framework can unlock more cost-efficient solutions to the generation problem. First, we reiterate the most general and relevant cost metrics in terms of circuit parameters. Next, by leveraging the knowledge of the necessary and sufficient requirements (the eligibility conditions) for the intermediate states throughout the generation process, we identify available degrees of freedom in our algorithm. This allows us to develop a versatile optimization framework that provides a suite of tools to target various cost metrics. The flexibility in the optimization is of significance as not all photon emitting platforms face the same challenges, e.g., while coherence time of the emitter qubits might be the main bottleneck for the state fidelity in one platform, two-qubit interactions might be the dominant source of noise in another. 
Finally, we discuss a range of optimization strategies enabled by the introduced tools.
\subsection{Cost Metrics}
In emitter-based photonic graph generation circuits, common cost metrics, from the point of view of both quality of the output state and the feasibility of the process, include the number of photon sources (emitter qubits), the circuit depth, and the number of two-qubit gates between emitters \cite{li_photonic_2022, kaur_resource-efficient_2024, ghanbari_optimization_2023, Takou2024}. Below, we discuss each of these metrics in turn.

The emitter qubit count is desired to be as low as possible to minimize the technological and engineering overhead in both fabrication and maintaining coherent control of each emitter. Besides, providing all-to-all \cnot[cnot]-connectivity which allows arbitrary Clifford operation on emitter qubits as required, gets increasingly more challenging as the emitter set grows in size. To this end, our algorithm, by default, uses the minimum possible number of emitters (\cref{eq:min_emitter}) to generate the target state. Furthermore, the new algorithm ensures that at each intermediate step, the number of active emitters is also minimal (see condition \ref{I})---introducing new emitters only when required and immediately identifying and decoupling unnecessary emitters. One the other hand, we remark that allowing additional emitters can be beneficial, for instance, in reducing circuit depth or number of two-qubit gates. 
As a result, minimizing emitters may not always be the optimal method; this trade-off is discussed in detail in \cref{sec:morethanminimal}. Notably, our algorithm can be easily generalized to handle such more-than-minimal emitter cases. 

Another important cost factor in the generation process is the number of two-qubit gates between emitter qubits. Not only is implementing an interaction channel to couple a pair of distant emitter qubits experimentally challenging, but two-qubit gates are also in general the primary sources of noise and error in quantum circuits and require longer execution times than single-qubit operations. Therefore, reducing their use in the circuit is advantageous. As shown before, the proposed scheme offers--on average--significant reductions in using such gates to prepare a target state. Nevertheless, there is still room for additional optimizations to further improve this performance metric using the degrees of freedom available in our generation algorithm. 

The depth of the generation circuit is another key metric affecting output‐state quality. Because emitter qubits have finite coherence times, in deeper circuits they accumulate more noise, and photonic qubits emitted later in the active-time of an emitter would inherit this decoherence. Crucially, we can consider “circuit depth” here to concern only the emitter registers of the circuit. This is because photonic operations are restricted to single‐qubit unitaries and all gates acting on a photon can be merged into a single effective unitary, fixing the depth of the photonic registers at 1. Furthermore, single‐qubit gates' execution times are negligible compared to two‐qubit ones. It is therefore more meaningful to instead use the quantity \textit{two‐qubit depth}, e.g, \cz[cz] depth or \cnot[cnot] depth---defined as the number of two‐qubit gates applied on each emitter between each initialization and subsequent measurement.

\subsection{Cost of Graphical Transformations}

The proposed elementary set of graphical operations introduced in \cref{sec:graphical} is developed with cost considerations in mind. In particular, the batch edge-toggle operations in \cref{sec:batch} each require only a single \cz[cz] gate to create or remove multiple edges simultaneously and hence contribute minimally to the two-qubit gate count and the \cz[cz]-depth. In contrast, note that a naive edge-creating approach based on \cref{eq:graph_state} requires one \cz[cz] gate per \textit{every} edge created/removed. Besides, local complementations that are utilized in many of the introduced operation in our the graphical framework in \cref{sec:graphical}, only require single qubit gates, which can be considered to have little overhead. Additionally, the introduced emission modes rely solely on single-qubit gates to create multiple edges at once for the newly emitted photon. Note that emitter–photon \cnot[cnot] gates are used to model the emission of each photon and do not count toward the generation cost.

When it comes to combining these operations in each individual generation step where a new photon is added to the system, our algorithm---by design---avoids redundancies, ensuring that no two operations cancel the effect of each other. However, when all steps in the generation process are viewed as a whole, some \cz[cz] gate on a pair of emitters at the end of one step may be followed by another \cz[cz] gate on the same pair in later steps, indicating a chance for gate cancellation. This is an example of redundancy that can be avoided by considering the whole sequence of prescribed operations across all generation steps and, when possible, simplifying such repetitive occurrences. We have implemented such a recipe‐simplification process directly at the graphical level, without transitioning to the quantum‐circuit representation to avoid complexities of constrained circuit simplification tasks. By identifying and exploiting the commutation relations among our elementary graphical transformations, we cancel/merge eligible operations across all steps of the generation, and as a result minimize the redundancy in use of two-qubit gates. Employing the minimal-cost elementary set of graphical operations together with the recipe-simplification process leads to a cost-efficient generation circuit. Appendix \ref{app: simplification} provides more details on the recipe simplification algorithm. 

 So far, the internal degrees of freedom of the generation algorithm are considered to be fixed. The next section explores these additional available degrees of freedom that can be leveraged for further optimization.

\subsection{Optimization Degrees of Freedom}

The objective of a generation algorithm in each step is to prepare the state such that after each emission, the intermediate state still belongs to the generative set of the target graph. However, the transformation from a given intermediate graph with $n$ photons, $G'(n)$, to any graph $G'(n+1) \in$ GenSet($n+1$), is not always unique. The freedom in choosing operations to complete this transformation---if utilized properly---can steer the generation process toward lower cost routes. In what follows, we enumerate the complete set of available degrees of freedom and advise on how to address each when encountered, in order to reduce cost with respect to specific metrics. An exhaustive search over all these optimization parameters would in principle guarantee the optimal result. Nevertheless, such brute-force search approaches are not practical (except for small sized graphs) due to the sheer size of the search space and the exponential growth of the size of the decision tree. Instead, targeted or greedy optimization strategies, enabled by the knowledge of the available degrees of freedom and their behavior, can prove more advantageous in general.

\subsubsection{Freedom in choice of the emitter}
For any given intermediate state $G'$, the \hyperref[sec:case_analysis]{case analysis} process uniquely determines to which case the $G'$ belongs. In each case, the first action is selecting the emitter of the next photon. The set of eligible emitters is given for each case. When more than one emitter is eligible, we can exploit this freedom by selecting the best emitter with respect to some metric of choice. For example, one can rank the emitters according to their depth or active time in the circuit and always choose the youngest (least noisy) option for the next photon emission. 

\subsubsection{Freedom in assigning emitter rows}
The choice of the set of emitter rows $\{R_{e_j}\}$ which forms a basis for the row space of the biadjacency matrix is not unique if the matrix lacks full row rank. Since the required connectivity between photon ``$k$'' and the emitters (according to condition \ref{II}) is determined by the decomposition of the corresponding row ($R_k$) in the biadjacency matrix in terms of the emitter rows, the freedom in choice of emitter row set can affect the required operations. 

For example, in the case ``column effect $=-1$, row effect $=0$'' at least one prior emitter row \(R_{e_i}\) becomes linearly dependent on the others:
\begin{align}\label{eq:e_row_freedom}
    R_{e_i} = \sum_{{e_{m\neq i}}\in \mathcal{M}} R_{e_m}
\end{align}
\noindent and can be removed from the set. Now according to the requirement stated in \cref{J}, the photons connected to ${e_i}$ must get connected to all emitters $e_m \in$ $\mathcal{M}$ instead. This is done by using the e-to-inside operation between the emitters $e_m$ and $e_i$. Since each e-to-inside operation needs one \cz[cz] gate, the cost scales with \(\lvert \mathcal{M}\rvert\) in \cref{eq:e_row_freedom}. A suitable change of basis can reduce the number of rows present in the decomposition of $R_i$ and thus can be used as a mean for optimization. This is especially important when the constraint on using the minimum number of emitters is relaxed and the number of possible configurations for the emitter rows' set is increased (due to an increased rank deficiency).

More generally, similar to the set $\mathcal{M}$, the choice of emitter rows affects the composition of the sets \hyperref[j_set]{$\mathcal{J}$} and \hyperref[k_set]{$\mathcal{K}$} as well, and since the number of costly operation are determined by these sets in our algorithm, one can form a well-defined cost function accordingly and optimize the choice of emitter rows based on it.

\subsubsection{Freedom in choice of emitter to decouple}
When the column effect is \(-1\) and the row effect is $0$, one emitter can be decoupled from the system at the end of the step. In this case, one can always identify a set of linearly dependent emitter rows such that:
\begin{align}\label{m_eq}
\sum_{e_m\in \mathcal{M}} R_{e_m} = 0
\end{align}
Any emitter in $\mathcal{M}$ is a valid candidate to be decoupled. Similar to the case of selecting the next emitter, we can rank these candidates by the time since they have been in use or their corresponding depth in the circuit and prioritize the oldest (noisiest) emitters for decoupling.

\subsubsection{Local equivalence in intermediate steps}\label{subsec:lc_freedom}
Condition \ref{III} has an inherent degree of freedom: at step $n$, with $\mathcal{P}$ indicating the set of emitted photons, the emitted subgraph $G'[\mathcal{P}]$ in the intermediate state need only match the subgraph $\tilde{G}[\mathcal{P}]$ of \textit{any} graph $\tilde{G}$ in the partial local equivalence class of the target state PLC$(G,\mathcal{P})$. In the restricted version of this conditiond \hyperref[III-2]{($\text{III}'$)} we opted for the special case where $\tilde{G} = G$.  In practice, however, any alternative $\tilde{G}\in$ PLC$(G,\mathcal{P})$ is a valid choice.

To make use of this freedom, let us look back at \cref{k_eq}, where $[R]_0$ indicates the first element of the row vector $R$. The set \hyperref[k_set]{$\mathcal{K}$} consists of emitters $\{e_k\}$ associated with the emitter rows $R_{e_k}$ which have non-zero first elements. In order for condition (III) to hold, the newly emitted photon at each step should get connected to photons in the neighborhood of emitters in $\mathcal{K}$. These connections are implemented by performing an e-to-inside operation between the new photon's emitter and each \(e_k\in\mathcal{K}\), and then emitting with an appropriate emission mode so that those edges transfer onto the newly emitted photon. Since each e-to-inside uses one \cz[cz] gate, the two-qubit cost at step \(n\) depends on \(\lvert\mathcal{K}\rvert\).  Now in an arbitrary step $n$, let us temporarily switch the target state from $G$ to $\tilde G$. Since at the beginning of the step we have $G'(n)\in$ GenSet($G, n$), then the graph $\tilde G'(n)\in$ GenSet($\tilde G, n$) can be obtained from $G'(n)$; this is because, by definition, $G$ can be converted to $\tilde G$ under local operation on all nodes and non-local gates on the nodes $\not\in$ $\mathcal{P}$. The same transformation can be applied on $G'$ to convert it to $\tilde G'$ where any Clifford operation applied on the nodes $\not\in$ $\mathcal{P}$ is replaced by a set of operation on the appropriate emitter nodes. Having in mind that emitter rows each correspond to one of the rows in $B(n)$, they are transformed accordingly when the target's biadjacency matrix $B(n)$ is replaced by $\tilde B(n)$ of the PLC-equivalent graph $\tilde{G}$. The set $\mathcal{K}$ can now be presented as:
\begin{align}
  \tilde{\mathcal{K}} = \left\{\,e_k \; \middle| \; [\tilde R_{e_k}(n)]_0 = 1\right\}
\end{align}
 
Consequently, at each emission step, one can pick the PLC-equivalent graph \(\tilde G\) whose corresponding emitter rows \(\{\tilde R_{e_j}(n)\}\) have the fewest nonzero first entries—thus minimizing \(\lvert\tilde{\mathcal{K}}\rvert\) and reducing the number of two-qubit gates used in that step. The same procedure can be applied for all $N$ steps of the generation. 

Finally, we remark that the transformation of the intermediate state from $G'$ to $\tilde G'$ at the beginning of a step might require extra two-qubit operations on the emitters. So, there is a trade off between the cost of transformation and the reduction in the size of ${\mathcal{K}}$. This can be avoided if we restrict the search space for the alternative targets $\tilde G$ to ``Local Clifford" (LC) equivalence class of the target graph which is a subset of the PLC. This ensures that the transformation of the intermediate state $G'\to\tilde G'$ only requires local Clifford operations and thus no extra cost is inflicted on the generation process while reductions in \(\lvert\tilde{\mathcal{K}}\rvert\) are obtainable at the same time.

\subsubsection{Extra emitters}\label{sec:morethanminimal}
The number of simultaneous active emitters allowed in the system is another degree of freedom of the algorithm. Using extra emitters can result in reducing the depth and the two-qubit gate count of the generation circuit, as previously demonstrated in Ref. \cite{kaur_resource-efficient_2024}. Although our algorithm was designed to use the minimum number of emitters by default, the generalization to relax this constraint is straightforward. This is because eligibility conditions \ref{I} and \ref{II} only set the required minimum and are both compatible with allowing the number of active emitters to exceed the rank of \(B(n)\) at each generation step $n$ while keeping track of a basis set of emitter rows. The third condition does not concern the number of emitters as well. 

More explicitly, an extra active emitter can always be introduced/retained in the system in the sequential generation process by disregarding the case analysis at the beginning of a step and using the sub-routine introduced in \hyperref[sec:case_analysis]{case (A)} of the algorithm to emit the next photon. The rest of the cases (B to D) use the existing emitters to emit the next photon and cannot increase the number of active emitters. The only other way of keeping more emitters in the system is not to decouple a redundant emitter whenever we fall under the case (D). However, when  sub-routines of case (D) are utilized, the chosen emitter can no longer contribute to the entanglement generation for the rest of the state due to its position in the graph, i.e, it is connected to photons that have no remaining future edges and the corresponding emitter row would be a zero vector, so the emitter cannot be used without first being decoupled from its current neighbors. In other words, the chosen emitter becomes truly redundant.
 
The availability of extra emitters can lead to cost reduction if utilized with care. The first and most immediate effect of using a new emitter instead of the existing ones is reducing the two-qubit depth; instead of adding operations to an existing emitter and increasing its depth in the circuit, the operation are added to a new emitter that has a starting depth of zero. 

Furthermore, as demonstrated in the case analysis, the number of costly operations in each step depend on the size and overlap of the sets $\mathcal{J}$, $\mathcal{K}$, and $\mathcal{M}$. As seen in \cref{j_eq_1,k_eq,m_eq}, the composition of these sets is fully determined in terms of emitter rows. Since there is a corresponding emitter row for each active emitter in the system, the use of extra emitters allows one to have flexibility in making up the sets $\mathcal{J}$, $\mathcal{K}$, and $\mathcal{M}$, to minimize the number of costly operations in each step. 
\textcolor{black}{
\subsubsection{Emission order}
}
\textcolor{black}{
Scheduling the emission photons is a common practice to avoid unnecessary delay lines and photon storage. If the intended application of the generated state does not require a fixed emission order, the emission sequence can be treated as an additional degree of freedom with potentially significant effects on the generation circuit. Changing the ordering of nodes in a graph alters the corresponding adjacency matrix and the linear dependency relations between its rows and columns. As a consequence, the eligibility conditions \ref{I} and \ref{II}, which depend on these relations, are generally not invariant under node reordering. One immediate implication is a possible change in the number of emitters required for state generation. This can be exploited to identify generation circuits requiring fewer emitters and, potentially, a reduced number of emitter–emitter two-qubit gates when a search over different emission orders is performed. However, since the search space grows factorially with graph size, such an approach does not constitute a scalable optimization strategy. Indeed, Ref.~\cite{li_photonic_2022} argues that, in general, finding the optimal emission order that minimizes the number of emitters is an NP-hard problem.}
\textcolor{black}{
Nevertheless, heuristic methods can be employed to select suitable emission orders in specific cases. For example, for loop-free graphs (tree graphs), ordering nodes according to a depth-first traversal maximizes the continuity of entanglement along each branch, often resulting in a reduced emitter requirement. This strategy is the default ordering used to obtain the tree-graph results presented in \Cref{tab:1}. Moreover, for parameterized structured graphs, one may perform an exhaustive search over small instances to identify optimal orderings and subsequently generalize the observed patterns to larger cases. A similar approach was adopted in Ref.~\cite{ghanbari_optimization_2023}.
}

\subsection{Optimization Strategies}

As finding a generation recipe for photonic graph states is in the category of sequential decision-making problems, the optimization is naturally mapped onto a decision-tree search; at each step, multiple options in fixing the free parameters lead to the expansion of this tree with each branch having a different cost. In general, the final effect of a decision on the total cost cannot be known until the branch is carried through to completion  and a full-branch evaluation is necessary.

By identifying the degrees of freedom and their effect on cost at each step, we have provided an optimization toolset that is already tailored to greedy, stepwise optimization strategies, either by applying analytic and/or heuristic methods or by a brute-force search to minimize the immediate cost at each step.  
However, to guarantee a global optimum, an exhaustive search that explores every path to the final state is necessary. But this approach suffers from a combinatorial explosion of possibilities in such inherently sequential problems, in all but the smallest instances.

As a practical compromise, one can employ a \(k\)-step look-ahead strategy, i.e, for each current step \(n\), we temporarily expand the decision tree through steps \(n, n+1,\dots,n+k\), evaluate the resulting costs, and then commit only to the choice at step \(n\) that minimizes the projected cost $k$ step ahead. This finite-horizon planning and backtracking strategy balances improved foresight against manageable computational effort, while being perfectly compatible with the optimization tools we have introduced here.

\section{Proofs} \label{sec:proofs}
The necessity and sufficiency of conditions~\ref{I}, \ref{II}, and \ref{III} for intermediate graphs was assumed in the design and implementation of the generation algorithm and in identification of its degrees of freedom. We claim the following for any graph state $G'(n)$ with $n$ photons:
\begin{align*}
G'(n) \in \text{GenSet}(G, n) \Leftrightarrow 
\left\{
\begin{array}{l}
 \text{Condition \ref{I} holds, and} \\
\text{Condition \ref{II} holds, and} \\
\text{Condition \ref{III} holds}
\end{array}
\right.
\end{align*}
where \hyperref[gen_set]{GenSet} is the generative set defined in \cref{sec:gen_set}.
We begin by providing a proof of necessity for each of the three conditions individually and end this section with a proof of sufficiency for the collective set of conditions.

\subsection{Necessity of Condition I}
Condition (I) asserts the need for the the physical intermediate graph's biadjacency matrix $B'(n)$ to have at least the same rank as the $B(n)$. 
We aim to prove that condition (I) is necessary for any intermediate graph $G'(n)$ to be in the generative set of the target graph $G$, i.e.,
\begin{theorem}\label{theo:nes_I}
\[
G'(n)\in GenSet(G,n) \Rightarrow \text{rank}\left[B'(n)\right]\geq \text{rank}\left[B(n)\right].
\]
\end{theorem}

\begin{proofof}{theo:nes_I}
Using a proof by contradiction, let us assume the following: 
\begin{align} \label{assump1}
\exists \, \tau \in \mathcal{T}^* \; \text{such that} \; G'(n)\xrightarrow{\tau} G'(N) = G 
\end{align}
while
\begin{align} \label{assump1_2}
\text{rank}\, [B'(n)] < \text{rank} \, [B(n)]
\end{align}
where $G$ is the target graph and $\mathcal{T}^*$ is the set of all sequences of allowed transformations. 
According to \cref{prop1}, no $\tau \in \mathcal{T}^*$ applied on a graph state $\ket{G'(n)}$ can increase the bipartite von Neumann entanglement entropy $S_\mathcal{P}$ on the partition given at step $n$, i.e., $\mathcal{P}=\{0,\dots, n-1\}$. Then using \cref{theo1}, we can make the same statement for the rank of the biadjacency matrix of the physical intermediate state through its evolution, i.e., 
\begin{align} \label{assump2}
\text{rank}\, [B'(n)] \geq \text{rank}\, [B_N'(n)]
\end{align}
where $B_N'(n)$ is the biadjacency matrix of the evolved state $G'(N)$, after emission of the last photon, on the same partition $\mathcal{P}$, i.e., the first $n$ photons. Referring to \cref{assump1}, we find $G'(N) = G$ and thus $B_N'(n)=B(n)$. Now by using \cref{assump2}, we can write:
\begin{align}
\text{rank}\, [B'(n)] \geq \text{rank}\, [B(n)]
\end{align}
which contradicts our assumption in \cref{assump1_2}. 
\end{proofof}

Since \cref{prop1} and \cref{theo1} were used in the proof of \cref{theo:nes_I}, we provide proofs for these two statements as well.

Before proceeding to proof of \cref{theo1}, let us mention the following relation derived in Ref. \cite{nahum_quantum_2017}:
\begin{align}\label{theo0}
   S(\rho_\mathcal{A}) = \mathcal{I}_\mathcal{A} - |\mathcal{A}|
\end{align} 
where $S(\rho_\mathcal{A})$ is the von Nuemann entropy of the reduced state $\rho_\mathcal{A}$ of subsystem $\mathcal{A}$ in bipartition:
\begin{align}
    -\text{Tr}(\rho_\mathcal{A} \log \rho_\mathcal{A})\,
\end{align}
and $\mathcal{I}_\mathcal{A}$ is the maximum size of a set of independent stabilizers restricted to the subspace $\mathcal{A}$. A set of stabilizers is considered independent if no member of the set can be expressed as a product of the other members.
\begin{proofof}{theo1}
In a graph state, the adjacency vector for the node $i$ is written as 
\begin{align}
    V_i = (v_{i0}, \dots, v_{ij}, \dots); \quad 0\leq j<N
\end{align} with $a_{ij}$ equal to 1 if the two nodes are connected with an edge and 0 otherwise.
The stabilizer operator $g_i$ corresponding to each node $i$ is the tensor product of the operators
\begin{align}
    g_i = Z_0^{v_{i0}}  \cdots  X_{j=i}  \cdots   Z_j^{v_{ij}} \cdots
\end{align}
\noindent Here the indices indicate on which qubit the Pauli operators act. When restricting to subsystem A including the nodes 0 to n-1 (i.e., $|\mathcal{A}|=n$), we truncate the operator $g_i$ and only keep the part acting on these qubits. The restricted stabilizer $\bar g_i$ is then equal to

\begin{align}
    \bar g_i &= Z_0^{v_{i0}}  \cdots  X_{j=i}  \cdots   Z_j^{v_{i j}} \cdots Z_n^{v_{i n}} &\text{if} \hspace{3mm} i< n \\
    \bar g_i &= Z_0^{v_{i0}} \cdots  Z_j^{v_{i j}}  \cdots   Z_n^{v_{i n}} &\text{if} \hspace{3mm} i\geq n
\end{align}
For all nodes $i< n$ the stabilizers $\bar g_i$ has an X operator in the tensor product acting on different qubits. As a result, no product of $\bar g_i$ can cancel the Pauli X in each case, making them all independent. The size of this independent set is the same as the size of the subsystem $|\mathcal{A}|=n$.

For the rest of $\bar g_i$ with $i\geq n$, the stabilizers only have Z operators and are only independent if the corresponding truncated adjacency vectors $\bar V_i$ are independent (over $\mathbb{Z}_2$). This is because a product of stabilizers is equivalent to addition of the respective adjacency vectors. The number of independent stabilizers for $\bar g_{i\geq n}$ is thus equal to the number of independent adjacency vectors $V_{i\geq n}$ when truncated to the first subsystem. But this corresponds to the same region in the adjacency matrix as we used to define the biadjacency matrix $B(n)$. As a result, the rank of $B$ equals to the number of independent stabilizers $\bar g_{i\geq n}$. 

Bringing together the two values for number of independent operators $\bar g_i$ for the cases of $i\geq n$ and $i< n$, we can write
\begin{align}
   \mathcal{I}_\mathcal{A} = |\mathcal{A}| + \text{rank}(B)
\end{align} 
and using \cref{theo0} we get
\begin{align}
    S(\rho_\mathcal{A}) = \mathcal{I}_\mathcal{A} - |\mathcal{A}| = \text{rank}(B)
\end{align}
\end{proofof}
\noindent Therefore, we have demonstrated that for a graph state, the number of linearly independent row (= rank) in the biadjacency matrix is equal to the entanglement entropy for each bipartition.

In order to prove \cref{prop1} we make use of \cref{theo1} and the following lemma:

\begin{lemma}\label{lem1} For any bipartition of a graph state, the rank of the corresponding biadjacency matrix is invariant under \textbf{local complementation} (LC) transformations, i.e., the number of linearly independent rows remains constant.
\end{lemma}
\begin{proofof}{lem1} The effect of applying a local complementation transformation on a graph's adjacency matrix is limited to a set of row/column addition operations as seen in \cref{eq:LC_row} and (\ref{eq:LC_col}). For a partitioned graph, the biadjacency sub-matrix is thus affected by row/column additions operations only, none of which can change the linear dependencies of the rows/columns. 
\end{proofof}

\begin{proofof}{prop1} Any allowed sequence of operations on a graph state during its generation can be viewed as a series of LC operations divided by two‑qubit interactions between emitters, photon emission events, or measurements. Since we allow no two-qubit gates on emitted photons, the edge creation/removal between the two subsystems of the state is only possible via applying local Clifford operations, whose effects on the edge structure of a graph state are shown to be reproducible with a sequence of local complementation (LC) transformations. Therefore, by using \cref{lem1}, one can deduce that such local operations cannot change the rank of a biadjacency matrix. The allowed two-qubit gates do not change the rank either since they are limited to only the second subsystem and create no edge across the bipartition, thus leaving the biadjacency matrix unchanged. Besides, photons emitted in future would not be a member of the first subsystem and are each introduced to the second subsystem as a leaf qubit attached to one emitter, as seen in Fig.~\ref{fig:0}(c), thus not affecting the rank of the biadjacency matrix. Lastly, it is evident that  measurements cannot increase the bipartite entanglement either. Having proved that rank cannot increase under the allowed operations for a fixed bipartition, one can use \cref{theo1}, relating rank to entanglement entropy, and complete the proof of the proposition. 
\end{proofof}
\subsection{Necessity of Condition II}

A trivial necessary condition for achieving the target graph starting from an intermediate state at step $n$ is to be able to produce all the required edges between each of the current photons and the future ones. This future connectivity requirement is encoded in the target graph's biadjacency matrix $B(n)$. In particular, at the beginning of step $n$, a future photon $j\geq n$ needs to be connected to the emitted photons in the set 
\begin{align}
    \left\{ i \mid B(n)_{ij} = 1 \right\}
\end{align}
which includes photons that corresponds to the non-zero elements in the column $j$ of the $B(n)$. Let us denote this column as $C_j(n)$. Similarly, the elements in the columns $C'_j(n)$ of the physical biadjacency matrix $B'(n)$ shows whether each of emitted photons are connected to the emitter $e_j$.

We now introduce the following:
\begin{proposition}\label{prop2} Let $G'(n)$ be a system of $n$ photons ($\mathcal{P}$) and $m$ emitters ($\mathcal{E}$). We define the adjacency vector $\mathcal{C}_h(r, \mathcal{P})$ to be a column vector of size $n$, representing the connectivity of photons in $\mathcal{P}$ to a node $h\not \in \mathcal{P}$ in any of the future intermediate states $G'(r\geq n)$. Then the vector $\mathcal{C}_h(r, \mathcal{P})$ lies in the space spanned by the columns $\{C'_j(n)\}$ of the current physical biadjacency matrix $B^\prime(n)$.

\end{proposition}
\noindent Note that the node $h$ can be either an emitter or a future photon emitted in any of the steps between $n$ and $r$. 

As a consequence of \cref{prop2}, we introduce the following:
\begin{lemma}\label{lem:nes_II}
If $G'(n) \in \text{GenSet}(G, n)$, then for all columns $C_i(n)$ in $B(n)$, we have $C_i(n) \in \text{span of }\{C'_j\}$.
\end{lemma}
\begin{proofof}{lem:nes_II}
    Obtaining the target graph $G$ at the final step means that $G'(N) = G$. Since $B(n)$ stores the adjacency relations between the nodes according to the target graph, each of its columns $C_i(n)$ represents the required adjacency vector, $\mathcal{C}_i(n, \mathcal{P})$, between the first $n$ photons and a future photonic node $i$ in the target graph. Using \cref{prop2}, this means that the columns of $B(n)$ must be in the space spanned by the columns of $B'(n)$ at each step. 
\end{proofof}

We first prove \cref{prop2} and then show how using it alongside \cref{lem:nes_II} leads to the formation of condition \ref{II}. 

\begin{proofof}{prop2} First, consider the case of $G'(r=n)$, for which the only nodes in the system apart from the emitted photons are the emitters. Each of the columns $C'_j(n)$ of $B'(n)$ are equal to the adjacency vector $\mathcal{C}_{e_j}(r, \mathcal{P})$ showing the connection between the emitted photons and emitter $e_j$. Therefore, \cref{prop2} holds for the step $n$. We now need to show it also holds true as $G'(n)$ is evolved, through the allowed transformations, to any $G'(r\geq n)$.

Let $B'_r(n)$ be the biadjacency matrix of the intermediate state $G'(r)$ on the partition $\mathcal{P}=\{0, ..., n-1\}$, i.e., the first $n$ photons make up the first subsystem and the rest of the photons $\{n, ..., r-1\}$ plus the emitter nodes in $G'(r)$ are in the other subsystem.
We now consider the adjacency vector $\mathcal{C}_{h'}(r,\mathcal{P})$ where $h'$ is assumed to be an emitter node in $G'(r)$. The allowed operations include LC transformations, two-qubit gates on emitters, photon emissions, and measurements. The action of LC operations on a graph, as shown in \cref{eq:LC_col}, is a subset of column addition operations in the adjacency matrix. As a result, after applying LC operations, an emitter's connectivity to the photonic system $\mathcal{P}=\{0, ..., n-1\}$, represented by its adjacency column $\mathcal{C}_{h'}(r,\mathcal{P})$, can only be a member of the space spanned by all columns of $B'_r(n)$ before the LC operations. Two-qubit operations on emitter qubits, and emission of new photons both only affect the adjacency relations within the second subsystem, therefore there is no change inflicted on $\mathcal{C}_{h'}(r,\mathcal{P})$. The measurement of an emitter decouples it from the rest of the nodes, and thus makes the corresponding $\mathcal{C}_{h'}(r,\mathcal{P})$ a zero vector, which is in principle an element of every vector space. As a result, when evolving the intermediate state $G'(r)\to G'(r+1)$, none of the allowed transformations that can move the vector $\mathcal{C}_{h'}(r,\mathcal{P})$ out of the space spanned by the columns of $B'_r(n)$. As a result, if \cref{prop2} is valid at the beginning of the step $r$, this is also the case at the beginning of step $r+1$, and since we had demonstrated the validity for $r=n$, then by induction \cref{prop2} holds true for any $r\geq n$ when considering the adjacency vectors for emitter nodes. 

We now show that any future photon's connectivity to the current set of photons $\mathcal{P}$ is also limited in the same way as the emitters. As dictated by the emission model, without loss of generality, any emitted photon can always be considered to start as a leaf node attached to its emitter node (the parent). For a leaf-parent pair of nodes, let us define the external neighborhood ($\mathcal{N}^{\mathrm{ex}}$) as the neighborhood excluding the nodes $L$ and $P$, i.e.,
\begin{equation}
\begin{split}
\mathcal{N}^{\mathrm{ex}}(L) &= \mathcal{N}(L) - \{P\} \\
\mathcal{N}^{\mathrm{ex}}(P) &= \mathcal{N}(P) - \{L\}
\end{split}
\end{equation}
where $\mathcal{N}(h)$ denotes the neighborhood of the node $h$. We can now state the following:
\begin{lemma}\label{lem2} 
Let \( L \) and \( P \) be a leaf--parent pair in a graph. Then, under any sequence of LC transformations, the external neighborhood \( \mathcal{N}^{\mathrm{ex}}(L) \) either becomes equal to \( \mathcal{N}^{\mathrm{ex}}(P) \) or remains unchanged.

\end{lemma}
\begin{proofof}{lem2} Since by definition, an LC operation applied on a node only affect the connectivity among its neighbors, then $\mathcal{N}(L)$ cannot change until an LC is applied on the parent node $P$, otherwise $L$ remains a leaf node. If an LC is applied on $P$, the node $L$ gets connected to the neighborhood of $P$ and we have $\mathcal{N}^{\mathrm{ex}}(L)=\mathcal{N}^{\mathrm{ex}}(P)$. From this point forward, the graph is invariant with respect to swapping the labels of the nodes $L$ and $P$. Any LC operation on nodes $\notin \{L, P\}$ that affects the neighborhood of $L$, changes the neighborhood of $P$ in the same way, so we always have $\mathcal{N}^{\mathrm{ex}}(L)=\mathcal{N}^{\mathrm{ex}}(P)$. 

Next, we consider the effect of LC on nodes $\{L, P\}$ starting from a case where $\mathcal{N}^{\mathrm{ex}}(L)=\mathcal{N}^{\mathrm{ex}}(P)$. The effect can be categorized into three scenarios:
\begin{enumerate}
    \item If the two nodes are not connected, an LC operation on one of the nodes results in no change in the neither of the external neighborhoods.
    \item If the external neighborhood is empty, the LC has no effect.
    \item If neither of the previous cases are true, i.e., the two nodes are connected and the external neighborhood is not empty, after an LC on the node $L$ ($P$), the external neighborhood of the other node $P$ ($L$) becomes empty.  
\end{enumerate}

With this, we deduce that after any sequence of LC operations on a graph with an initial leaf--parent pair, the final external neighborhood of the leaf is either equal to that of the parent or is empty. \Cref{fig:lem2} visualizes the limitation imposed by \cref{lem2} in a most general case. 
\end{proofof}
\begin{figure}[]
\centering
\includegraphics[width=0.75\columnwidth]{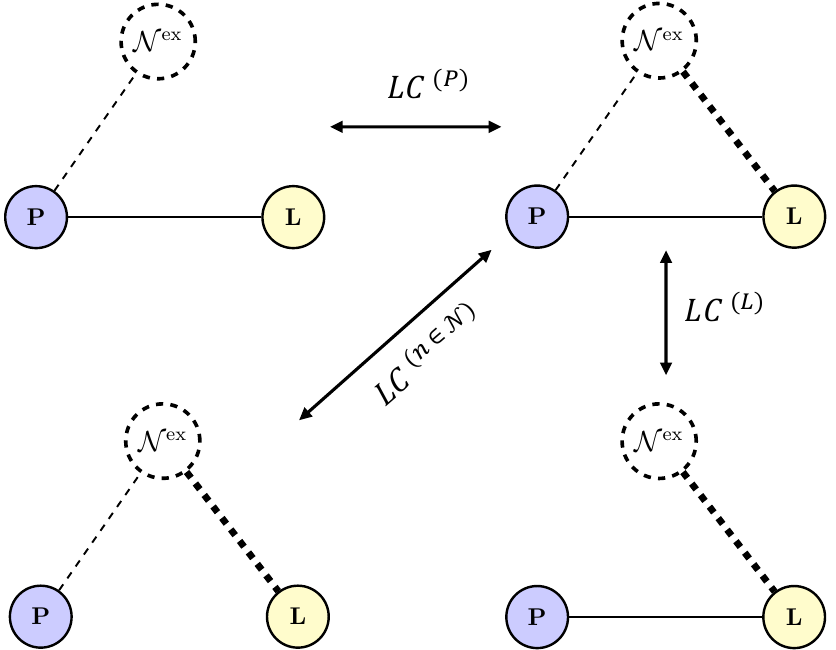}%
\caption{All possibilities for a parent--leaf (P--L) pair in a graph that is transformed under local complementations. Superscripts on arrows specify the node on which the LC is applied. The starting configuration is on top left, with $\mathcal{N}^{\mathrm{ex}}$ representing the external neighbors of $P$. As evident, the external neighborhood of $L$ is either equal to $\mathcal{N}^{\mathrm{ex}}$, or else empty.}
\label{fig:lem2}
\end{figure}
Using \cref{lem2}, and noting that every photon starts as a leaf to one emitter, we can conclude that the connectivity of a future photon to some part of the graph is limited to the connectivity of its emitter to the same part. Since \cref{prop2} was partially proved for the case of adjacency vectors $\mathcal{C}_{h'}(r,\mathcal{P})$, where $h'$ is an emitter, the recent argument extends its validity for future photonic nodes as well and hence completes the full proof of the proposition.
\end{proofof}

In \cref{lem:nes_II}, we used \cref{prop2} to show that in order to have $G'(n)\in \mathrm{GenSet}(G,n)$ , it is necessary to have
\begin{align}\label{nec1}
\{C_j(n)\}\in \text{span of} \; \{C'_j(n)\}
\end{align}
where $\{C_j(n)\}$ and $\{C'_j(n)\}$ are the set of columns of $B(n)$ and $B'(n)$ respectively. We are now in a place to demonstrate that the specific connectivity configuration between the emitters and photon required by condition \ref{II} is necessary for the statement \ref{nec1} to hold.

\begin{proposition}\label{prop3} Let $R_i$ and $R'_i$ denote row vectors associated with photon $i$ in the biadjacency matrix $B(n)$ and the physical biadjacency matrix $B'(n)$, respectively. If we have
\begin{align}\label{eq:prop3-rows}
R_i = \sum_{j\in S(i)} R_j
\end{align}
then, in order for columns of $B'(n)$ to span the columns space of $B(n)$, it is necessary to have,
\begin{align}\label{eq:prop3}
R'_i = \sum_{j\in S(i)} R'_j
\end{align}
\end{proposition}
\noindent Since by definition, each row vector $R'_i$ shows the connections between photon $i$ and all emitters in the system, \cref{eq:prop3} dictates the allowed connectivity configurations between emitted photons and the emitters at each step.

\begin{proofof}{prop3} 
It is enough to show that for any basis for the column space of the biadjacency matrix $B(n)$, matches the columns of $B'(n)$ when the matrix $B'(n)$ is constructed according to \cref{eq:prop3}. From now on, we suppress the step index $n$ for notational simplicity.

If the rank of $B$ is $m$, then without loss of generality, let us select a (non-unique) set of $m$ linearly independent rows that form a basis for the row space of the biadjacency matrix and set them as the first $m$ rows in $B$. The row swap operations that are possibly needed for replacing the first $m$ rows with the basis rows corresponds to changing the order in which the emitted nodes are represented in the adjacency matrix and have no physical implications. As a result, when taking the transpose of the biadjacency matrix ($B^T$), the first $m$ columns (equal to $R_i^T$ for $i < m$) will be linearly independent. Since the rank of $B^T$ is also $m$, these columns form a basis for the columns of $B^T$. As a result, any column $C_j$ in $B^T$ can be written as a linear combination of its first $m$ columns,
\begin{align}\label{eq:calpha}
    C_j &\ = \sum_{i < m} \alpha^j_i\; C_i,
    &\ \alpha^j_i\in\{0,1\}
\end{align}
Additionally, because the columns of $\text{B}^T$ are equal to the rows of $\text{B}$, it immediately follows that for the corresponding row $R_j$ in $B$, the same coefficients $\{\alpha^j\}$ satisfy
\begin{align}\label{eq:ralpha}
    R_j &\ = \sum_{i < m} \alpha^j_i R_i, 
    &\ \alpha^j_i\in\{0,1\}
\end{align}
Next, using Gaussian elimination, we bring the transposed matrix \( B^T \) into its reduced row echelon form, denoted \( B_{\mathrm{echelon}}^T \), where the leading nonzero entries in the rows occupy the first \( m \) diagonal positions. As a result, the top-left \( m \times m \) submatrix of \( B_{\mathrm{echelon}}^T \) becomes the identity matrix, with the first \( m \) columns being unit vectors. Besides, all element below the $m$-th row become zero in the reduced row echelon form.
Using the fact that the row operations of Gaussian elimination do not alter linear dependencies among columns, \cref{eq:calpha} holds for columns of \( B^T_{\mathrm{echelon}} \) as well. But since the columns used for the expansion are unit vectors in this case, we know that the $k$-th element of the $l$-th column is equal to 1 if and only if that column has the $k$-th column in its expansion, i.e., using the notation in \cref{eq:calpha},
\begin{align} \label{eq:iff1}
\alpha^l_k=1 \Leftrightarrow [B^T_{\mathrm{echelon}}]_{k\,l}=1
\end{align}
where we used the notation $[A]_{ij}$ to denote the element on the i-th row and j-th column of matrix A. Now looking at \cref{eq:ralpha}, by definition, $\alpha^l_k=1$ if and only if $R_l$ has $R_k$ in its expansion in the basis rows, i.e., 
\begin{align} \label{eq:iff2}
\alpha^l_k=1 \Leftrightarrow \text{$R_l$ has $R_k$ in its expansion}
\end{align}
where $R_k$ is a member of the chosen basis for row space of $B$.
Consequently, 
\begin{align} \label{eq:iff3}
[B^T_{\mathrm{echelon}}]_{k\,l}=1\Leftrightarrow \text{$R_l$ has $R_k$ in its expansion}
\end{align}
We remark the fact that the set of the first $m$ rows of $B^T_{\mathrm{echelon}}$, denoted as $\mathcal{R}$, span the row space of the original matrix $B^T$. As a result, the set of column vectors formed by taking the transpose of these rows ($\mathcal{C}=\mathcal{R}^T$) must span the column space of the original matrix $B$. Additionally, using \cref{eq:iff1} we can determine each member of the spanning set $\mathcal{R}$, i.e., for $\mathcal{R}_k = (r^k_0, \dots, r^k_l, \dots)$ we have:
\begin{equation}\label{eq:falpha}
    r^k_l = \alpha^l_k 
\end{equation} 
where $\alpha^l_k$ are the same coefficients used in \cref{eq:iff2}. We can also write the corresponding column vector $\mathcal{C}_k$ as:
\begin{equation}\label{eq:c_cols}
\mathcal{C}_k = \mathcal{R}_k^T =
\begin{pmatrix}
\alpha^0_k \\
\vdots \\
\alpha^l_k \\
\vdots \\
\end{pmatrix}
\end{equation}
So far, we have proved that for any choice of a basis set among the rows of the matrix $B$, the set of columns $\mathcal{C}$ that is formed according to \cref{eq:c_cols} spans the column space of $B$. In order to prove \cref{prop3}, one needs to assume that the columns of $B'$ span the columns of $B$, and then show that for the matrix $B'$, the linear dependencies between rows must agree with \cref{eq:prop3}. 

If columns of $B'$ span the column space of $B$, then in the most general case, $B'$ must be a matrix that is formed by putting together columns $\mathcal{C}_k$ for $0\leq k<m$ where $m=$rank$(B)$. According to \cref{eq:c_cols}, the element on the $l$-th row and $k$-th column of $B'$ is:
\begin{align}\label{elements}
    [B']_{l\,k} = \alpha^l_k
\end{align}
Let us denote the elements of $B$ in with a similar notation:
\begin{align}
    [B]_{l\,k} = \beta^l_k
\end{align}
Using the matrix elements, we can rewrite \cref{eq:ralpha} (with a change of dummy index $i\to i'$ to avoid confusion) :
\begin{align}\label{eq:ralpha-rewrite}
     \beta^j_k = \sum_{i'< m} \alpha^j_{i'} \beta^{i'}_k
\end{align}

Also, using the same notation, we can rewrite the assumed \cref{eq:prop3-rows} as:
\begin{align}\label{eq:prop3-rows-rewrite}
     \beta^i_k = \sum_{j\in S(i)} \beta^j_k
\end{align}
By combining \cref{eq:ralpha-rewrite,eq:prop3-rows-rewrite} we get:
\begin{align}
    \beta^i_k = \sum_{j\in S(i)} \sum_{i'< m} \alpha^j_{i'} \beta^{i'}_k
\end{align}
Also, using \cref{eq:ralpha-rewrite} alone and by setting $j\to i$,
\begin{align}
    \beta^i_k = \sum_{i'< m} \alpha^i_{i'} \beta^{i'}_k
\end{align}
We now equate the two expressions above for $\beta^i_k$,
\begin{align}
\sum_{j \in S(i)} \sum_{i'< m} \alpha^j_{i'} \beta^{i'}_k = \sum_{i'< m} \alpha^i_{i'} \beta^{i'}_k
\end{align}
exchange the summation order on the left-hand side:
\begin{align}
\sum_{j \in S(i)} \sum_{i'< m} \alpha^j_{i'} \beta^{i'}_k 
&= \sum_{i'< m} \sum_{j \in S(i)} \alpha^j_{i'} \beta^{i'}_k \\
&= \sum_{i'< m} \left( \sum_{j \in S(i)} \alpha^j_{i'} \right) \beta^{i'}_k
\end{align}
So the equation becomes:
\begin{align}
\sum_{i'< m} \left( \sum_{j \in S(i)} \alpha^j_{i'} \right) \beta^{i'}_k = \sum_{i'< m} \alpha^i_{i'} \beta^{i'}_k = \beta^i_k
\end{align}
which must be valid for all elements $\beta^i_k$ in the matrix $B$. Therefore, the equality holds if and only if
\begin{align}\label{eq:iff4} 
 \alpha^i_{i'}=\sum_{j \in S(i)} \alpha^j_{i'}  \quad \text{for all } i' < m
\end{align}
Now let us write \cref{eq:prop3} in terms of matrix elements:
\begin{align}
     \alpha^i_k = \sum_{j\in S(i)} \alpha^j_k \quad \text{for all } k < m
\end{align}
which is equal to \cref{eq:iff4} by a change of an index name $i'\leftrightarrow k$. Note that the biadjacency matrix has $m$ columns and thus $0\leq k<m$.

This completes the proof of \cref{prop3}.
\end{proofof}

To complete the proof of the necessity of condition \ref{II}, we introduce another lemma:
\begin{lemma}\label{lem:necII_2}
    The connectivity between photons and emitters described in condition \ref{II} matches the configuration of \cref{eq:prop3} established in \cref{prop3}.
\end{lemma}
 
\begin{proofof}{lem:necII_2}
By definition, the set of emitter rows forms a basis for the row space of $B$. Then, similar to \cref{eq:ralpha} one can write:
\begin{align}\label{expansion}
R_i = \sum_{j<m} \alpha^i_j R_{e_j}
\end{align}
Following the same reasoning used between \cref{eq:ralpha,elements}, the elements of $B'$ can be written as:
\begin{align}\label{bicond}
[B']_{ij} = \alpha^i_j = 1 \Leftrightarrow j\in S(i)
\end{align}
where we used \cref{eq:iff1,eq:iff2,eq:iff3} in order to get to the last biconditional. Here the set $S(i)$ is defined by the expansion:
\begin{align}   
R_i = \sum_{j\in S(i)} R_j
\end{align} which is the starting assumption in \cref{prop3}. 
Using the notation utilized in statement of condition \ref{II} we can write the same expansion as:
\begin{align}
R_i = \sum_{j\in\mathcal{N}(i)}R_{e_{j}}
\end{align}
and comparing with \cref{expansion}, we can find that
\begin{align}
j\in \mathcal{N}(i) \Leftrightarrow \alpha^i_j=1 
\end{align}
adding this result to \cref{bicond}, leads to the following:
\begin{align}
[B']_{ij} = 1 \Leftrightarrow j\in \mathcal{N}(i)
\end{align}
which is the statement made in condition \ref{II} regarding the adjacency between nodes $i$ and $j$. Therefore, we have shown the equivalency of the given structure for the matrix $B'$ in \cref{prop3} and condition \ref{II}. 
\end{proofof}

Finally, we establish the necessity of condition \ref{II} with the following theorem:
\begin{theorem}\label{theo:nes_II}
If $G'(n)\in GenSet(G,n)$, then condition \ref{II} holds.
\end{theorem}
\begin{proofof}{theo:nes_II}    
From \cref{lem:nes_II} we know that if \( G' \in \mathrm{GenSet}(G) \), then the columns of \( B' \) span the column space of \( B \). \Cref{prop3} states that if the columns of \( B' \) span the column space of \( B \), then the connectivity configuration must be as specified in \cref{eq:prop3}. Lastly, we use \cref{lem:necII_2} to show the equivalence between \cref{eq:prop3} and condition \ref{II}. As a result, condition \ref{II} is a necessary eligibility condition for \( G' \in \mathrm{GenSet}(G) \). 
\end{proofof}

\subsection{Necessity of Condition III}
\begin{theorem}\label{theo:nes_III}
If $G'(n)\in GenSet(G,n)$, then condition \ref{III} holds.
\end{theorem}

We start by assuming $G'(n) \in$ GenSet$(G,n)$ and show that for the set of nodes $\mathcal{P}=\{0,\dots,n-1\}$, there exists a $\tilde G \in$ PLC$(G, \mathcal{P})$ such that $G'(n)[\mathcal{P}]=\tilde G[\mathcal{P}]$, which ensures that condition III holds. 

\begin{proofof}{theo:nes_III}
As $G'(n)$ is a member of the generative set, it can be evolved into the target state $G$ using a sequence of operations in \hyperref[T]{$\mathcal{T}$}. Previously, in \cref{sec:protocol} we demonstrated that the measurement operations are needed to decouple some of the emitter qubits, that have become redundant, at the end of certain generation steps. Let us assume we postpone all these measurement to the end of the last step. The only implication of this is an increase in the total number of emitters used for the generation since no emitter can be recycled to be used more than once. As a result, after the emission of the last photon, we end up with a graph state $H$ that is made up of the target graph $G$ with $N$ photonic nodes, in addition to $M$ emitter nodes each attached to some photonic nodes of $H$. We can then measure all emitters to end up with the target graph $G$. In summary, by definition of $G'(n)\in$ GenSet$(G,n)$, we can write:
\begin{align}
G'(n) \xrightarrow{\tau} G, \quad \tau \in \mathcal{T}^*
\end{align}
where $\mathcal{T}^*$ denotes any finite sequence of operations in $\mathcal{T}$. Also, by postponing measurements we get:
\begin{align}\label{tau1}
G'(n) \xrightarrow{\tau_1} H \xrightarrow{\tau_2} G 
\end{align}
where $\tau_1$ includes no measurement operation and $\tau_2$ encompassed the measurements needed to decouple all emitters from the rest of the system. Let us introduce the following proposition:

\begin{proposition}\label{prop4}
    Let graph \( H \) be defined as above; that is, the graph obtained at the final step of the generation when no emitter is yet decoupled from the system. Let $\tilde H$ denote a graphs that is a member of the set $\mathrm{PLC}(H, \mathcal{P})$. We can now state the following:

\begin{align}\label{eq:prop4}
\forall\, \tilde{H} ,\ \exists\, \tilde{G}\in\mathrm{PLC}(G,\mathcal{P})\; \text{such that}\ \tilde{G}[\mathcal{P}] = \tilde{H}[\mathcal{P}]
\end{align}

\end{proposition}
\begin{proofof}{prop4}
    
    According to \cref{prop2,prop3}, and having established the necessity of condition~\ref{II}, we know that the adjacency vector \( \mathcal{C}_{h}(N,\mathcal{P}) \) of any node \( h \notin \mathcal{P}\) of \( H \), representing its connections to the set of photons \( \mathcal{P} = \{0, \dots, n-1\} \), lies in the span of the adjacency column vectors of the emitters in the graph \( G'(n) \in \mathrm{GenSet}(G, n) \). This span is also equal to the column space of \( B(n) \). We can then write:
    \begin{align}\label{S}
    \mathcal{C}_{h}(N,\mathcal{P}) = \sum_{j\in S(h)} C_j(n)
    \end{align}
    where $C_j(n)$ are columns of $B(n)$.
    
    Furthermore, \cref{prop2} ensures that such an expansion can be written for the adjacency vectors of the corresponding nodes in any graph \( \tilde{H} \in \mathrm{PLC}(H, \mathcal{P}) \). This is because the allowed operations in the transformation \( H \to \tilde{H} \) form a subset of those considered for the emitters in \cref{prop2} and thus the result holds.

    In addition, since two-qubit operations are not allowed on any node in \( \mathcal{P} \), no direct edge creation or removal can occur within \( \mathcal{P} \), or between nodes in \( \mathcal{P} \) and the rest of the graph. However, such structural changes can still occur indirectly via LC operations. The effect of LC operations on the adjacency relations involving nodes in \( \mathcal{P} \) is limited to column addition operations in the adjacency matrix, as described in \cref{eq:LC_col}.
    As a result, in the transformation $H \to \tilde H$, any LC operation on a node $h$ that affects the internal structure of the subgraph induced on $\mathcal{P}$,
    \[
    H\xrightarrow{\text{LC on $h$}} \tilde H
    \]
    can be replaced by LCs on the nodes of $G$ that are in the set $S(h)$ according to \cref{S}:
    \[
    G\xrightarrow{\text{LC on all $j\in S(h)$} }\tilde G 
    \]
    whenever $h\notin \mathcal{P}$. For the case \( h \in \mathcal{P} \), one can simply apply the LC operation on the same node in the graph \( G \). We can thus always obtain a graph \( \tilde{G} \) satisfying \cref{eq:prop4}, i.e., we get
    \begin{align}
    \tilde G[\mathcal{P}]= \tilde H[\mathcal{P}]
    \end{align}    
    
\end{proofof}
We now argue that $G'(n)$ is a member of the $\mathrm{PLC}(H, \mathcal{P})$. According to \cref{tau1}, $H$ is obtained from $G'(n)$ by the transformation $\tau_1$ which includes photon emissions for photonic nodes $\notin \mathcal{P}$, two-qubit gates on emitter nodes (also $\notin \mathcal{P}$), and LC operations on all nodes. Therefore, if the reverse transformation, denoted as $\tau^{-1}_1$, exists, it takes $H$ to $G'(n)$. In order to prove $G'(n)\in \mathrm{PLC}(H, \mathcal{P})$, we only need to show that the operations of $\tau_1$ are all (i) reversible and (ii) members of the allowed transformations in forming the set $\mathrm{PLC}(H, \mathcal{P})$ according to its \hyperref[PLC]{definition}.

As discussed before, photon emissions are basically two-qubit operations between an emitter node and an isolated photonic node. In the case of $\tau_1$, these operations are between emitter nodes and photons $\notin \mathcal{P}$, and since the $\mathrm{PLC}(H, \mathcal{P})$ allows for any two-qubit operation on all nodes $\notin \mathcal{P}$ in the graph $H$, then the reverse operation of a photon emission is basically the same two-qubit gate ($\cnot[cnot]^{-1}=\cnot[cnot]$) applied the same emitter-photon pair of nodes. Moreover, the rest of the operations in $\tau_1$ are trivially both reversible and compatible with the PLC requirements.

Having established that $G'(n)$ is equal to some $\tilde H \in \mathrm{PLC}(H, \mathcal{P})$, using \cref{eq:prop4} in \cref{prop4} we can state that for any graph $G'(n)\in$ GenSet$(G,n)$:
\begin{align}
\exists\, \tilde{G}\in\mathrm{PLC}(G,\mathcal{P})\; \text{such that}\ \tilde{G}[\mathcal{P}] = {G'(n)}[\mathcal{P}]
\end{align}
which is the statement we needed to prove for condition\ref{III} to be a necessary eligibility condition. 
\end{proofof}

\subsection{Proof of Sufficiency}
We claim that the proposed eligibility conditions are sufficient to ensure an intermediate state is a member of the generative set:
\begin{theorem}\label{theo:suff0}For any $G'(n)$ if
\begin{align*}
\left.
\begin{array}{l}
 \text{Condition \ref{I} holds,} \\
\text{Condition \ref{II} holds,} \\
\text{Condition \ref{III} holds,}
\end{array}
\right\} \Rightarrow  G'(n) \in \text{GenSet}(G, n)
\end{align*}
\end{theorem}

\noindent We begin by proving a key lemma:
\begin{lemma}\label{lem:suff00}If Condition \ref{III} holds for $G'(n)$ at the final step ($n=N$), then the target graph is obtainable. 
\end{lemma}

\begin{proofof}{lem:suff00}
By definition of condition \ref{III} we have 
\begin{align}\label{eq:suff1}
   \exists \, \tilde G \in PLC(G, \mathcal P) \, \text{ such that }\,  G'[\mathcal{P}]=\tilde G[\mathcal P] 
\end{align}
where in the $n=N$ case, $\mathcal P$ is the set of all $N$ photons in the target state. By definition of the \hyperref[PLC]{$PLC$}, the graph $\tilde G$ can be obtained from $G$ by local Clifford operations on nodes in $\mathcal P$ and arbitrary Clifford unitaries on the rest of the nodes, but $\mathcal P$ now includes all node of $G$; therefor, (i) $\tilde G=\tilde G[\mathcal P]$, and (ii) $\tilde G$ is related to $G$ by local operations. Consequently, by using \cref{eq:suff1} we see that $G'[\mathcal{P}]$ is also locally equivalent to $G$. Therefore, if condition \ref{III} holds for the final step $n=N$, the target state can be obtained by decoupling (measuring) all the emitters in the physical graph $G'$ to get to the subgraph $G'[\mathcal{P}]$, and then applying the required local gates on the photons to reach $G$. This completes the proof of lemma.
\end{proofof}

Next, we propose the following:
\begin{proposition}\label{prop:suff2}
    If all three conditions hold true for some arbitrary $n<N$, it is possible to move from $n$ to $n+1$ and have all conditions satisfied with the extra photon added to the system, i.e., 
\begin{align*}
& \text{Conditions \ref{I}, \ref{II}, \ref{III} hold}
 \Rightarrow 
\exists \, \tau\in\mathcal T^* \\ &\text{ such that: } G'(n)\xrightarrow[]{\tau} G'(n+1), \\ &\text{ for which \ref{I}, \ref{II}, \ref{III} still hold}
\end{align*}
\end{proposition}
\noindent Here \hyperref[T]{$\mathcal T^*$} is, as previously defined, the set of all finite sequences of the allowed operations in the generation process. 
\begin{proofof}{prop:suff2}
Consider the result of the validity of condition \ref{III} as mentioned in \cref{eq:suff1}, this time for $\mathcal P=\{i\mid 0 \leq i < n\}$. By definition of \hyperref[PLC]{$PLC$}, $\tilde G$ can be obtained from $G$ by Clifford operations with two-qubit gates restricted to the nodes outside $\mathcal P$. As a result, is it also possible to transform any $\tilde G$ to $G$ by applying the inverse operations. 
Let $\tau' = (t_i)_{i=1}^l$ be a finite sequence of Clifford operations of length $l$, comprised of LC and two-qubit operations on the node of the graph $\tilde G$ corresponding to this transformation: 
\begin{align}
    \tilde G \xrightarrow[]{\tau'} G
\end{align}
Having $\tau'$, we can find an explicit transformation $\tau''$ on the physical graph $G'(n)$ such that its photonic subgraph $G'[\mathcal P]$---initially equal to $\tilde G[\mathcal P]$---becomes equal to the corresponding subgraph of the target state $G[\mathcal P]$: 
\begin{align}\label{eq:suff_tau''}
    G'[\mathcal P] \xrightarrow[]{\tau''} G[\mathcal P]
\end{align}
To find such transformation, we note that there is no direct two-qubit gate on nodes in $\mathcal P$ in $\tau'$. So when transforming $\tilde G$ to $G$, the changes in the edge structure of the subgraph restricted to $\mathcal P$ is limited to the effect of LC operations applied on the nodes of $\tilde G$. With this in mind, we iterate over all operations $t_i$ in $\tau'$, applying them to the graph $\tilde G$ on the go and denoting the updated graph after $t_i$ by $\tilde G_i$. 

Assuming the equality between the initial subgraphs, $G'[\mathcal{P}]=\tilde G[\mathcal P]$, for an LC operation on some node $h$ of $\tilde G$, if $h\in \mathcal P$, then applying an LC on the same node in the physical graph $G'$ would change the subgraph $G'[\mathcal P]$ in the same way that $\tilde G[\mathcal P]$ changes. Now consider the case where the next operation, $t_{i+1}$, is an LC on $h\notin \mathcal P$. By examining $\tilde G_i$, we find the nodes in $\mathcal P$ that are in the neighborhood of $h$. This neighborhood can be represented by an adjacency vector $C_h(i)$, where the $j$-th entry in $C_h(i)$, for $0\leq j<n$, is 1 if node $j$ is connected to $h$ in the graph $\tilde G_i$ and 0 otherwise. Since conditions \ref{I} and \ref{II} are assumed to be holding, and as a result of \cref{prop2}, this adjacency vector belongs to the column space of the matrix $B(n)$, which is equal to the column space of the physical biadjacency matrix $B'(n)$. As a result, one can define $S(h, i)$ as the set of indices of the columns that make up $C_h(i)$, such that: 
\begin{align}\label{eq:suff3}
    C_h(i) = \sum_{k\in S(h, i)} C'_k
\end{align}
Since each column $C'_k$ of the matrix $B'(n)$ represents the neighborhood of the emitter $e_k$ in $\mathcal P$, \cref{eq:suff3} states that the neighborhood of the node $h$ can be constructed by taking the symmetric difference of the neighborhoods of the emitters in the set $\{e_k\mid k\in S(h, i)\}$ with $S(h, i)$ defined by \cref{eq:suff3}. As a consequence, any effect on the edge structure of the subgraph $\tilde G_i[\mathcal P]$ that will be resulted by an LC on node $h\notin \mathcal P$ can be reproduced on the subgraph $G'[\mathcal P]$ by applying LC on all the emitters in $\{e_k\mid k\in S(h,i)\}$ in the physical graph $G'$, given that the two subgraphs were equal before $t_i$. 

Consequently, one can obtain the transformation, $\tau''$ of \cref{eq:suff_tau''}, by iterating over $\tau' = (t_i)_{i=1}^l$ and adding operations to $\tau''$ according to the following rules:
\begin{enumerate}
    \item If $t_i \in \tau'$ is an LC on node $h$, it is transformed into the composite LC operation on $G'$ on the nodes:
    \[
\begin{cases}
  h & \text{if } h \in \mathcal P \\
  \{e_k\mid \, k\in S(h,i)\}   & \text{if } h \notin \mathcal P
\end{cases}
    \]
    This new operation is included in $\tau''$.
    \item If an operation $t_i \in \tau'$ is a two-qubit operation, it is omitted.
\end{enumerate}
The order of the resulting operations in $\tau''$ is preserved from the original sequence $\tau'$. Note that $\tau''$ is in $\mathcal T^*$ as it is comprised of LC operations only.

Applying $\tau''$ on the graph $G'$, ensures that the subgraph $G'[\mathcal P]$ evolves exactly the same as $\tilde G[\mathcal P]$ under $\tau'$. After completing the transformation on both $\tilde G$ and $G'$, we get
\begin{align}
    G'[\mathcal P] = \tilde G[\mathcal P] = G[\mathcal P]
\end{align}
which gives us the requirement for the restricted version of the eligibility condition \hyperref[III-2]{$(\text{III}')$} introduced in \cref{sec:gen_set}. But for the restricted case of condition III we have already provided an explicit algorithms (see \cref{sec:protocol}) that prescribes the required operations to emit the next photon while keeping all the three conditions satisfied. Let us call this transformation $\tau''' \in \mathcal T^*$: 
\begin{align}
G'\xrightarrow{\tau'''}G
\end{align}
Then the required $\tau\in \mathcal T^*$ as defined in \cref{prop:suff2} will be:
\begin{align}
    \tau = \tau''' \circ \tau''
\end{align}
The proof of \cref{prop:suff2} is then complete. 
\end{proofof}

\begin{proofof}{theo:suff0}
    \Cref{lem:suff00} states that having condition III satisfied for the final step ($n=N$) is sufficient to obtain the target graph. Once the conditions are satisfied for $G'(n)$ in any arbitrary step $n<N$, one can use \cref{prop:suff2} repeatedly to reach the final step $n=N$ and thus acquire the target graph according to \cref{lem:suff00}. The statement $G'(n) \in \text{GenSet}(G, n)$ by definitions means that one can reach the target $G$ starting from the physical graph $G'(n)$, the proof of \cref{theo:suff0} is thus complete.
\end{proofof}

\section{Discussion and conclusion}
 
In this work, we introduced a novel graphical framework for the deterministic, emitter-based generation of arbitrary photonic graph states, providing an intuitive yet mathematically rigorous scheme for deriving generation recipes based on  a set of necessary and sufficient conditions and elementary graph transformations. By leveraging this framework, we presented \textit{Graph Builder}, a cost-efficient algorithm that achieves substantial, often order-of-magnitude, reductions in the number of required two-qubit gates for state generation compared to alternative methods, as demonstrated on both random graphs and structured states such as ring and RHG graphs, which are critical resources for fault-tolerant quantum computing. This gain in efficiency is intrinsic to the framework's design, achieved without additional optimization and while using the minimum possible number of emitters for a fixed photon-emission sequence.

In our approach, the graph state generation problem is guided by a set of necessary and sufficient eligibility conditions on the adjacency matrix of the evolving graph. The correctness of the chosen quantum operations that are employed between emission events is thus reframed as compliance with these conditions. This provides useful insight by distinguishing the operations that are necessary, from those that are replaceable or optimizable in a generation circuit. We also introduced a set of elementary graph operations to evolve the state under generation in a graphical picture. Recognizing commutation relations among these operations allows for a recipe-simplification analysis, merging or canceling redundant two-qubit gates across all steps of the generation before the recipe is transpiled into a quantum circuit, thereby avoiding the complexities of a constrained quantum circuit optimization problem.

Beyond raw performance, systematically identifying the degrees of freedom in the generation process allows us to characterize the optimization landscape and transform the search for resource-efficient recipes from a black-box optimization into a structured, sequential decision-making problem. The eligibility conditions impose constraints but do not uniquely determine the intermediate graph shapes as the state evolves toward the target. The operations used to realize each intermediate state from its predecessor are likewise non-unique. As a result, multiple decision points arise, namely, the choice of the next emitter, assignment of emitter rows (or equivalently the basis selection for the row space of a biadjacency matrix), emitter decoupling priorities, and the use of local equivalence freedom at intermediate steps. Each of these can be naturally associated with cost functions, such as two-qubit gate count, circuit depth, or emitter decoherence, enabling targeted or greedy strategies for cost reduction in place of intractable exhaustive searches. We also showed that the minimal-emitter constraint can be relaxed in a controlled way to trade additional emitters for reduced depth and gate count. This extension is accommodated within the same eligibility-based generation framework with only minor modifications.

\textit{Limitations and outlook}---
As cost optimization for graph generation is conjectured to be an inherently sequential decision problem, challenges such as limited parallelizability and the difficulty of predicting how different choices in the generation process affect the final cost are often unavoidable. While finding the global optimum circuit remains an open problem in general, this further underscores the importance of having greater control over the decision points in a generation algorithm and understanding their impact on different cost metrics, enabling the development of better optimization strategies.

It is also worth noting that our analysis assumes a fixed emission order and all possible optimizations are subject to that choice; extending the framework to include and handle joint optimization over emission order, emitter scheduling, and local-equivalence selection is a natural next step. For a given emission order, although we default to the minimum emitter count to limit engineering overhead, some platforms may benefit from trading off the number of emitters against other costs, making the generalized algorithm with additional emitters worth exploring.

Finally, the framework can also serve as a bridge to probabilistic or fusion-based generation schemes, for instance by providing guidelines and conditions for decomposing a target state into smaller subgraphs that can be merged together to reconstruct the target state.

In conclusion, the graphical framework and the \textit{Graph~Builder} algorithm presented here offer not just an incremental improvement but a foundation for devising new photonic graph state generation protocols. By providing a cost-aware, mathematically rigorous, and optimization-friendly approach, we expect this work to become a versatile tool for the theoretical planning and experimental realization of large-scale photonic entangled states, paving the way for measurement-based or fusion-based quantum computing and communication platforms.

\section*{ACKNOWLEDGEMENT}
We thank Jie Lin and Luc Robichaud for helpful discussions during the development of this work.
This work was supported in part by MITACS, NSERC Discovery Grant, and National Research Council of Canada Applied Quantum Computing Challenge program \#AQC-104-1. Hoi-Kwong Lo acknowledges research support from National Research Council of Canada High Throughout Secure Networks (HTSN) program \#HTSN-352-2, Canadian Foundaton for Innovation (CFI) \#43072, and the National University of Singapore Start-up grant.

\appendix
\section{Recipe Simplification Formalism}
\label{app: simplification}
The output of the generation algorithm, referred to as the generation recipe, is a list of instructions in terms of graphical operations that are needed to be done in order to obtain a target graph. The recipe can be partitioned into steps, each corresponding to the operations that prepare the state for the emission of a next photon. Although the operations within each step are selected to have no redundancy, i.e., no two operations cancel the effect of each other, some may be simplifiable across different steps when considering the full picture. As a result, a simplification process is necessary to avoid cross-step redundant operations with added cost. 

The two category of simplification processes considered here are merging and cancellations. Two or more operations are merge-able if an equivalent graphical operation with the same effect can be found such that it has a lower cost (in terms of number of emitter-emitter \cz[cz] gates) than the sum of the initial ones.  The cancellation is applied when two subsequent operations leave the graph unchanged, removing both from the recipe. 

In terms of elementary operations, the allowed merging operations include: 
\begin{enumerate}
    \item Direct edge-toggle (CZ) + e-to-inside = e-to-inside-connect
    \item Direct edge-toggle (CZ) + e-to-inside-connect = e-to-inside
    \item  e-to-inside + e-to-inside-connect = Direct edge-toggle (CZ)
\end{enumerate}
\noindent through which the number of used \cz[cz] gates is reduced by one. The cancellations are:
\begin{enumerate}
    \item Direct edge-toggle (CZ) + Direct edge-toggle (CZ) 
    \item e-to-inside + e-to-inside
    \item e-to-inside-connect + e-to-inside-connect
\end{enumerate}
\noindent through which the number of used \cz[cz] gates is reduced by two.
\\

One requirement for simplification is that both operations are applied on the same two emitters, e.g., a \cz[cz] between $e_i$~and~$e_j$~followed by a~$e_i\text{-to-inside-}e_j$ operation. A less obvious condition is that the operations should either be applied right after one another, \textit{or}, they need to commute with the rest of the operations in between them, such that we can consider them to occur in immediate succession.
\begin{figure}[]
\centering
\includegraphics[width=0.9\columnwidth]{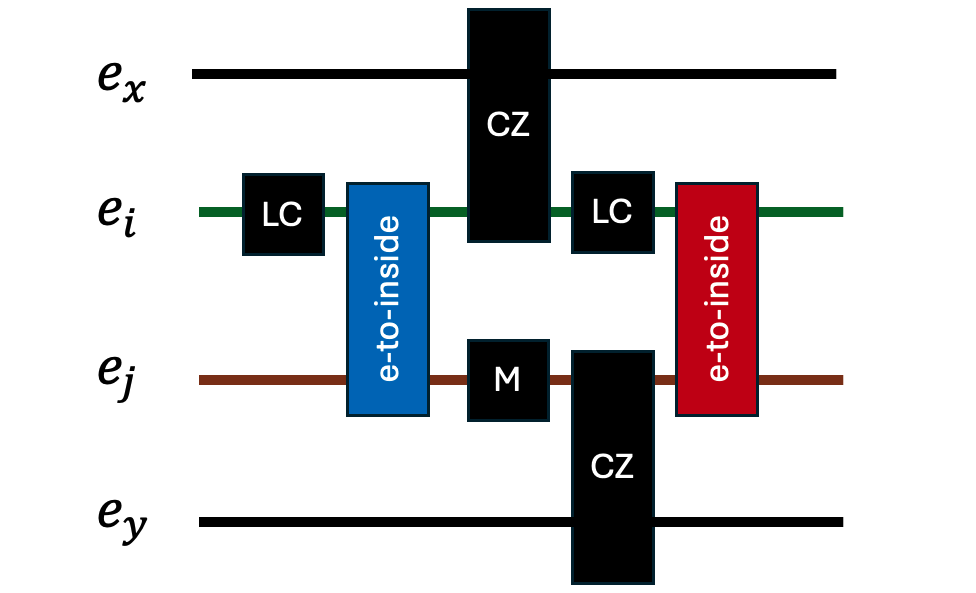}%
\caption{A circuit representation of the graphical operations on emitter nodes. The used gates are: Local complementation (LC), Measurement (M), Controlled Z (\cz[cz]), and e-to-inside.}
\label{fig:simp-circ}
\end{figure}
A set of graphical commutation relations are then necessary to assess the possibility of simplification. For the relevant operations that involve two emitters, e.g., $e_i\text{-to-inside-}e_j$, let us call the emitter that comes first the control and the second one the target. Note that commutation should be considered on each of the two nodes separately, e.g., operation A may commute with B on the control qubit but not on the target qubit. The commutation relations can be derived as follows:
\begin{itemize}
[itemindent=10pt, leftmargin=0pt]
\item[--] For \textbf{$e_i\text{-to-inside-}e_j$} operation, edges are made between the control node $e_i$ and $N(e_j)$, the neighbors of the target node $e_j$. So any operation that changes the neighborhood of $e_j$ should not be allowed to commute with $e_i\text{-to-inside-}e_j$ on the target qubit. This includes: 
\begin{itemize}
    \item[$\bullet$] Emission by $e_j$ in (L), (CS), and (SS) modes
    \item[$\bullet$] Direct edge-toggle (CZ)
    \item[$\bullet$] ``e-to-inside" with $e_j$ as control qubit
    \item[$\bullet$] Measurement on $e_j$
\end{itemize}

Furthermore, since the control node $e_j$'s connections are changed in this operation, any operation that \textit{depends} on the neighborhood of $e_i$ is on the non-commuting list as well: 
\begin{itemize}
    \item[$\bullet$] Emission by $e_i$ in (L), (S), and (CS) modes
    \item[$\bullet$] ``e-to-inside" with $e_i$ as target qubit
    \item[$\bullet$] Measurement on $e_i$
\end{itemize}

\item[--]\textbf{Direct edge-toggle (CZ)} between $e_i$ and $e_j$ does not depend on the neighborhood of the two nodes, but it changes both $N(e_i)$ and $N(e_j)$. So any operation that depends on these neighborhoods does not commute with the \cz[cz] on the node $e_i$ ($e_j$): 
\begin{itemize}
    \item[$\bullet$] Emission by $e_i$ ($e_j$) in (L), (S), and (CS) modes
    \item[$\bullet$] ``e-to-inside" with $e_i$ ($e_j$) as target qubit
    \item[$\bullet$] Measurement on $e_i$ ($e_j$) 
\end{itemize}       

\item[--]\textbf{$e_i\text{-to-inside-}e_j\text{-connect}$} is a combination of e-to-inside and \cz[cz], therefore, an operation commutes with it on either of the nodes only if it commutes with both e-to-inside and \cz[cz] on the same node.
\end{itemize}

\begin{algorithm}[H]
\caption{Simplification Procedure}
\label{simp:alg}
\begin{algorithmic}[1]
\State Initialize $i \gets 0$.
\If{$i$ is last emitter}
\State \textbf{end} procedure.
\EndIf
\State Find the next two-node operation on emitter $e_i$, denote it as $\alpha_{ij}$ with second emitter $e_{j > i}$.

\If {no such $\alpha_{ij}$ is found}
    \State $i \gets i+1$
    \State Go to line 2.
\EndIf
\State Find the next two-node operation on $e_i$ acting on the same pair $(e_i,e_j)$, denote it as $\beta_{ij}$.

\If {no such $\beta_{ij}$ is found}
    \State Go to line 4.
\EndIf

\If{$\alpha_{ij}$ and $\beta_{ij}$ do not form a simplification case}
    \State Go to line 8.
\EndIf

\State Find the first operation on $e_i$ after $\alpha_{ij}$ that does not commute with $\alpha_{ij}$; denote it $\Gamma_i$.

    \If{$\Gamma_i$ occurs before $\beta_{ij}$ on $e_i$}
        \If{$\beta_{ij}$ does not commute with all operations on $e_i$ between (and including) $\Gamma_i$ and $\beta_{ij}$}
            \State Go to line 8.
        \EndIf
    \EndIf

    \State Find the first operation on $e_j$ after $\alpha_{ij}$ that does not commute with $\alpha_{ij}$; denote it $\Gamma_j$.

    \If{$\Gamma_j$ occurs before $\beta_{ij}$ on $e_j$}
        \If{$\beta_{ij}$ does not commute with all operations on $e_j$ between (and including) $\Gamma_j$ and $\beta_{ij}$}
            \State Go to line 8.
        \EndIf
    \EndIf

    \State Move both $\alpha_{ij}$ and $\beta_{ij}$ to immediately before $\Gamma_i$ on $e_i$ and $\Gamma_j$ on $e_j$ respectively.

    \State Update the DAG structure of sequence of operations and check for cycles.
    \If{a cycle is detected}
        \State Go to line 8.
    \EndIf

    \State Cancel or merge $\alpha_{ij}$ with $\beta_{ij}$.

\State Go to line 4.

\end{algorithmic}
\end{algorithm}

\begin{figure}[]
\centering
\includegraphics[width=0.9\columnwidth]{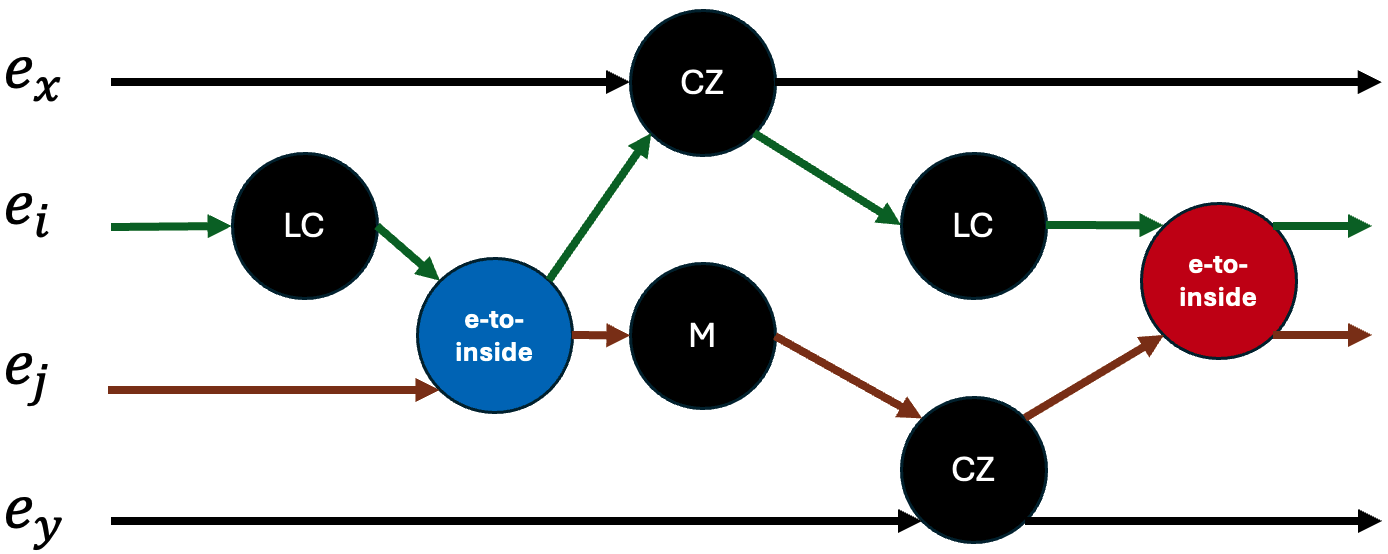}%
\caption{A DAG representation of the same piece of circuit depicted in Fig.~\ref{fig:simp-circ}.}
\label{fig:simp-dag}
\end{figure}

To have a consistent simplification process we introduce a formalism to represent the graphical generation recipe similar to that of a quantum circuit, i.e, a set of wires each representing a node instead of a qubit, with graphical transformation gates that are applied on one or two nodes and change the shape of the graph in the process (see Fig.~\ref{fig:simp-circ}). An underlying Directed-Acyclic-Graph (DAG) representation can also be extracted from this circuit, showing the topological order of the operations on the nodes. Each node in the DAG is an operation and directed edges indicate the input and output for those operation (see Fig.~\ref{fig:simp-dag}). 
A DAG representation guarantees a consistent topological order of operations and vice versa, as a result, if commuting two operations on a qubit causes a cycle (loop) in the initial DAG, the circuit is no longer executable due to ordering discrepancies such as needing the output of an operation to prepare its own input. One must have this in mind when considering the commutation of operation on different emitters. 

\begin{figure*}[]
\centering
\includegraphics[width=0.9\textwidth]{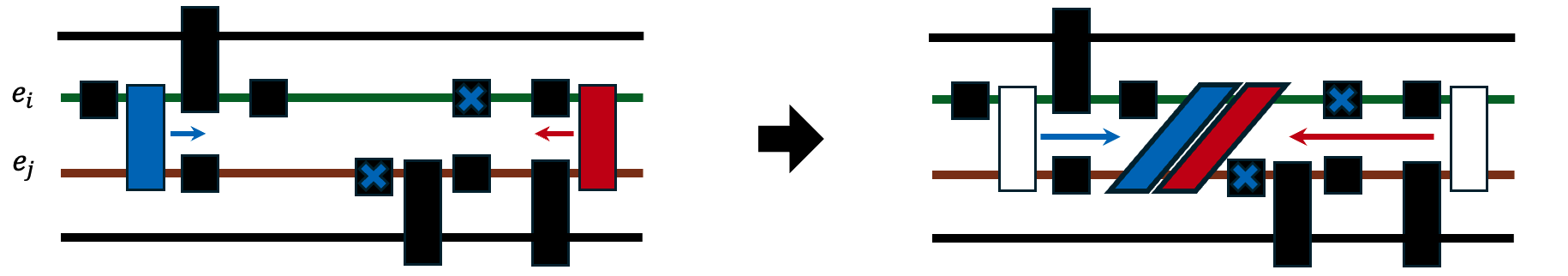}%
\caption{Example of simplification process. The gates marked by (x) indicate the boundary up-to which the blue gate (on the left) can commute. For simplification to be possible, the red gate (on the right) must be able to commute with all gate to reach the same spot. For the commuting process to be allowed, it should not induce any loops in the DAG structure of the circuit. The simplification can then be applied, replacing the two operations with a lower cost gate (merging) or removing both (cancellation), reducing the overall cost in the circuit. Graph operations on emitters (except for LC operations) are implemented using \cz[cz] gates. If two such operations can be merged, the number of \cz[cz] gates (and consequently the circuit depth) can be reduced. If the gates are identical and the operation is an involution, the gates cancel.}
\label{fig:simp-commute}
\end{figure*}

The simplification algorithm can now be written as \cref{simp:alg}. \Cref{fig:simp-commute} illustrates this process for an example circuit, where the eligibility of two gates for simplification is first checked through commutations to make them adjacent, and then they are merged or canceled if possible.

For maximum cost reduction \cref{simp:alg} is first employed to find and apply all possible cancellations, and then for merging. The reason is that the two-qubit gate count reduction associated with cancellation is two, while for all merging cases this number is one, hence the prioritization of the canceling simplifications.

\section{Case Analysis Rationale}
\label{app: rationale}
We provide rationale for each possible scenario discussed in the case analysis in \cref{sec:protocol}. The rationale for case A was given in the main text (see \cref{rationale: A}), so we start from case B.

\subsection*{Case: B1}
\begin{figure}[h]
\centering
\includegraphics[width=0.99\columnwidth]{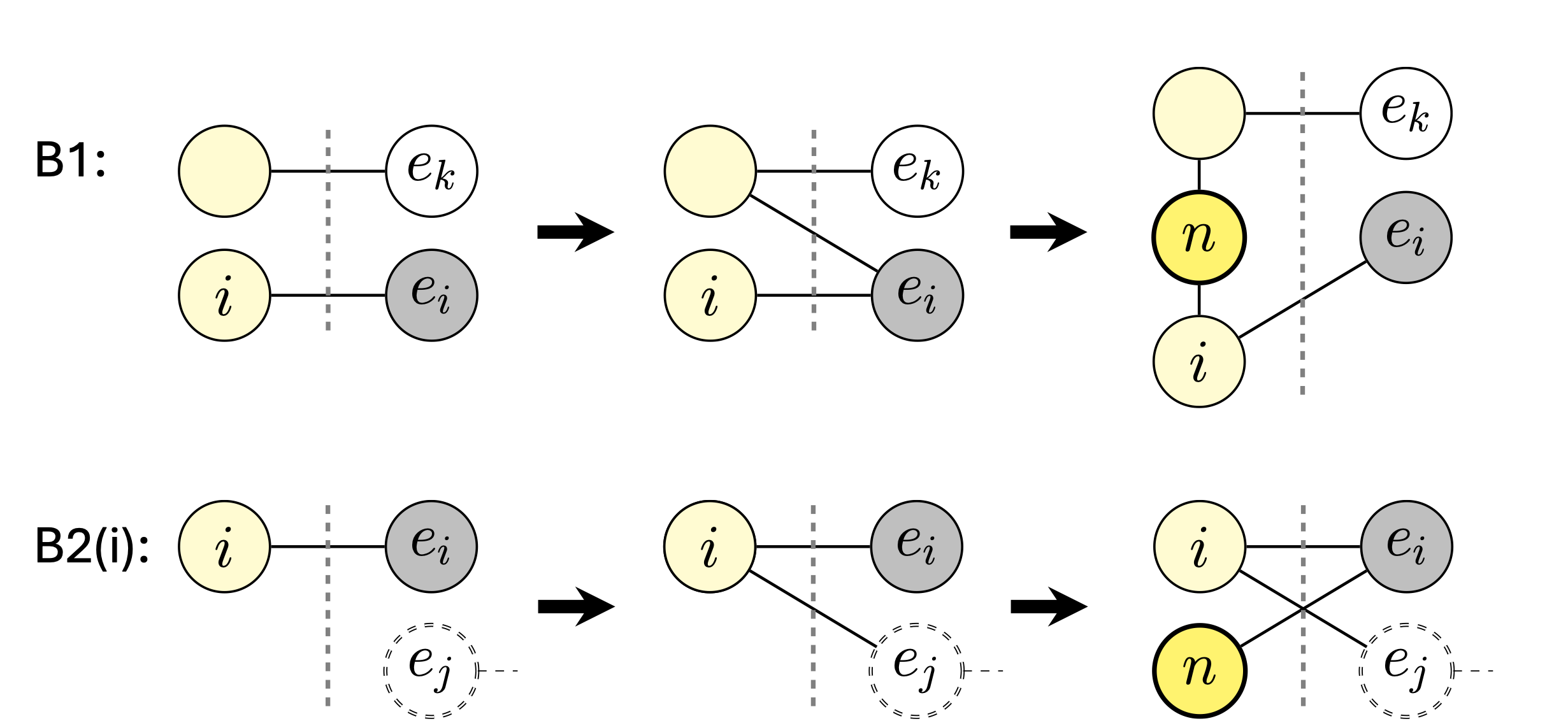}%
\caption{Cases B1 and B2(i). From left to right, the state of the physical graph at the beginning of the step (only showing the affected region), just before emission, and after emitting the new photon. The node $e_i$ is the chosen emitter and node $i$ represents its initial neighborhood. $e_k$ and $e_j$ are representatives of the emitters in the sets $\mathcal K$ and $\mathcal J$, respectively. Node $n$ is the new emitted photon. 
}
\label{fig:B1B2i}
\end{figure}

The set of emitter rows need no change as they still form a basis for $B$ matrix at the end of the step (condition \ref{I} satisfied). Since the new photon's row $R_{new}$ is zero, no connection between this node and any of the emitters is required (condition \ref{II} satisfied). The inside connections of the new photon are also established by connecting $e_i$ to inside of all $e_k\in \mathcal{K}$ (Condition \ref{III} satisfied). See Fig.~\ref{fig:B1B2i} for the process.

\subsection*{Case: B2 (i)}

The (SS) emission mode ensures the new photon is only connected to its emitter. In other words, the updated row of the chosen emitter $R_{e_i}(n+1)$ can be considered to be equal to $R_{new}$ and future edges of the new photon can be handled by $e_i$. The inside photons that were connected to this emitter before this step get connected to an equivalent set of emitters $e_j$ such that $\Sigma R_{e_j} = R_{e_i}(n)$, so their future connections can still be established (conditions \ref{I} \& \ref{II} satisfied). In other words, the photons that relied on the chosen emitter $e_i$ to build their future connections, are now connected to a combination of emitters that has the same future edge creation capacity, because $R_{e_i} = \Sigma R_{e_j}$ for members of the \textit{$\mathcal{J}$}. The \textit{$\mathcal{K}$} is empty in this scenario so condition \ref{III} is satisfied by default. See Fig.~\ref{fig:B1B2i} for the process.

\subsection*{Case: B2 (ii)}

\begin{figure*}[]
\centering
\includegraphics[width=0.9\textwidth]{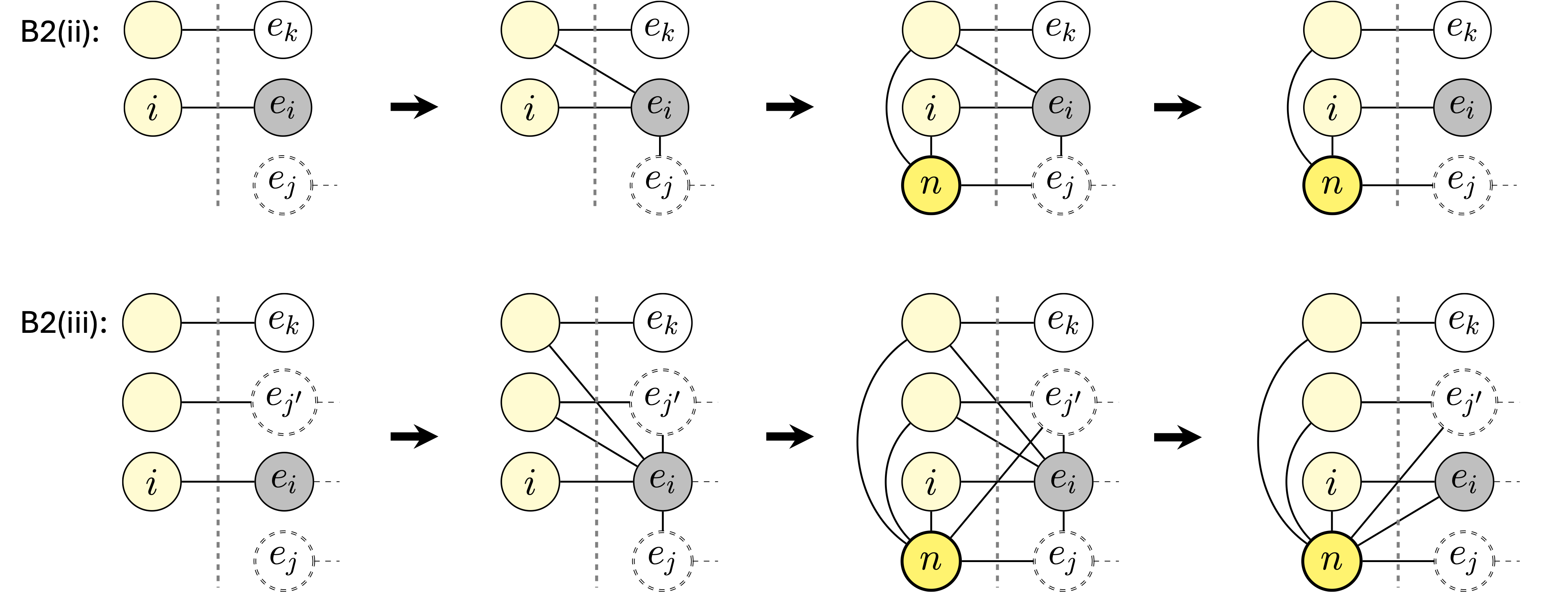}%
\caption{Cases B2(ii) and B2(iii). From left to right, the state of the physical graph at the beginning of the step (only showing the affected region), just before emission, after emitting the new photon, and lastly the disconnection step to undo the unnecessary edges.
The node $e_i$ is the chosen emitter and node $i$ represents its initial neighborhood. $e_k$ and $e_j$ are representatives of the emitters in the sets $\mathcal K$ and $\mathcal J$, respectively, while $e_{j'}\in \mathcal K \cap\mathcal J$. Node $n$ is the new emitted photon. 
}
\label{fig:B2iiB2iii}
\end{figure*}

The (S) emission mode makes the created photon connected to the neighbors of the emitter $e_i$ before the emission. Therefore, in this case the new photon will be connected to the emitters $e_j\in\mathcal{J}$ whose rows make up the new row, i.e., $R_{new} = \Sigma R_{e_j}$ (condition \ref{II} satisfied). The inside connections are also handled as the emitter $e_i$ was chosen from the \textit{$\mathcal{K}$} and was connected to the neighborhood of the rest of the members of \textit{$\mathcal{K}$} before emitting. As a result, the new photon has its inside connections inherited from its emitter (condition \ref{III} satisfied.) The set of emitter rows remains unchanged, so condition \ref{I} is satisfied. See Fig.~\ref{fig:B2iiB2iii} for the process. 

\subsection*{Case: B2 (iii)}

For the emitters only in one of \textit{$\mathcal{J}$} or \textit{$\mathcal{K}$}, the required edges to $e_i$ are made with the same reasoning as of the case B2(ii). For all emitters $e\in \mathcal{J}\cap\mathcal{K}$, the new photon should be connected to $e$ and also its neighborhood. This also includes the chosen emitter $e_i$ itself. In the last step, each disconnection operation between $e_i$ and neighbors of $e_{}\in \textit{$\mathcal{K}$} \cap \textit{$\mathcal{J}$}$ results in toggling the connection between the emitted photon and $e_i$ as the photon is already in the neighborhood of all such $e_{}$. Since the edge between the new photon and $e_i$ is required at the end, the emission mode is chosen considering the size of the intersection set being odd or even, to decide whether the photon and its emitter must start as connected or disconnected, ensuring the existence of that edge (conditions \ref{II} \& \ref{III} satisfied). There is no change in the emitter rows, and $R_{new}$ is linearly dependent, so condition \ref{I} is already satisfied in this case. See Fig.~\ref{fig:B2iiB2iii} for the process.

\subsection*{Case: C}

\begin{figure}[h]
\centering
\includegraphics[width=0.99\columnwidth]{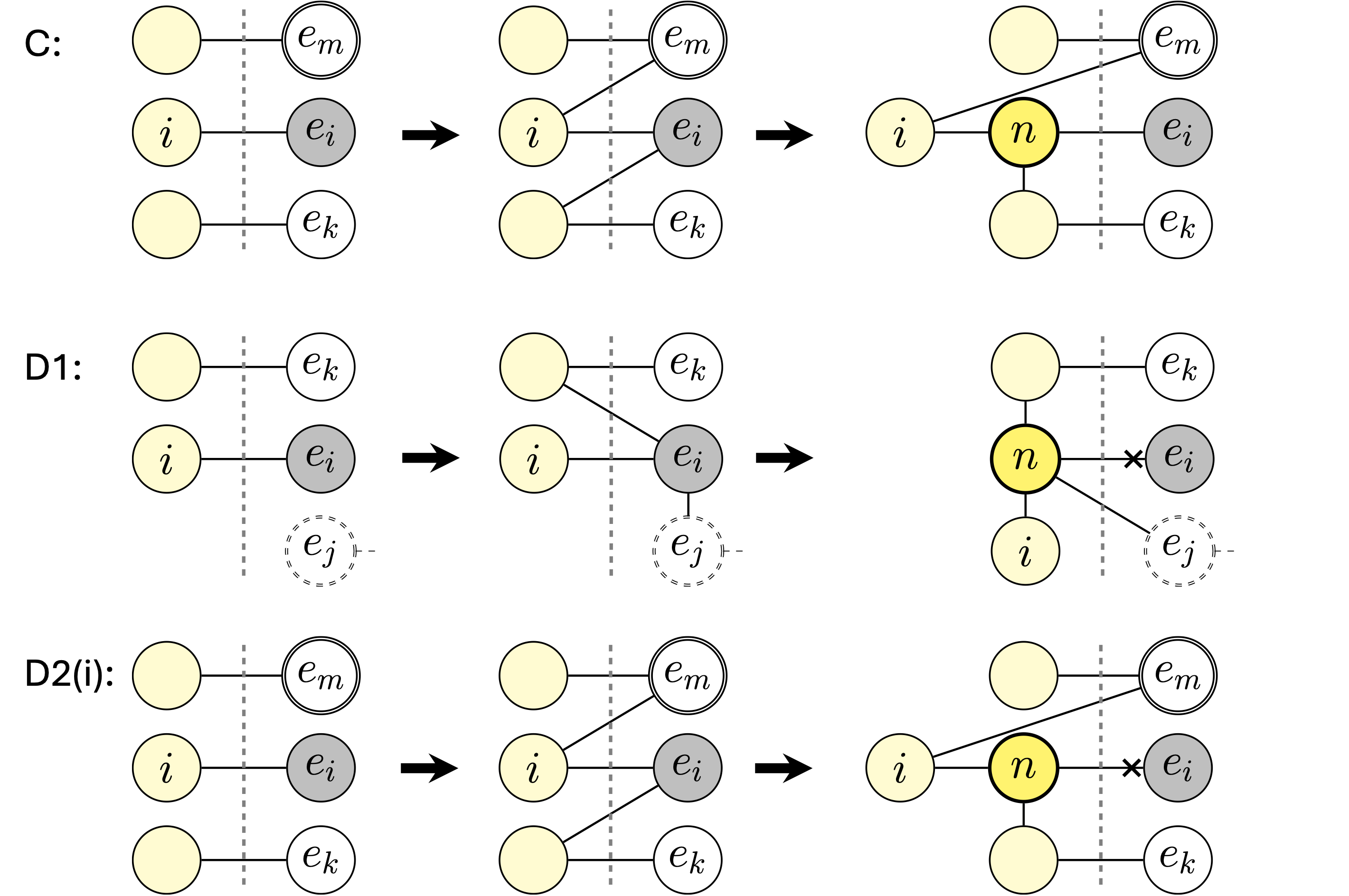}%
\caption{Cases C, D1, and D2(i). From left to right, the state of the physical graph at the beginning of the step (only showing the affected region), just before emission, and after emitting the new photon. 
The node $e_i$ is the chosen emitter and node $i$ represents its initial neighborhood. $e_k$, $e_j$, and $e_m$ are representatives of the emitters in the sets $\mathcal K$, $\mathcal J$, and $\mathcal M$ respectively. Node $n$ is the new emitted photon. 
}
\label{fig:CD1D2i}
\end{figure}

In the beginning of the step, emitter $e_i$, with the emitter row $R_{e_i}(n)$, can be responsible for handling the future edges of some photonic nodes that are connected to this emitter. After updating the $B$ matrix, the same row in $B$---corresponding to $R_{e_i}(n)$---becomes linearly dependent on the some other rows, i.e, it can be written as $\sum R_{e_m\neq i}$. As a result, any photon relying on $e_i$ for future edge can still establish its future edge if we it is instead connected to the set of emitters $\mathcal{M}$. The chosen $e_i$ is then free to take the new independent row $R_{new}$ as its new emitter row $R_{e_i}(n+1)$. At the end of the step, emitter $e_i$ is connected to the new photon “$n$” and is able to handle its future connections as describe in $R_{new}$ (conditions \ref{I} \& \ref{II} satisfied). The emission mode ensures that the new photon inherits all the neighbors of $e_i$ just before emission, which are the the combination of photons connected to emitters in $\mathcal{K}$ (condition \ref{III} satisfied). See Fig.~\ref{fig:CD1D2i} for the process.

\subsection*{Case: D1}

In this case $R_{e_i}(n)$ becomes zero after updating the $B$ matrix, so after the emission of the new photon, $e_i$ becomes free and is not responsible for no future edge creation. The inside connections of the new photon are ensured with first connecting the emitter to the required photons (inside neighbors of emitters in $\mathcal{K}$) and later using the emission in (L) mode (condition \ref{III} satisfied). Furthermore, since the row effect is 0, the future connections of the new photon ($R_{new} = \Sigma R_j$) can be handled if the new photon is connected to the respective emitters in \textit{$\mathcal{J}$}. This is done by first connecting $e_i$ to these emitter, then with mode (L) emission, the neighborhood of the $e_i$ transfers to the new photon and the required emitter connections are established (condition \ref{II} satisfied). At the end, $R_{e_i}$ is no longer part of the basis for the rows of $B$ matrix and $e_i$ is no longer needed to be attached to any photon, so it can be measured to be disconnected from the rest of the graph, allowing us to reset it and move it back in the pool of isolated/inactive emitters (condition \ref{I} satisfied). See Fig.~\ref{fig:CD1D2i} for the process.

\subsection*{Case: D2 (i)}

Since $R_{e_i}$ can be written as $\Sigma R_{e_m}$ for $e_{m\neq i} \in \mathcal{M}$, first we ensure the inside neighbors of $e_i$ are connected to the new set of emitters $\mathcal{M} - \{e_i\}$ instead of $e_i$ for future edge creations (condition \ref{II} satisfied). Next, the required inside neighbors of the next photon are connected to the emitter. And finally with (L) mode emission, the new photon takes the place of $e_i$, inheriting its neighbors (condition \ref{III} satisfied). As $R_{new}$ is a zero vector, future connections are not needed for the new photon so the emitter is free and can be measured and reset and its emitter row is removed from the basis set (condition \ref{I} satisfied). See Fig.~\ref{fig:CD1D2i} for the process.

\subsection*{Case: D2 (ii) (a)}

\begin{figure}[h]
\centering
\includegraphics[width=0.99\columnwidth]{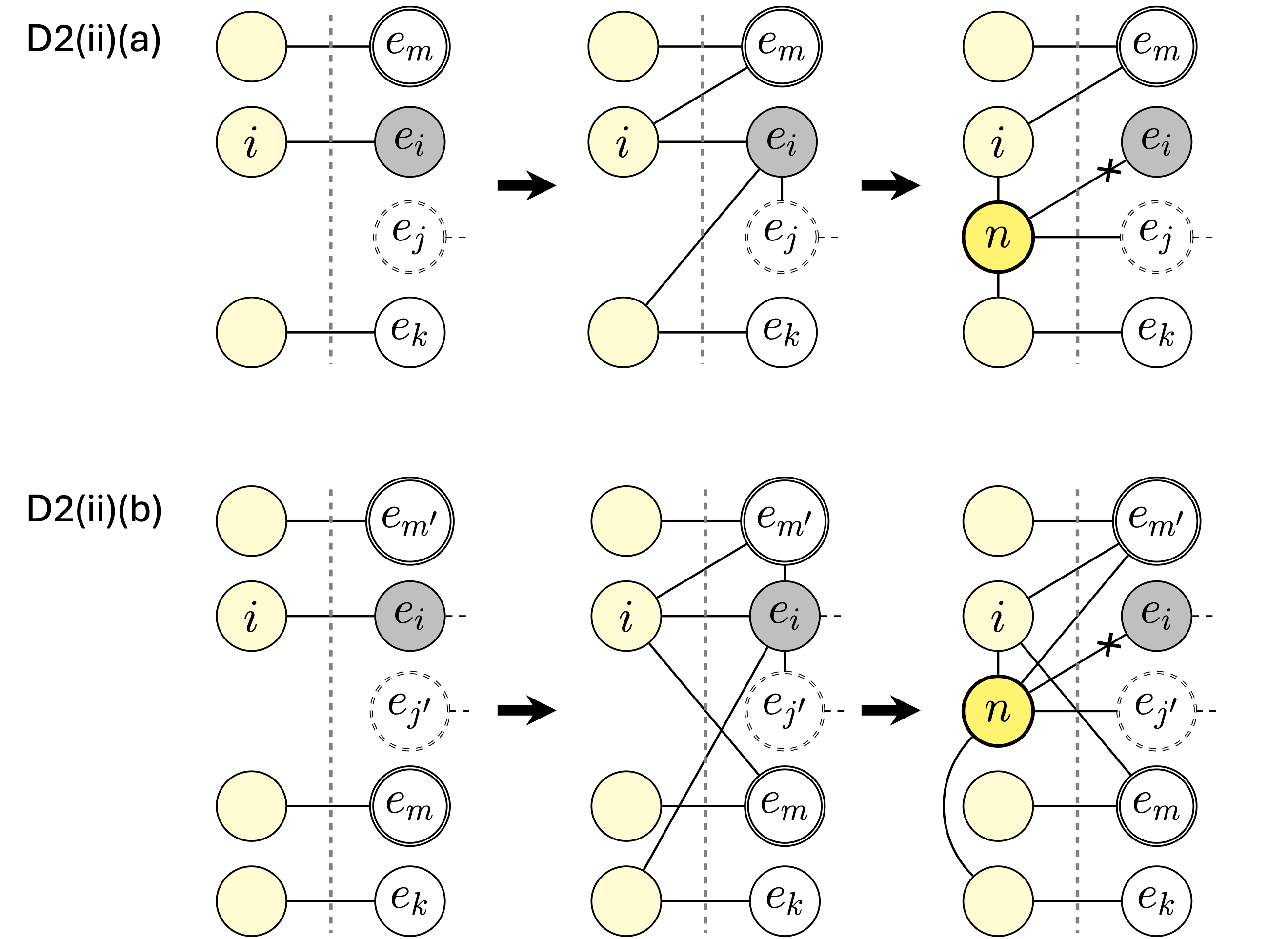}%
\caption{Cases D2(ii)(a) and D2(ii)(b). From left to right, the state of the physical graph at the beginning of the step (only showing the affected region), just before emission, and after emitting the new photon. 
The node $e_i$ is the chosen emitter and node $i$ represents its initial neighborhood. $e_k$, $e_j$, and $e_m$ are representatives of the emitters in the sets $\mathcal K$, $\mathcal J$, and $\mathcal M$ respectively, while $e_{j'}$ and $e_{m'}$ nodes represent the emitters that are exclusive members of the \textit{$\mathcal{J}$} and $\mathcal{M}$ sets, respectively. Node $n$ is the new emitted photon. 
}
\label{fig:D2iiaD2iib}
\end{figure}

The argument is the same as the previous case, except that the new photon row $R_{new}$ is now handled by making sure the new photon is connected to the emitters in the \textit{$\mathcal{J}$} as required by condition \ref{II}. See Fig.~\ref{fig:D2iiaD2iib} for the process.

\begin{figure*}[]
\centering
\includegraphics[width=1\textwidth]{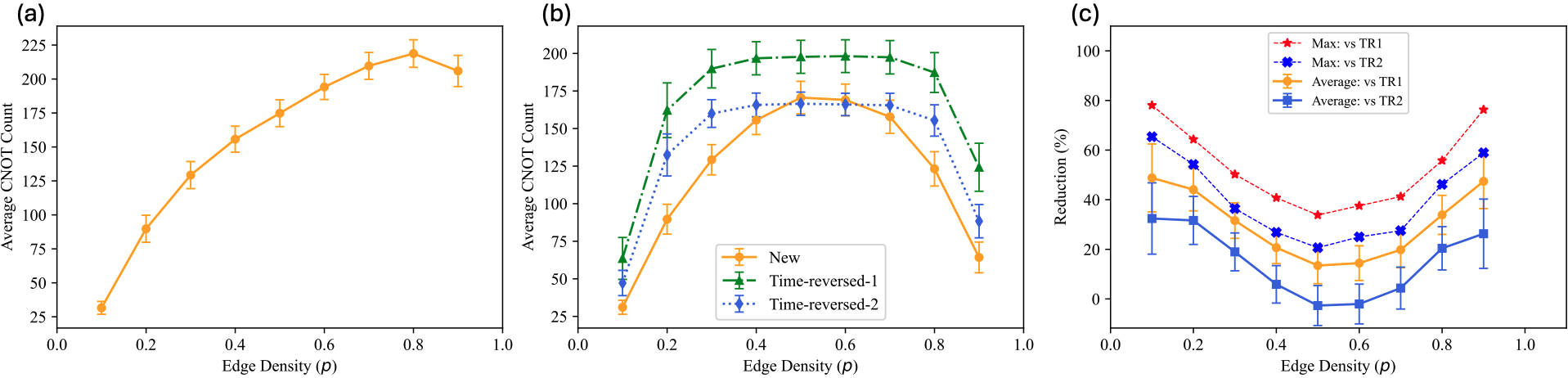}%
\caption{Behavior of \cnot[cnot] cost as a function of edge density. \textbf{(a)} Average number of \cnots[cnots] used by the new algorithm versus edge density $p$ (ranging from 10\% to 90\%) averaged over sample of 1000 random graphs with $N=30$ nodes for each $p$. Error bars represent one standard deviation. \textbf{(b)} Average \cnots[cnots] count comparison for the same set of samples for the three algorithms under study, \textit{New, Time-reversed-1, and Time-reversed-2}. The edge reduction is integrated to the New algorithm in this case. \textbf{(c)} The average reduction in \cnot[cnot] gates obtained by using the New method relative to the other two alternatives, \textit{Time-reversed-1 (TR1), and Time-reversed-2 (TR2)}.}
\label{fig:cnot_vs_p}
\end{figure*}

\subsection*{Case: D2 (ii) (b)}
Once the emitter is chosen, the next prescribed operation handles any future edge creations that were relying on the chosen emitter $e_i$ by connecting the emitters in the set $\mathcal{M}$ to the photons in the inside neighborhood of $e_i$ (condition \ref{II} satisfied). Next, $e_i$ gets connected to \textit{$\mathcal{K}$}, which ensures the inside connections for the next photon after the emission (condition \ref{III} satisfied). To be able to create $R_{new} = \Sigma R_j$ in future, one needs to establish connections to the emitters corresponding to each $R_j$. In this special case, we know one of such emitters is the chosen $e_i$ itself, but since $R_{e_i} = \Sigma R_{e_m}$, the connection between the new photon and $e_i$, can be substituted with a set of connections to all other $e_m \in \mathcal{M}$. This is in accordance with condition \ref{II}. Therefore, before the emission, $e_i$ needs to be connected to all $e_j \in \text{\textit{$\mathcal{J}$}}$, and all $e_m \in \mathcal{M}$, and since a double attempt in connecting two emitters cancels itself ($\cz[cz] \times \cz[cz] = \text{Identity}$), for common members of $\mathcal{M}$ and \textit{$\mathcal{J}$}, $e_i$ gets acted upon twice by \cz[cz] gates and the action is canceled. Consequently, it is enough to only make connections between $e_i$ and the nodes in the symmetric difference between the two sets. See Fig.~\ref{fig:D2iiaD2iib} for the process.

\section{Edge-density Dependence in Random Graphs}  \label{app:edge_dep}
Random graphs used in our analysis have a uniform probability for each pair of nodes to be connected by an edge (similar to the Erdos-Renyi model \cite{Erdos2022OnRG}). The total number of edges $|E|$ in a graph is determined by the edge density,  
\begin{align}
p = {|E|}/{\binom{N}{2}},
\end{align}
which quantifies the connectivity relative to a complete graph with all-to-all connections. The average number of two-qubit gates required to generate random graphs of a given size varies with edge density. Figure~\ref{fig:cnot_vs_p}(a) illustrates this behavior for random graphs with $N=30$ nodes, where each data point is averaged over 1000 random graphs. The \cnot[cnot] count increases with edge density, reaches a maximum, and then decreases for highly connected graphs.

Because the \cnot[cnot] count generally increases with edge density, we utilize the method introduced in Ref.~\cite{ghanbari_optimization_2023}, which replaces the target graph with a locally equivalent alternative having fewer edges, thereby reducing the \cnot[cnot] cost. To identify such alternatives, we use the deterministic edge reduction strategy described in Appendix~C of Ref.~\cite{ghanbari_optimization_2023}, which is an efficient process that finishes in at most $\mathcal{O}(N^2)$ steps. This edge reduction is applied as a built-in first step before executing the main algorithm to construct a generation circuit. Figure~\ref{fig:cnot_vs_p}(b) shows the resulting \cnot[cnot] suppression, shifting the \cnot[cnot] cost peak from approximately 80\% edge density to around 50\%, and improving performance over a broader range of $p$ values.

\section{Runtime}\label{app:runtime}
\begin{figure}[H]
    \centering
\includegraphics[width=0.85\columnwidth]{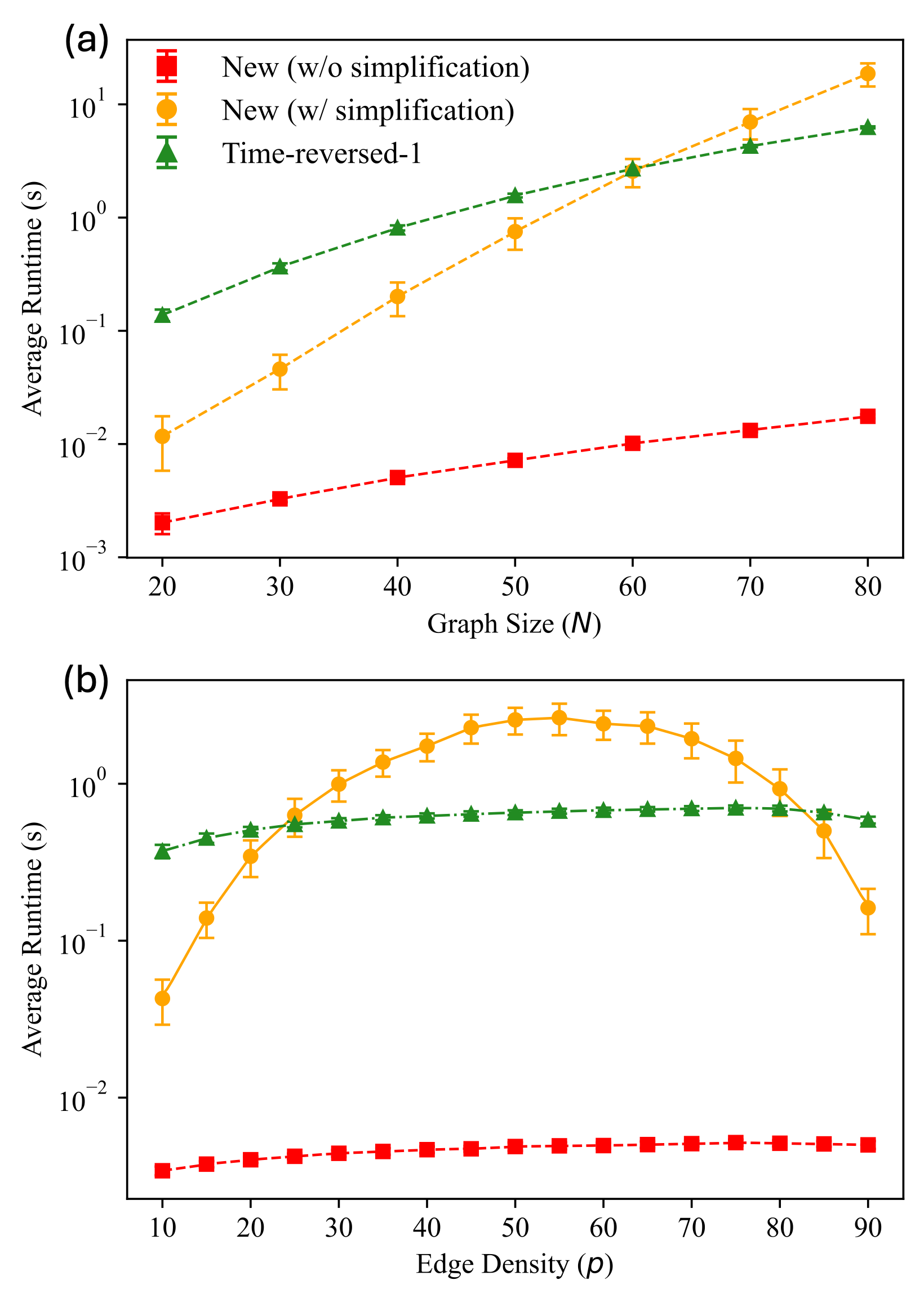}
    \caption{\textbf{(a)} Average runtime versus graph size $N$ (ranging from 20 to 80) for random graphs with edge density $p=10\%$. Results are shown for the proposed algorithm, both with and without recipe simplification, and for the \textit{time-reversed-1} algorithm as defined in the text. Each data point is averaged over 100 random graphs, with error bars representing one standard deviation. \textbf{(b)} Runtime analysis for random graphs of size $N=30$ with varying edge density from $p=10\%$ to $90\%$.}

    \label{fig:runtimes}
\end{figure}

We present runtime and scaling data for the new algorithm, applied to random graphs of varying sizes and edge densities. Figure~\ref{fig:runtimes}(a) shows the runtime of the base algorithm, with and without the recipe simplification procedure introduced in appendix \ref{app: simplification}. The runtime of the \textit{time-reversed-1} algorithm is also included for reference. All runtimes are averaged over 100 random graphs for each graph size with an edge density of $p=0.1$. The data are based on a Python implementation executed on a commercial Apple M1 Pro chip, without claiming optimality of the implementation.
Figure~\ref{fig:runtimes}(b) shows the runtime comparison for a fixed graph size of $N=30$ and varying edge densities. Similar to the \cnot[cnot] count, the runtime initially increases, reaches a maximum, and then decreases for highly connected graphs. This trend is more pronounced when using the simplification procedure. 

As evident (and without claiming optimality of the implementation), the runtime remains within a practical range for reasonably large graphs. The scaling and absolute runtime of the base algorithm (without simplifications) show significant improvement over the \textit{Time-reversed-1} method (also implemented in Python). Although the runtime of the simplification process grows faster than other shown algorithms, it yields more cost-efficient solutions as a trade-off. This becomes a concern only for very large graphs, in which case the simplification can be restricted or cut-off after a certain time, depending on the available computational resources.

\bibliography{Refs.bib}

@article{li_photonic_2022,
	title = {Photonic resource state generation from a minimal number of quantum emitters},
	volume = {8},
	copyright = {2022 The Author(s)},
	issn = {2056-6387},
	url = {https://www.nature.com/articles/s41534-022-00522-6},
	doi = {10.1038/s41534-022-00522-6},
	 
	number = {1},
	urldate = {2022-12-13},
	journal = {npj Quantum Inf},
	author = {Li, Bikun and Economou, Sophia E. and Barnes, Edwin},
	month = feb,
	year = {2022},
	pages = {1--7},

}

@article{van_den_nest_graphical_2004,
	title = {Graphical description of the action of local {Clifford} transformations on graph states},
	volume = {69},
	url = {https://link.aps.org/doi/10.1103/PhysRevA.69.022316},
	doi = {10.1103/PhysRevA.69.022316},
	number = {2},
	urldate = {2022-12-13},
	journal = {Phys. Rev. A},
	author = {Van den Nest, Maarten and Dehaene, Jeroen and De Moor, Bart},
	month = feb,
	year = {2004},
	pages = {022316},
}

@article{browne_resource-efficient_2005,
	title = {Resource-{Efficient} {Linear} {Optical} {Quantum} {Computation}},
	volume = {95},
	issn = {0031-9007, 1079-7114},
	url = {https://link.aps.org/doi/10.1103/PhysRevLett.95.010501},
	doi = {10.1103/PhysRevLett.95.010501},
	 
	number = {1},
	urldate = {2023-02-18},
	journal = {Phys. Rev. Lett.},
	author = {Browne, Daniel E. and Rudolph, Terry},
	month = jun,
	year = {2005},
	pages = {010501},

}

@article{hein_multiparty_2004,
	title = {Multiparty entanglement in graph states},
	volume = {69},
	issn = {1050-2947, 1094-1622},
	url = {https://link.aps.org/doi/10.1103/PhysRevA.69.062311},
	doi = {10.1103/PhysRevA.69.062311},
	 
	number = {6},
	urldate = {2023-02-18},
	journal = {Phys. Rev. A},
	author = {Hein, M. and Eisert, J. and Briegel, H. J.},
	month = jun,
	year = {2004},
	pages = {062311},

}

@article{nahum_quantum_2017,
	title = {Quantum {Entanglement} {Growth} under {Random} {Unitary} {Dynamics}},
	volume = {7},
	issn = {2160-3308},
	url = {http://link.aps.org/doi/10.1103/PhysRevX.7.031016},
	doi = {10.1103/PhysRevX.7.031016},
	 
	number = {3},
	urldate = {2023-03-02},
	journal = {Phys. Rev. X},
	author = {Nahum, Adam and Ruhman, Jonathan and Vijay, Sagar and Haah, Jeongwan},
	month = jul,
	year = {2017},
	pages = {031016},

}

@article{schon_sequential_2005,
	title = {Sequential {Generation} of {Entangled} {Multiqubit} {States}},
	volume = {95},
	issn = {0031-9007, 1079-7114},
	url = {https://link.aps.org/doi/10.1103/PhysRevLett.95.110503},
	doi = {10.1103/PhysRevLett.95.110503},
	 
	number = {11},
	urldate = {2023-03-04},
	journal = {Phys. Rev. Lett.},
	author = {Schön, C. and Solano, E. and Verstraete, F. and Cirac, J. I. and Wolf, M. M.},
	month = sep,
	year = {2005},
	pages = {110503},

}

@article{schwartz_deterministic_2016,
	title = {Deterministic generation of a cluster state of entangled photons},
	volume = {354},
	url = {https://www.science.org/doi/full/10.1126/science.aah4758},
	doi = {10.1126/science.aah4758},
	number = {6311},
	urldate = {2023-03-15},
	journal = {Science},
	author = {Schwartz, I. and Cogan, D. and Schmidgall, E. R. and Don, Y. and Gantz, L. and Kenneth, O. and Lindner, N. H. and Gershoni, D.},
	month = oct,
	year = {2016},
	note = {},
	pages = {434--437},

}

@article{yang_sequential_2022,
	title={Sequential generation of multiphoton entanglement with a {Rydberg} superatom},
	volume={16},
	issn={1749-4885, 1749-4893},
	url={https://www.nature.com/articles/s41566-022-01054-3},
	doi={10.1038/s41566-022-01054-3},
	number={9},
	urldate={2023-05-02},
	journal={Nat. Photon.},
	author={Yang, Chao-Wei and Yu, Yong and Li, Jun and Jing, Bo and Bao, Xiao-Hui and Pan, Jian-Wei},
	month=sep,
	year={2022},
	pages={658--661},
}

@article{thomas_efficient_2022,
	title = {Efficient generation of entangled multiphoton graph states from a single atom},
	volume = {608},
	issn = {0028-0836, 1476-4687},
	url = {https://www.nature.com/articles/s41586-022-04987-5},
	doi = {10.1038/s41586-022-04987-5},
	 
	number = {7924},
	urldate = {2023-05-02},
	journal = {Nature},
	author = {Thomas, Philip and Ruscio, Leonardo and Morin, Olivier and Rempe, Gerhard},
	month = aug,
	year = {2022},
	pages = {677--681},

}

@article{chu_independent_2023,
	title = {Independent {Electrical} {Control} of {Two} {Quantum} {Dots} {Coupled} through a {Photonic}-{Crystal} {Waveguide}},
	volume = {131},
	url = {https://link.aps.org/doi/10.1103/PhysRevLett.131.033606},
	doi = {10.1103/PhysRevLett.131.033606},
	number = {3},
	urldate = {2023-07-25},
	journal = {Phys. Rev. Lett.},
	author = {Chu, Xiao-Liu and Papon, Camille and Bart, Nikolai and Wieck, Andreas D. and Ludwig, Arne and Midolo, Leonardo and Rotenberg, Nir and Lodahl, Peter},
	month = jul,
	year = {2023},
	pages = {033606},

}

@article{cogan_deterministic_2023,
	title = {Deterministic generation of indistinguishable photons in a cluster state},
	volume = {17},
	issn = {1749-4885, 1749-4893},
	url = {https://www.nature.com/articles/s41566-022-01152-2},
	doi = {10.1038/s41566-022-01152-2},
	 
	number = {4},
	urldate = {2023-07-25},
	journal = {Nat. Photon.},
	author = {Cogan, Dan and Su, Zu-En and Kenneth, Oded and Gershoni, David},
	month = apr,
	year = {2023},
	pages = {324--329},

}

@article{economou_optically_2010,
	title = {Optically {Generated} 2-{Dimensional} {Photonic} {Cluster} {State} from {Coupled} {Quantum} {Dots}},
	volume = {105},
	issn = {0031-9007, 1079-7114},
	url = {https://link.aps.org/doi/10.1103/PhysRevLett.105.093601},
	doi = {10.1103/PhysRevLett.105.093601},
	 
	number = {9},
	urldate = {2023-08-16},
	journal = {Phys. Rev. Lett.},
	author = {Economou, Sophia E. and Lindner, Netanel and Rudolph, Terry},
	month = aug,
	year = {2010},
	pages = {093601},

}

@article{buterakos_deterministic_2017,
	title = {Deterministic {Generation} of {All}-{Photonic} {Quantum} {Repeaters} from {Solid}-{State} {Emitters}},
	volume = {7},
	issn = {2160-3308},
	url = {https://link.aps.org/doi/10.1103/PhysRevX.7.041023},
	doi = {10.1103/PhysRevX.7.041023},
	 
	number = {4},
	urldate = {2023-08-16},
	journal = {Phys. Rev. X},
	author = {Buterakos, Donovan and Barnes, Edwin and Economou, Sophia E.},
	month = oct,
	year = {2017},
	pages = {041023},

}

@article{pant_rate-distance_2017,
	title = {Rate-distance tradeoff and resource costs for all-optical quantum repeaters},
	volume = {95},
	issn = {2469-9926, 2469-9934},
	url = {https://link.aps.org/doi/10.1103/PhysRevA.95.012304},
	doi = {10.1103/PhysRevA.95.012304},
	 
	number = {1},
	urldate = {2023-08-16},
	journal = {Phys. Rev. A},
	author = {Pant, Mihir and Krovi, Hari and Englund, Dirk and Guha, Saikat},
	month = jan,
	year = {2017},
	pages = {012304},

}

@article{briegel_persistent_2001,
	title = {Persistent {Entanglement} in {Arrays} of {Interacting} {Particles}},
	volume = {86},
	issn = {0031-9007, 1079-7114},
	url = {https://link.aps.org/doi/10.1103/PhysRevLett.86.910},
	doi = {10.1103/PhysRevLett.86.910},
	 
	number = {5},
	urldate = {2023-09-02},
	journal = {Phys. Rev. Lett.},
	author = {Briegel, Hans J. and Raussendorf, Robert},
	month = jan,
	year = {2001},
	pages = {910--913},

}

@article{zwerger_measurement-based_2012,
	title = {Measurement-based quantum repeaters},
	volume = {85},
	issn = {1050-2947, 1094-1622},
	url = {https://link.aps.org/doi/10.1103/PhysRevA.85.062326},
	doi = {10.1103/PhysRevA.85.062326},
	 
	number = {6},
	urldate = {2023-09-02},
	journal = {Phys. Rev. A},
	author = {Zwerger, M. and Dür, W. and Briegel, H. J.},
	month = jun,
	year = {2012},
	pages = {062326},

}

@article{zwerger_universal_2013,
	title = {Universal and {Optimal} {Error} {Thresholds} for {Measurement}-{Based} {Entanglement} {Purification}},
	volume = {110},
	issn = {0031-9007, 1079-7114},
	url = {https://link.aps.org/doi/10.1103/PhysRevLett.110.260503},
	doi = {10.1103/PhysRevLett.110.260503},
	 
	number = {26},
	urldate = {2023-09-02},
	journal = {Phys. Rev. Lett.},
	author = {Zwerger, M. and Briegel, H. J. and Dür, W.},
	month = jun,
	year = {2013},
	pages = {260503},

}

@article{knill_scheme_2001,
	title = {A scheme for efficient quantum computation with linear optics},
	volume = {409},
	copyright = {2001 Macmillan Magazines Ltd.},
	issn = {1476-4687},
	url = {https://www.nature.com/articles/35051009},
	doi = {10.1038/35051009},
	 
	number = {6816},
	urldate = {2023-09-02},
	journal = {Nature},
	author = {Knill, E. and Laflamme, R. and Milburn, G. J.},
	month = jan,
	year = {2001},
	pages = {46--52},

}

@article{grice_arbitrarily_2011,
	title = {Arbitrarily complete {Bell}-state measurement using only linear optical elements},
	volume = {84},
	issn = {1050-2947, 1094-1622},
	url = {https://link.aps.org/doi/10.1103/PhysRevA.84.042331},
	doi = {10.1103/PhysRevA.84.042331},
	 
	number = {4},
	urldate = {2023-09-02},
	journal = {Phys. Rev. A},
	author = {Grice, W. P.},
	month = oct,
	year = {2011},
	pages = {042331},

}

@article{russo_generation_2019,
	title = {Generation of arbitrary all-photonic graph states from quantum emitters},
	volume = {21},
	url = {https://doi.org/10.1088/1367-2630/ab193d},
	doi = {10.1088/1367-2630/ab193d},
	number = {5},
	journal = {New Journal of Physics},
	author = {Russo, Antonio and Barnes, Edwin and Economou, Sophia E.},
	month = may,
	year = {2019},
	pages = {055002},

}

@incollection{hein_entanglement_2006-1,
	place = {NL},
  title = {Entanglement in graph states and its applications},
  volume = {162},
  ISSN = {0074-784X},
  url = {https://doi.org/10.3254/978-1-61499-018-5-115},
  DOI = {10.3254/978-1-61499-018-5-115},
  abstractNote = {11. Summary},
  number = {Quantum Computers,  Algorithms and Chaos},
  journal = {Proceedings of the International School of Physics &ldquo;Enrico Fermi&rdquo;},
  publisher = {IOS Press},
  author = {Hein, M and Dür, Wolfgang and Eisert, Jens and Raussendorf, Robert and Van den Nest, Maarten and Briegel, Hans.-J},
  year = {2006},
  pages = {115–218}
}

@phdthesis{GottesmanThesis,
author={Gottesman,Daniel},
year={1997},
title={Stabilizer codes and quantum error correction},
journal={ProQuest Dissertations and Theses},
pages={114},
isbn={978-0-591-49468-6},

url={http://myaccess.library.utoronto.ca/login?qurl=https%3A%2F%2Fwww.proquest.com%2Fdissertations-theses%2Fstabilizer-codes-quantum-error-correction%2Fdocview%2F304364982%2Fse-2%3Faccountid%3D14771},
}

@article{zwerger_measurement-based_2016,
	title = {Measurement-based quantum communication},
	volume = {122},
	issn = {1432-0649},
	url = {https://doi.org/10.1007/s00340-015-6285-8},
	doi = {10.1007/s00340-015-6285-8},
	 
	number = {3},
	urldate = {2023-09-02},
	journal = {Appl. Phys. B},
	author = {Zwerger, M. and Briegel, H. J. and Dür, W.},
	month = mar,
	year = {2016},
	pages = {50},

}

@article{azuma_all-photonic_2015,
	title = {All-photonic quantum repeaters},
	volume = {6},
	issn = {2041-1723},
	url = {https://www.nature.com/articles/ncomms7787},
	doi = {10.1038/ncomms7787},
	 
	number = {1},
	urldate = {2023-09-02},
	journal = {Nat Commun},
	author = {Azuma, Koji and Tamaki, Kiyoshi and Lo, Hoi-Kwong},
	month = apr,
	year = {2015},
	pages = {6787},

}

@article{briegel_measurement-based_2009,
	title = {Measurement-based quantum computation},
	volume = {5},
	issn = {1745-2473, 1745-2481},
	url = {http://www.nature.com/articles/nphys1157},
	doi = {10.1038/nphys1157},
	 
	number = {1},
	urldate = {2023-03-14},
	journal = {Nature Phys},
	author = {Briegel, H. J. and Browne, D. E. and Dür, W. and Raussendorf, R. and Van den Nest, M.},
	month = jan,
	year = {2009},
	pages = {19--26},

}

@article{lindner_proposal_2009,
	title = {Proposal for {Pulsed} {On}-{Demand} {Sources} of {Photonic} {Cluster} {State} {Strings}},
	volume = {103},
	issn = {0031-9007, 1079-7114},
	url = {https://link.aps.org/doi/10.1103/PhysRevLett.103.113602},
	doi = {10.1103/PhysRevLett.103.113602},
	 
	number = {11},
	urldate = {2023-09-02},
	journal = {Phys. Rev. Lett.},
	author = {Lindner, Netanel H. and Rudolph, Terry},
	month = sep,
	year = {2009},
	pages = {113602},

}

@article{Erdos2022OnRG,
  title={On random graphs. {I}.},
  author={Paul Erd\"{o}s and Alfr{\'e}d R{\'e}nyi},
  journal={Publicationes Mathematicae Debrecen},
  volume = {6},
  year={1959},
  pages = {290-297},

}

@article{GraphiQ,
  title = {GraphiQ: Quantum circuit design for photonic graph states},
  volume = {8},
  ISSN = {2521-327X},
  url = {http://dx.doi.org/10.22331/q-2024-08-28-1453},
  DOI = {10.22331/q-2024-08-28-1453},
  journal = {Quantum},
  publisher = {Verein zur Forderung des Open Access Publizierens in den Quantenwissenschaften},
  author = {Lin,  Jie and MacLellan,  Benjamin and Ghanbari,  Sobhan and Belleville,  Julie and Tran,  Khuong and Robichaud,  Luc and Melko,  Roger G. and Lo,  Hoi-Kwong and Roztocki,  Piotr},
  year = {2024},
  month = aug,
  pages = {1453}
}

@article{ghanbari_optimization_2023,
  title = {Optimization of deterministic photonic-graph-state generation via local operations},
  volume = {110},
  ISSN = {2469-9934},
  url = {http://dx.doi.org/10.1103/PhysRevA.110.052605},
  DOI = {10.1103/physreva.110.052605},
  number = {5},
  journal = {Physical Review A},
  publisher = {American Physical Society (APS)},
  author = {Ghanbari,  Sobhan and Lin,  Jie and MacLellan,  Benjamin and Robichaud,  Luc and Roztocki,  Piotr and Lo,  Hoi-Kwong},
  year = {2024},
  month = nov 
}

@misc{kaur_resource-efficient_2024,
	title = {Resource-efficient and loss-aware photonic graph state preparation using an array of quantum emitters, and application to all-photonic quantum repeaters},
	url = {http://arxiv.org/abs/2402.00731},
	doi = {10.48550/arXiv.2402.00731},
	urldate = {2024-02-07},
	publisher = {arXiv},
	author = {Kaur, Eneet and Patil, Ashlesha and Guha, Saikat},
	month = feb,
	year = {2024},
	note = {arXiv:2402.00731 [quant-ph]},

}

@article{huet_deterministic_2025,
  title = {Deterministic and reconfigurable graph state generation with a single solid-state quantum emitter},
  volume = {16},
  ISSN = {2041-1723},
  url = {http://dx.doi.org/10.1038/s41467-025-59693-3},
  DOI = {10.1038/s41467-025-59693-3},
  number = {1},
  journal = {Nature Communications},
  publisher = {Springer Science and Business Media LLC},
  author = {Huet,  H. and Ramesh,  P. R. and Wein,  S. C. and Coste,  N. and Hilaire,  P. and Somaschi,  N. and Morassi,  M. and Lemaître,  A. and Sagnes,  I. and Doty,  M. F. and Krebs,  O. and Lanco,  L. and Fioretto,  D. A. and Senellart,  P.},
  year = {2025},
  month = may 
}

@article{Azuma2023,
  title = {Quantum repeaters: From quantum networks to the quantum internet},
  volume = {95},
  ISSN = {1539-0756},
  url = {http://dx.doi.org/10.1103/RevModPhys.95.045006},
  DOI = {10.1103/revmodphys.95.045006},
  number = {4},
  journal = {Reviews of Modern Physics},
  publisher = {American Physical Society (APS)},
  author = {Azuma,  Koji and Economou,  Sophia E. and Elkouss,  David and Hilaire,  Paul and Jiang,  Liang and Lo,  Hoi-Kwong and Tzitrin,  Ilan},
  year = {2023},
  month = dec 
}

@misc{cesa_hierarchical_2025,
	title = {Hierarchical generation and design of quantum codes for resource-efficient loss-tolerant quantum communications},
	url = {http://arxiv.org/abs/2501.18693},
	doi = {10.48550/arXiv.2501.18693},
	urldate = {2025-02-08},
	publisher = {arXiv},
	author = {Cesa, Francesco and Feri, Tommaso and Bassi, Angelo},
	month = feb,
	year = {2025},
	note = {arXiv:2501.18693 [quant-ph]},
}

@article{Chan2025,
  title = {Tailoring Fusion-Based Photonic Quantum Computing Schemes to Quantum Emitters},
  volume = {6},
  ISSN = {2691-3399},
  url = {http://dx.doi.org/10.1103/PRXQuantum.6.020304},
  DOI = {10.1103/prxquantum.6.020304},
  number = {2},
  journal = {PRX Quantum},
  publisher = {American Physical Society (APS)},
  author = {Chan,  Ming Lai and Bell,  Thomas J. and Pettersson,  Love A. and Chen,  Susan X. and Yard,  Patrick and Sørensen,  Anders S. and Paesani,  Stefano},
  year = {2025},
  month = apr 
}

@article{Pettersson2025,
  title = {Deterministic Generation of Concatenated Graph Codes from Quantum Emitters},
  volume = {6},
  ISSN = {2691-3399},
  url = {http://dx.doi.org/10.1103/PRXQuantum.6.010305},
  DOI = {10.1103/prxquantum.6.010305},
  number = {1},
  journal = {PRX Quantum},
  publisher = {American Physical Society (APS)},
  author = {Pettersson,  Love A. and Sørensen,  Anders S. and Paesani,  Stefano},
  year = {2025},
  month = jan 
}

@misc{wein_minimizing_2024,
  doi = {10.48550/ARXIV.2412.08611},
  url = {https://arxiv.org/abs/2412.08611},
  author = {Wein,  Stephen C. and de Brugière,  Timothée Goubault and Music,  Luka and Senellart,  Pascale and Bourdoncle,  Boris and Mansfield,  Shane},
  title = {Minimizing resource overhead in fusion-based quantum computation using hybrid spin-photon devices},
  publisher = {arXiv},
  year = {2024},
  copyright = {arXiv.org perpetual,  non-exclusive license}
}

@article{Bell2023,
  title = {Optimizing Graph Codes for Measurement-Based Loss Tolerance},
  volume = {4},
  ISSN = {2691-3399},
  url = {http://dx.doi.org/10.1103/PRXQuantum.4.020328},
  DOI = {10.1103/prxquantum.4.020328},
  number = {2},
  journal = {PRX Quantum},
  publisher = {American Physical Society (APS)},
  author = {Bell,  Thomas J. and Pettersson,  Love A. and Paesani,  Stefano},
  year = {2023},
  month = may 
}

@article{Inlek2017,
  title = {Multispecies Trapped-Ion Node for Quantum Networking},
  volume = {118},
  ISSN = {1079-7114},
  url = {http://dx.doi.org/10.1103/PhysRevLett.118.250502},
  DOI = {10.1103/physrevlett.118.250502},
  number = {25},
  journal = {Physical Review Letters},
  publisher = {American Physical Society (APS)},
  author = {Inlek,  I. V. and Crocker,  C. and Lichtman,  M. and Sosnova,  K. and Monroe,  C.},
  year = {2017},
  month = jun 
}

@article{Drmota2023,
  title = {Robust Quantum Memory in a Trapped-Ion Quantum Network Node},
  volume = {130},
  ISSN = {1079-7114},
  url = {http://dx.doi.org/10.1103/PhysRevLett.130.090803},
  DOI = {10.1103/physrevlett.130.090803},
  number = {9},
  journal = {Physical Review Letters},
  publisher = {American Physical Society (APS)},
  author = {Drmota,  P. and Main,  D. and Nadlinger,  D. P. and Nichol,  B. C. and Weber,  M. A. and Ainley,  E. M. and Agrawal,  A. and Srinivas,  R. and Araneda,  G. and Ballance,  C. J. and Lucas,  D. M.},
  year = {2023},
  month = mar 
}

@article{Takou2024,
  title = {Optimization complexity and resource minimization of emitter-based photonic graph state generation protocols},
  volume = {11},
  ISSN = {2056-6387},
  url = {http://dx.doi.org/10.1038/s41534-025-01056-3},
  DOI = {10.1038/s41534-025-01056-3},
  number = {1},
  journal = {npj Quantum Information},
  publisher = {Springer Science and Business Media LLC},
  author = {Takou,  Evangelia and Barnes,  Edwin and Economou,  Sophia E.},
  year = {2025},
  month = jul 
}

@article{Aqua2025,
  title = {Atom-Mediated Deterministic Generation and Stitching of Photonic Graph States},
  volume = {6},
  ISSN = {2691-3399},
  url = {http://dx.doi.org/10.1103/PRXQuantum.6.010340},
  DOI = {10.1103/prxquantum.6.010340},
  number = {1},
  journal = {PRX Quantum},
  publisher = {American Physical Society (APS)},
  author = {Aqua,  Ziv and Dayan,  Barak},
  year = {2025},
  month = mar 
}

@article{Lee2023,
  title = {Graph-theoretical optimization of fusion-based graph state generation},
  volume = {7},
  ISSN = {2521-327X},
  url = {http://dx.doi.org/10.22331/q-2023-12-20-1212},
  DOI = {10.22331/q-2023-12-20-1212},
  journal = {Quantum},
  publisher = {Verein zur Forderung des Open Access Publizierens in den Quantenwissenschaften},
  author = {Lee,  Seok-Hyung and Jeong,  Hyunseok},
  year = {2023},
  month = dec,
  pages = {1212}
}

@article{kwiat_new_1995,
  title = {New High-Intensity Source of Polarization-Entangled Photon Pairs},
  author = {Kwiat, Paul G. and Mattle, Klaus and Weinfurter, Harald and Zeilinger, Anton and Sergienko, Alexander V. and Shih, Yanhua},
  journal = {Phys. Rev. Lett.},
  volume = {75},
  issue = {24},
  pages = {4337--4341},
  numpages = {0},
  year = {1995},
  month = {Dec},
  publisher = {American Physical Society},
  doi = {10.1103/PhysRevLett.75.4337},
  url = {https://link.aps.org/doi/10.1103/PhysRevLett.75.4337}
}

@article{varnava_how_2008,
  title = {How Good Must Single Photon Sources and Detectors Be for Efficient Linear Optical Quantum Computation?},
  author = {Varnava, Michael and Browne, Daniel E. and Rudolph, Terry},
  journal = {Phys. Rev. Lett.},
  volume = {100},
  issue = {6},
  pages = {060502},
  numpages = {4},
  year = {2008},
  month = {Feb},
  publisher = {American Physical Society},
  doi = {10.1103/PhysRevLett.100.060502},
  url = {https://link.aps.org/doi/10.1103/PhysRevLett.100.060502}
}

@article{zhang_demonstration_2008,
  title = {Demonstration of a scheme for the generation of ``event-ready'' entangled photon pairs from a single-photon source},
  author = {Zhang, Qiang and Bao, Xiao-Hui and Lu, Chao-Yang and Zhou, Xiao-Qi and Yang, Tao and Rudolph, Terry and Pan, Jian-Wei},
  journal = {Phys. Rev. A},
  volume = {77},
  issue = {6},
  pages = {062316},
  numpages = {4},
  year = {2008},
  month = {Jun},
  publisher = {American Physical Society},
  doi = {10.1103/PhysRevA.77.062316},
  url = {https://link.aps.org/doi/10.1103/PhysRevA.77.062316}
}

@article{akopian_entangled_2006,
  title = {Entangled Photon Pairs from Semiconductor Quantum Dots},
  author = {Akopian, N. and Lindner, N. H. and Poem, E. and Berlatzky, Y. and Avron, J. and Gershoni, D. and Gerardot, B. D. and Petroff, P. M.},
  journal = {Phys. Rev. Lett.},
  volume = {96},
  issue = {13},
  pages = {130501},
  numpages = {4},
  year = {2006},
  month = {Apr},
  publisher = {American Physical Society},
  doi = {10.1103/PhysRevLett.96.130501},
  url = {https://link.aps.org/doi/10.1103/PhysRevLett.96.130501}
}

@article{ewert_34-efficient_2014,
  title = {$3/4$-Efficient Bell Measurement with Passive Linear Optics and Unentangled Ancillae},
  author = {Ewert, Fabian and van Loock, Peter},
  journal = {Phys. Rev. Lett.},
  volume = {113},
  issue = {14},
  pages = {140403},
  numpages = {5},
  year = {2014},
  month = {Sep},
  publisher = {American Physical Society},
  doi = {10.1103/PhysRevLett.113.140403},
  url = {https://link.aps.org/doi/10.1103/PhysRevLett.113.140403}
}

@article{dhara_entangling_2023,
  title = {Entangling quantum memories via heralded photonic Bell measurement},
  author = {Dhara, Prajit and Englund, Dirk and Guha, Saikat},
  journal = {Phys. Rev. Res.},
  volume = {5},
  issue = {3},
  pages = {033149},
  numpages = {19},
  year = {2023},
  month = {Sep},
  publisher = {American Physical Society},
  doi = {10.1103/PhysRevResearch.5.033149},
  url = {https://link.aps.org/doi/10.1103/PhysRevResearch.5.033149}
}

@article{levonian_optical_2022,
  title = {Optical Entanglement of Distinguishable Quantum Emitters},
  author = {Levonian, D. S. and Riedinger, R. and Machielse, B. and Knall, E. N. and Bhaskar, M. K. and Knaut, C. M. and Bekenstein, R. and Park, H. and Lon\ifmmode \check{c}\else \v{c}\fi{}ar, M. and Lukin, M. D.},
  journal = {Phys. Rev. Lett.},
  volume = {128},
  issue = {21},
  pages = {213602},
  numpages = {7},
  year = {2022},
  month = {May},
  publisher = {American Physical Society},
  doi = {10.1103/PhysRevLett.128.213602},
  url = {https://link.aps.org/doi/10.1103/PhysRevLett.128.213602}
}

@article{hurst_generating_2019,
  title = {Generating Maximal Entanglement between Spectrally Distinct Solid-State Emitters},
  author = {Hurst, David L. and Joanesarson, Kristoffer B. and Iles-Smith, Jake and M\o{}rk, Jesper and Kok, Pieter},
  journal = {Phys. Rev. Lett.},
  volume = {123},
  issue = {2},
  pages = {023603},
  numpages = {6},
  year = {2019},
  month = {Jul},
  publisher = {American Physical Society},
  doi = {10.1103/PhysRevLett.123.023603},
  url = {https://link.aps.org/doi/10.1103/PhysRevLett.123.023603}
}

@article{wei_cavity-assisted_2025,
  title = {Cavity-assisted efficient interaction between two nonidentical emitters via a dressed resonance},
  author = {Wei, Yan and Liao, Zeyang and Xing, Fan and Lu, Yu-wei and Wang, Xue-Hua},
  journal = {Phys. Rev. B},
  volume = {111},
  issue = {16},
  pages = {165303},
  numpages = {7},
  year = {2025},
  month = {Apr},
  publisher = {American Physical Society},
  doi = {10.1103/PhysRevB.111.165303},
  url = {https://link.aps.org/doi/10.1103/PhysRevB.111.165303}
}

@article{carlson_theory_2019,
  title = {Theory and experiments of coherent photon coupling in semiconductor nanowire waveguides with quantum dot molecules},
  author = {Carlson, Chelsea and Dalacu, Dan and Gustin, Chris and Haffouz, Sofiane and Wu, Xiaohua and Lapointe, Jean and Williams, Robin L. and Poole, Philip J. and Hughes, Stephen},
  journal = {Phys. Rev. B},
  volume = {99},
  issue = {8},
  pages = {085311},
  numpages = {18},
  year = {2019},
  month = {Feb},
  publisher = {American Physical Society},
  doi = {10.1103/PhysRevB.99.085311},
  url = {https://link.aps.org/doi/10.1103/PhysRevB.99.085311}
}

@article{qiao_coherent_2020,
  title = {Coherent Multispin Exchange Coupling in a Quantum-Dot Spin Chain},
  author = {Qiao, Haifeng and Kandel, Yadav P. and Deng, Kuangyin and Fallahi, Saeed and Gardner, Geoffrey C. and Manfra, Michael J. and Barnes, Edwin and Nichol, John M.},
  journal = {Phys. Rev. X},
  volume = {10},
  issue = {3},
  pages = {031006},
  numpages = {10},
  year = {2020},
  month = {Jul},
  publisher = {American Physical Society},
  doi = {10.1103/PhysRevX.10.031006},
  url = {https://link.aps.org/doi/10.1103/PhysRevX.10.031006}
}

@article{caterpillar,
title = {The number of caterpillars},
journal = {Discrete Mathematics},
volume = {6},
number = {4},
pages = {359-365},
year = {1973},
issn = {0012-365X},
doi = {https://doi.org/10.1016/0012-365X(73)90067-8},
url = {https://www.sciencedirect.com/science/article/pii/0012365X73900678},
author = {Frank Harary and Allen J. Schwenk}
}

@article{Ralph-loss-2005,
  title = {Loss-Tolerant Optical Qubits},
  author = {Ralph, T. C. and Hayes, A. J. F. and Gilchrist, Alexei},
  journal = {Phys. Rev. Lett.},
  volume = {95},
  issue = {10},
  pages = {100501},
  numpages = {4},
  year = {2005},
  month = {Aug},
  publisher = {American Physical Society},
  doi = {10.1103/PhysRevLett.95.100501},
  url = {https://link.aps.org/doi/10.1103/PhysRevLett.95.100501}
}

@article{Lee-parity-2023,
	author = {Lee, Seok-Hyung and Omkar, Srikrishna and Teo, Yong Siah and Jeong, Hyunseok},
	date = {2023/04/24},
	doi = {10.1038/s41534-023-00705-9},
	id = {Lee2023},
	isbn = {2056-6387},
	journal = {npj Quantum Information},
	number = {1},
	pages = {39},
	title = {Parity-encoding-based quantum computing with Bayesian error tracking},
	url = {https://doi.org/10.1038/s41534-023-00705-9},
	volume = {9},
	year = {2023}}

@article{bartolucci_fusion-based_2023,
  title = {Fusion-based quantum computation},
  volume = {14},
  ISSN = {2041-1723},
  url = {http://dx.doi.org/10.1038/s41467-023-36493-1},
  DOI = {10.1038/s41467-023-36493-1},
  number = {1},
  journal = {Nature Communications},
  publisher = {Springer Science and Business Media LLC},
  author = {Bartolucci,  Sara and Birchall,  Patrick and Bombín,  Hector and Cable,  Hugo and Dawson,  Chris and Gimeno-Segovia,  Mercedes and Johnston,  Eric and Kieling,  Konrad and Nickerson,  Naomi and Pant,  Mihir and Pastawski,  Fernando and Rudolph,  Terry and Sparrow,  Chris},
  year = {2023},
  month = feb 
}

@article{Raussendorf-fault-2006,
  title = {A fault-tolerant one-way quantum computer},
  volume = {321},
  ISSN = {0003-4916},
  url = {http://dx.doi.org/10.1016/j.aop.2006.01.012},
  DOI = {10.1016/j.aop.2006.01.012},
  number = {9},
  journal = {Annals of Physics},
  publisher = {Elsevier BV},
  author = {Raussendorf,  R. and Harrington,  J. and Goyal,  K.},
  year = {2006},
  month = sep,
  pages = {2242–2270}
}

@article{Raussendorf-topological-2007,
  title = {Topological fault-tolerance in cluster state quantum computation},
  volume = {9},
  ISSN = {1367-2630},
  url = {http://dx.doi.org/10.1088/1367-2630/9/6/199},
  DOI = {10.1088/1367-2630/9/6/199},
  number = {6},
  journal = {New Journal of Physics},
  publisher = {IOP Publishing},
  author = {Raussendorf,  R and Harrington,  J and Goyal,  K},
  year = {2007},
  month = jun,
  pages = {199–199}
}

@article{raussendorf_one-way_2001,
  title = {A One-Way Quantum Computer},
  author = {Raussendorf, Robert and Briegel, Hans J.},
  journal = {Phys. Rev. Lett.},
  volume = {86},
  issue = {22},
  pages = {5188--5191},
  numpages = {0},
  year = {2001},
  month = {May},
  publisher = {American Physical Society},
  doi = {10.1103/PhysRevLett.86.5188},
  url = {https://link.aps.org/doi/10.1103/PhysRevLett.86.5188}
}

@article{Raussendorf-measurement-based-2003,
  title = {Measurement-based quantum computation on cluster states},
  author = {Raussendorf, Robert and Browne, Daniel E. and Briegel, Hans J.},
  journal = {Phys. Rev. A},
  volume = {68},
  issue = {2},
  pages = {022312},
  numpages = {32},
  year = {2003},
  month = {Aug},
  publisher = {American Physical Society},
  doi = {10.1103/PhysRevA.68.022312},
  url = {https://link.aps.org/doi/10.1103/PhysRevA.68.022312}
}

@article{Varnava-loss-2006,
  title = {Loss Tolerance in One-Way Quantum Computation via Counterfactual Error Correction},
  volume = {97},
  ISSN = {1079-7114},
  url = {http://dx.doi.org/10.1103/PhysRevLett.97.120501},
  DOI = {10.1103/physrevlett.97.120501},
  number = {12},
  journal = {Physical Review Letters},
  publisher = {American Physical Society (APS)},
  author = {Varnava,  Michael and Browne,  Daniel E. and Rudolph,  Terry},
  year = {2006},
  month = sep 
}

@article{li_resource_2015,
  title = {Resource Costs for Fault-Tolerant Linear Optical Quantum Computing},
  author = {Li, Ying and Humphreys, Peter C. and Mendoza, Gabriel J. and Benjamin, Simon C.},
  journal = {Phys. Rev. X},
  volume = {5},
  issue = {4},
  pages = {041007},
  numpages = {15},
  year = {2015},
  month = {Oct},
  publisher = {American Physical Society},
  doi = {10.1103/PhysRevX.5.041007},
  url = {https://link.aps.org/doi/10.1103/PhysRevX.5.041007}
}

@article{Song2024,
  title = {Encoded-Fusion-Based Quantum Computation for High Thresholds with Linear Optics},
  author = {Song, Wooyeong and Kang, Nuri and Kim, Yong-Su and Lee, Seung-Woo},
  journal = {Phys. Rev. Lett.},
  volume = {133},
  issue = {5},
  pages = {050605},
  numpages = {6},
  year = {2024},
  month = {Aug},
  publisher = {American Physical Society},
  doi = {10.1103/PhysRevLett.133.050605},
  url = {https://link.aps.org/doi/10.1103/PhysRevLett.133.050605}
}

@article{Schlingemann_2001,
  title = {Quantum error-correcting codes associated with graphs},
  author = {Schlingemann, D. and Werner, R. F.},
  journal = {Phys. Rev. A},
  volume = {65},
  issue = {1},
  pages = {012308},
  numpages = {8},
  year = {2001},
  month = {Dec},
  publisher = {American Physical Society},
  doi = {10.1103/PhysRevA.65.012308},
  url = {https://link.aps.org/doi/10.1103/PhysRevA.65.012308}
}

@misc{ren2025,
  doi = {10.48550/ARXIV.2503.16346},
  url = {https://arxiv.org/abs/2503.16346},
  author = {Ren,  Xiangyu and Huang,  Yuexun and Liang,  Zhiding and Barbalace,  Antonio},
  title = {A Scalable and Robust Compilation Framework for Emitter-Photonic Graph State},
  publisher = {arXiv},
  year = {2025},
  copyright = {Creative Commons Attribution 4.0 International}
}

@article{Ferreira2024,
  title = {Deterministic generation of multidimensional photonic cluster states with a single quantum emitter},
  volume = {20},
  ISSN = {1745-2481},
  url = {http://dx.doi.org/10.1038/s41567-024-02408-0},
  DOI = {10.1038/s41567-024-02408-0},
  number = {5},
  journal = {Nature Physics},
  publisher = {Springer Science and Business Media LLC},
  author = {Ferreira,  Vinicius S. and Kim,  Gihwan and Butler,  Andreas and Pichler,  Hannes and Painter,  Oskar},
  year = {2024},
  month = feb,
  pages = {865–870}
}

@article{Pichler2017,
  title = {Universal photonic quantum computation via time-delayed feedback},
  volume = {114},
  ISSN = {1091-6490},
  url = {http://dx.doi.org/10.1073/pnas.1711003114},
  DOI = {10.1073/pnas.1711003114},
  number = {43},
  journal = {Proceedings of the National Academy of Sciences},
  publisher = {Proceedings of the National Academy of Sciences},
  author = {Pichler,  Hannes and Choi,  Soonwon and Zoller,  Peter and Lukin,  Mikhail D.},
  year = {2017},
  month = oct,
  pages = {11362–11367}
}

@misc{supp,
  note = "See Supplemental Materials for more information."
}

\pagebreak
\newpage
\widetext
\begin{center}
	\Large{\textbf{Supplemental Materials}}
\end{center}

\setcounter{section}{0}
\setcounter{equation}{0}
\setcounter{figure}{0}
\setcounter{table}{0}
\setcounter{algorithm}{0}
\makeatletter
\renewcommand{\thetable}{S\arabic{table}}   
\renewcommand{\thefigure}{S\arabic{figure}}
\renewcommand{\thealgorithm}{S\arabic{algorithm}}

\renewcommand*{\citenumfont}[1]{S#1}
\renewcommand*{\bibnumfmt}[1]{[S#1]}

\section*{A: emission order for the studied graphs}
Figure~\ref{fig1s} shows the graph structures along with their emission order, indicated by the node labels. A depth-first search was used to determine the node traversal order in cases where no natural or heuristic ordering was apparent.

\begin{figure*}[h]
	\centering
	\includegraphics[width=0.95\textwidth]{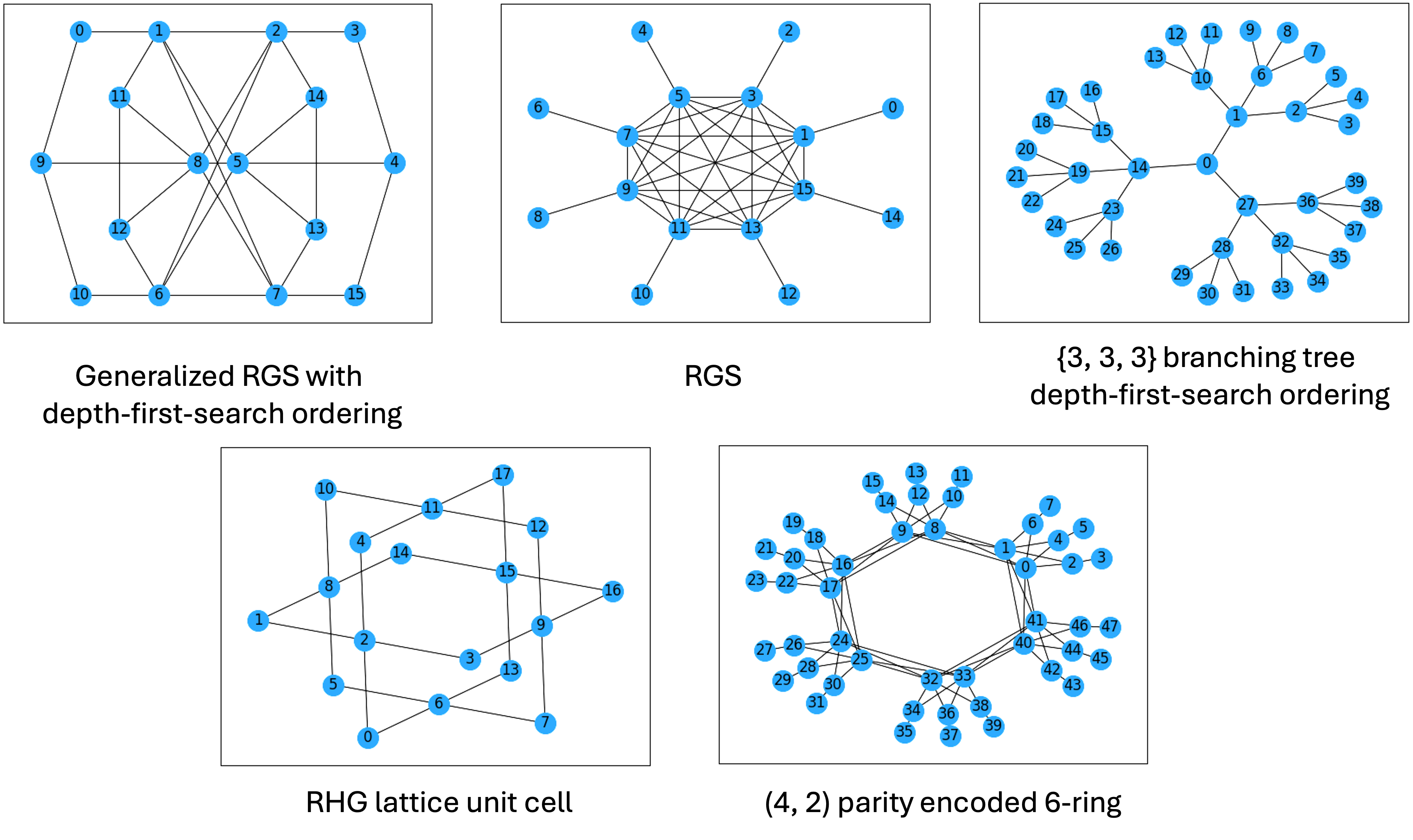}
	\caption{Emission order of photons represented as node labels for different types of graph.}
	\label{fig1s}
\end{figure*}

\section*{B: random graphs}
The following algorithm outlines the procedure used to generate the random connected graphs employed in this work.

\begin{algorithm}[H]
	\caption{Connected Random Graph Generation}
	\begin{algorithmic}[1]
		\Require $n$ (number of nodes), $p$ (edge probability parameter)
		\Ensure Connected random graph $G$
		\State Initialize $G$ as a random spanning tree with $n$ nodes
		\State $expected\_edges \gets \lfloor p \cdot n \cdot (n-1) / 2 \rfloor$
		\State $additional\_edges \gets \max(expected\_edges - (n-1), 0)$
		\While{$\#edges(G) < n - 1 + additional\_edges$}
		\State Randomly select two distinct nodes $(u, v)$
		\If{$(u, v) \notin E(G)$}
		\State Add edge $(u, v)$ to $G$
		\EndIf
		\EndWhile
		\State \Return $G$
	\end{algorithmic}
\end{algorithm}

\section*{C: encoded 6-ring graph elementary circuit blocks}
The parity-encoded 6-ring graphs are obtained by placing six logical nodes, each comprised of $m \times n$ physical qubits, in a ring configuration. The quantum circuit generating the state contains repeated circuit blocks due to the structural symmetries of these graphs and the repeated form of the entanglement within each logical node. First, let us employ the algorithm \textit{``time-reversed-2''} to determine the generation circuit. In Fig.~\ref{fig2s}, we show the elementary circuit blocks that are repeated $n$ times to generate each logical qubit. Each block includes $m$ photon emissions, yielding a total of $n \times m$ qubits per logical node. From the number of two-qubit gates used in each block, one can determine the overall scaling of two-qubit gate usage with graph size. Explicitly, in this case, $1, 2, 1, 2, 1,$ and $1$ CNOTs are used in the elementary circuit blocks corresponding to the six logical qubits, respectively. Accounting for the $n$ repetitions, the total scaling is $8n$. Note that the generation circuit can have a different structure when transitioning between logical nodes; such transition blocks introduce an offset in the exact number of two-qubit gates used in the full circuit. We have analyzed the generation circuits for a range of values of $n$ and $m$, summarized in Table~\ref{tab:parity-6ring}, and find that the exact number of CNOTs is $8n - 4$ for any $n,m > 2$. This is consistent with the expected $8n$ scaling in the asymptotic large-$n$ regime. Note that an increase in $m$ only affects the number of times the emission operations are repeated in each block and does not result in more emitter-emitter CNOTs.

We next consider the new \textit{``Graph Builder''} algorithm, whose corresponding elementary circuit blocks are shown in Fig.~\ref{fig3s}. In this construction, all six logical nodes are generated by circuit blocks that each contain a single two-qubit gate. Repeating these blocks $n$ times therefore leads to a total two-qubit gate scaling of $6n$. As before, transition blocks between logical nodes introduce a constant offset in the exact gate count. From the explicit generation circuits evaluated for a range of values of $n$ and $m$, summarized in Table~\ref{tab:parity-6ring}, we find that the total number of CZ gates is given by $6n+4$. This again agrees with the expected $6n$ scaling in the asymptotic large-$n$ limit.

Finally, we consider the \textit{``time-reversed-1''} algorithm. The generation circuit in this case can still be understood in terms of repeated circuit structures and the corresponding elementary blocks are presented in Fig.~\ref{fig4s}, showing the use of $2, 2, 1, 2, 1$, and $1$ CNOTs for each of the six logical node circuit blocks respectively. The overall scaling is thus $9n$. The exact two-qubit gate cost can also be extracted from the full generation circuits, with the results summarized in Table~\ref{tab:parity-6ring}. We find that the total number of two-qubit gates scales as $9n - 7$, confirming our prediction based on the circuit blocks.

\begin{figure*}[h]
	\centering
	\includegraphics[width=0.95\textwidth]{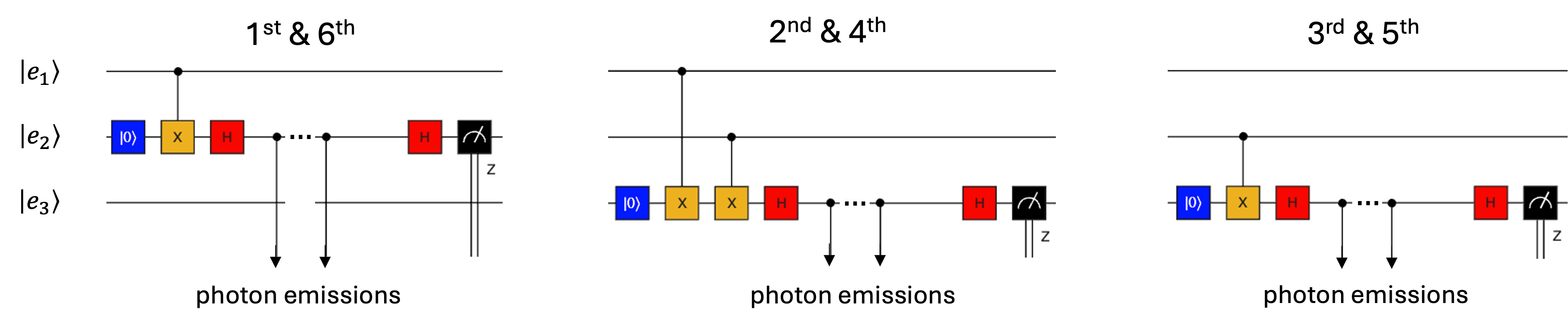}
	\caption{Elementary circuit blocks used in the generation of the parity-encoded 6-ring graph using \textit{``time-reversed-2''} algorithm. Three distinct circuit blocks are shown, corresponding to the logical nodes $(1,6)$, $(2,4)$, and $(3,5)$, as indicated above each block. Each block is repeated $n$ times to generate the corresponding logical node. Photon emissions are to be repeated $m$ times in each block and are represented by arrows originated from the emitters in the quantum circuit. Note that the first block and the third block are related by a symmetry, namely relabeling of the three emitters. However, the second block looks different from other two blocks and involves two (instead of one) CNOT gates.}
	\label{fig2s}
\end{figure*}

\begin{table*}[h]
	\caption{Number of two-qubit gates used for the generation of on 6-ring parity encoded graphs for different $(n, m)$ encoding configurations. The algorithms \textbf{TR1}, and \textbf{TR2}, are referring to the two \textit{time-reversed} generation algorithms defined in the main text. The number of emitters required ($n_e$), number of nodes ($N$), and number of edges ($|E|$) in each graph are also reported.}
	\label{tab:parity-6ring}
	\centering
	\small
	\begin{tabular*}{\textwidth}{l @{\extracolsep{\fill}} c c c c c c}
		\toprule
		\textbf{Encoding parameters $(n, m)$} & {\textbf{$n_e$}} & {\textbf{$N$}} & {\textbf{$\mid E \mid$}} & {\textbf{Graph Builder}} & {\textbf{TR2}} & {\textbf{\makecell{TR1}}} \\
		\midrule
		(4,2) & 3 & 48 & 78 & 28 & 28 & 29 \\
		(4,3) & 3 & 72 & 144 & 28 & 28 & 29 \\
		(5,3) & 3 & 90 & 174 & 34 & 36 & 38 \\
		(6,3) & 3 & 108 & 204 & 40 & 44 & 47 \\
		(5,4) & 3 & 120 & 264 & 34 & 36 & 38 \\
		(6,4) & 3 & 144 & 306 & 40 & 44 & 47 \\
		(7,4) & 3 & 168 & 348 & 46 & 52 & 56 \\
		(10,4) & 3 & 240 & 474 & 64 & 76 & 83 \\
		(9,6) & 3 & 324 & 744 & 58 & 68 & 73 \\
		(17,4) & 3 & 408 & 768 & 106 & 132 & 146 \\
		(12,7) & 3 & 504 & 1152 & 76 & 92 & 101 \\
		\bottomrule
	\end{tabular*}
\end{table*}

\begin{figure*}[h]
	\centering
	\includegraphics[width=0.95\textwidth]{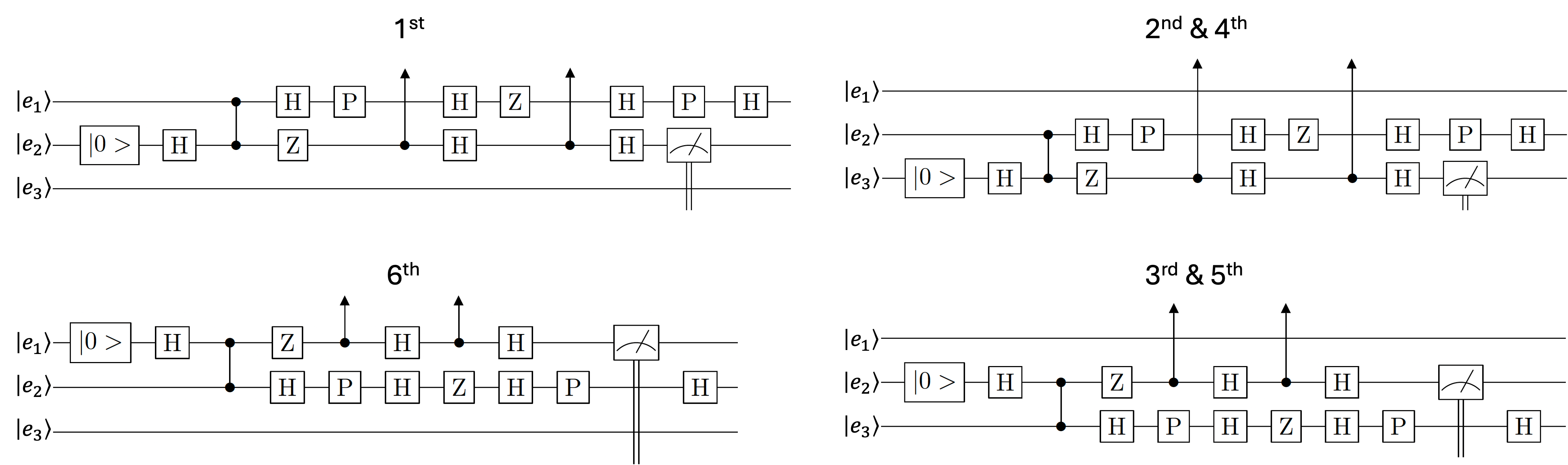}
	\caption{Elementary circuit blocks used in the generation of the parity-encoded 6-ring graph using the \textit{``Graph Builder''} algorithm. Four distinct circuit blocks are shown, corresponding to the logical nodes $(1)$, $(2,4)$, $(3,5)$, and $(6)$, as indicated above each block. Each block is repeated $n$ times to generate the corresponding logical node. Photon emissions are to be repeated $m$ times in each block (the shown example is for $m=2$) and are represented by arrows originated from the emitters in the quantum circuit. It worth noting that all four blocks are equivalent under a permutation of emitter qubit labels.}
	\label{fig3s}
\end{figure*}

\begin{figure*}[h]
	\centering
	\includegraphics[width=0.95\textwidth]{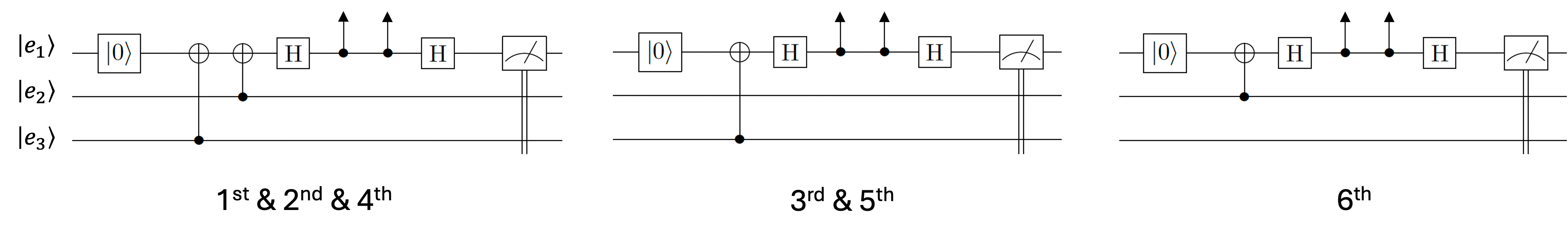}
	\caption{Elementary circuit blocks used in the generation of the parity-encoded 6-ring graph using the \textit{``time-reversed-1''} algorithm. Three distinct circuit blocks are shown, corresponding to the logical nodes $(1, 2, 4)$, $(3,5)$, and $(6)$, as indicated below each block. Each block is repeated $n$ times to generate the corresponding logical node. Photon emissions are to be repeated $m$ times in each block (the shown example is for $m=2$) and are represented by arrows originated from the emitters in the quantum circuit. Note that the second and third blocks are the same up to a relabeling of emitters. In contrast, the first block involves an extra CNOT.}
	\label{fig4s}
\end{figure*}

\section*{D: Explicit example for leveraging additional emitters}
In this section, we present a simple and explicit example illustrating the potential advantage of allowing additional emitters in the construction of photonic graph-state generation circuits. We first apply the default Graph Builder algorithm that constrains the construction to the minimum number of emitters required for the target graph. We then compare this with an alternative approach in which this constraint is relaxed and additional emitters are permitted.
The introduction of extra emitters follows a greedy strategy; before the emission of each photon, the option of introducing a new emitter is evaluated and is accepted only if it reduces the number of two-qubit gates required to prepare the system for the emission of the next immediate photon. Note that this strategy is local in nature and more sophisticated approaches, for example, those involving multi-step look-ahead, may yield further improvements.
The target graph is a random 8 node graph shown in Fig.~\ref{fig5s} and the resulting generation circuits are shown in Fig.~\ref{fig6s}. Due to the large circuit size, and to maintain visual clarity, we also provide a truncated version of each circuit showing all gates up to the final occurrence of a two-qubit gate; see Fig.~\ref{fig7s}. Allowing additional emitters leads to a reduction in the number of emitter–emitter CZ gates from 7 to 5, as well as a reduction in the two-qubit depth of the circuit from 7 to 4. Here, the two-qubit depth is defined in terms of the logical dependency structure of the circuit and is extracted from the directed acyclic graph (DAG) representation of the quantum circuit, considering emitters from their initializations to subsequent measurements.

\begin{figure}[h]
	\centering
	\includegraphics[width=0.35\textwidth]{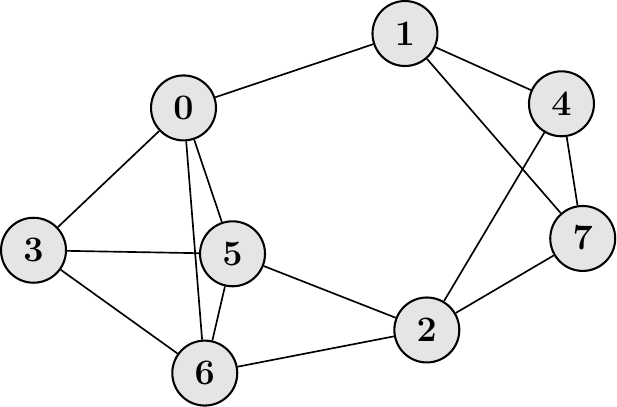}
	\caption{The target graph state.}
	\label{fig5s}
\end{figure}

\begin{figure*}[h]
	\centering
	\includegraphics[width=1\textwidth]{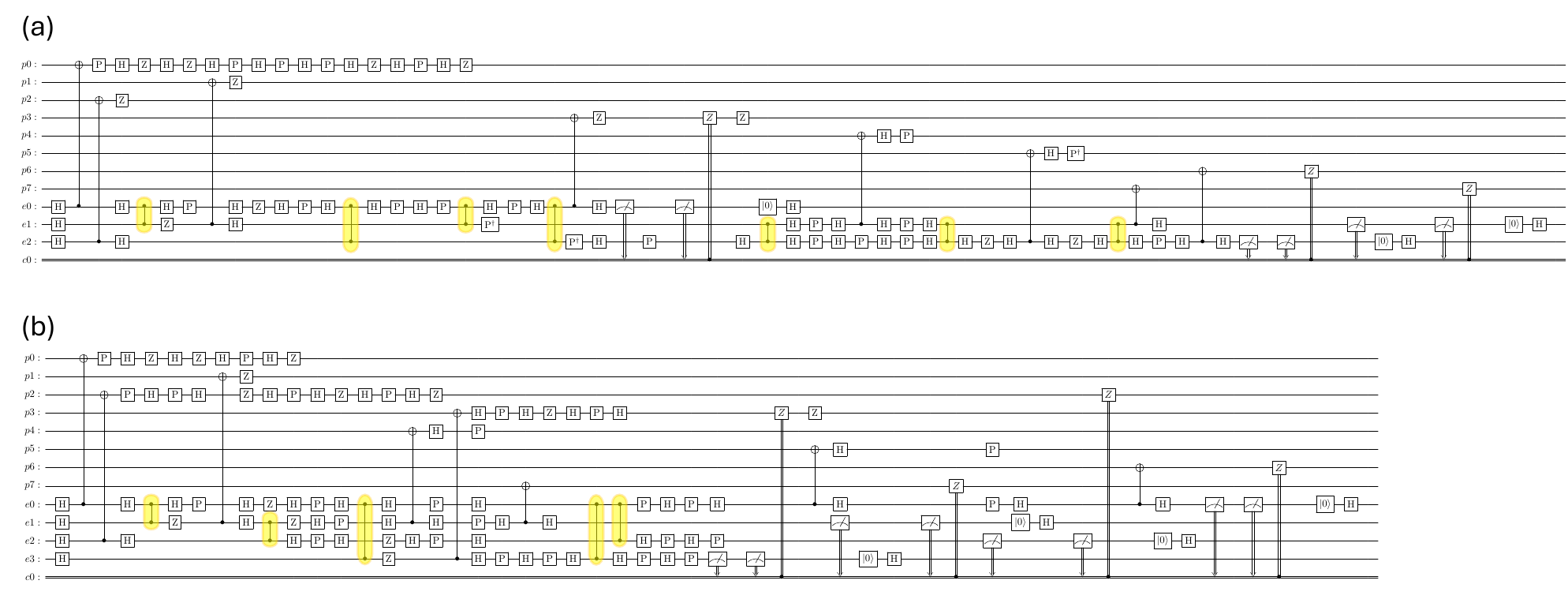}
	\caption{The generation circuits. (a) Using minimum number of emitters. (b) Allowing extra emitters. The advantage is clear in both number of two-qubit operations---highlighted in the circuit---and the depth of the circuit.}
	\label{fig6s}
\end{figure*}

\begin{figure}[h]
	\hspace*{-1cm}
	\includegraphics[width=1.07\textwidth]{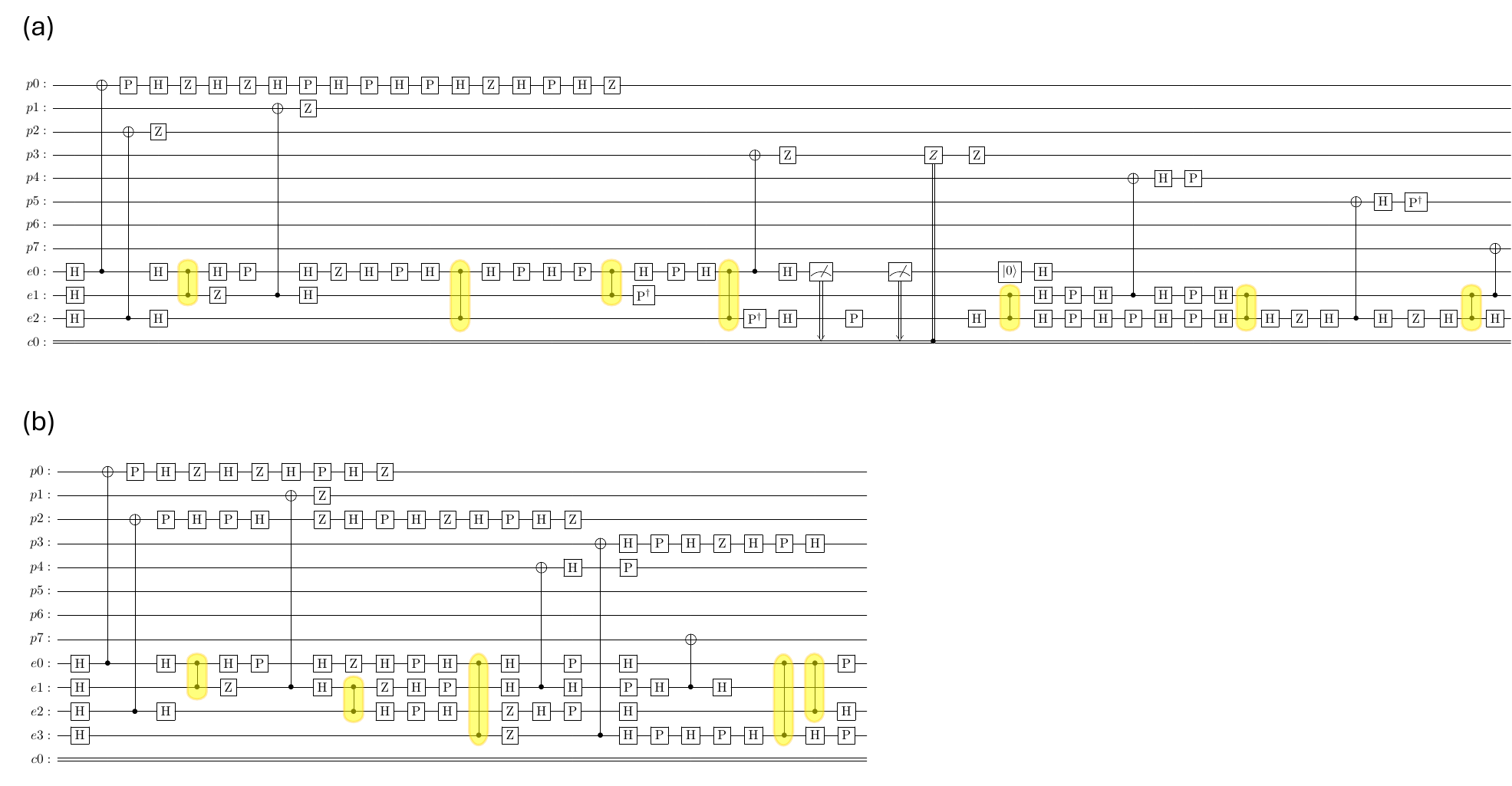}
	\caption{A truncated version of circuits in Fig.~\ref{fig6s}.}
	\label{fig7s}
\end{figure}

\end{document}